\documentclass[%
 aip,
 rsi,
 amsmath,amssymb,
 reprint,%
]{revtex4-2}

\usepackage{graphicx}
\usepackage{dcolumn}
\usepackage{bm}

\usepackage[utf8]{inputenc}
\usepackage[T1]{fontenc}
\usepackage{mathptmx}
\usepackage[colorlinks = true,
            linkcolor = blue,
            urlcolor  = blue,
            citecolor = blue,
            anchorcolor = blue]{hyperref}
\usepackage[pagewise,modulo]{lineno}

\usepackage{academicons}
\usepackage{tikz,xcolor,hyperref}

\definecolor{lime}{HTML}{A6CE39}
\DeclareRobustCommand{\orcidicon}{%
	\begin{tikzpicture}
	\draw[lime, fill=lime] (0,0) 
	circle [radius=0.16] 
	node[white] {{\fontfamily{qag}\selectfont \tiny ID}};
	\draw[white, fill=white] (-0.0625,0.095) 
	circle [radius=0.007];
	\end{tikzpicture}
	\hspace{-2mm}
}
\foreach \x in {A, ..., Z}{%
	\expandafter\xdef\csname orcid\x\endcsname{\noexpand\href{https://orcid.org/\csname orcidauthor\x\endcsname}{\noexpand\orcidicon}}
}

\begin{document}
\preprint{AIP/123-QED}

\title{A setup for grazing incidence fast atom diffraction}

\author{Peng Pan \orcidA{}}
\author{Jaafar Najafi Rad}
\author{Philippe Roncin\orcidB{}}

\affiliation{Universit\'{e} Paris-Saclay, CNRS, Institut des Sciences Mol\'{e}culaires d'Orsay (ISMO), 91405 Orsay, France}

\date{\today}

\begin{abstract}
We describe a UHV setup for grazing incidence fast atom diffraction (GIFAD) experiments. The overall geometry is simply a source of keV atoms facing an imaging detector. Therefore, It is very similar to the geometry of RHEED experiments, reflection high energy electron diffraction used to monitor growth at surfaces. Several custom instrumental developments are described making GIFAD operation efficient and straightforward. The difficulties associated with accurately measuring the small scattering angle and the related calibration are carefully analyzed.
\end{abstract}

\maketitle

\section{\label{sec:intro}Introduction}

\begin{figure}
\includegraphics[width=0.9\linewidth,angle =0,draft = false]{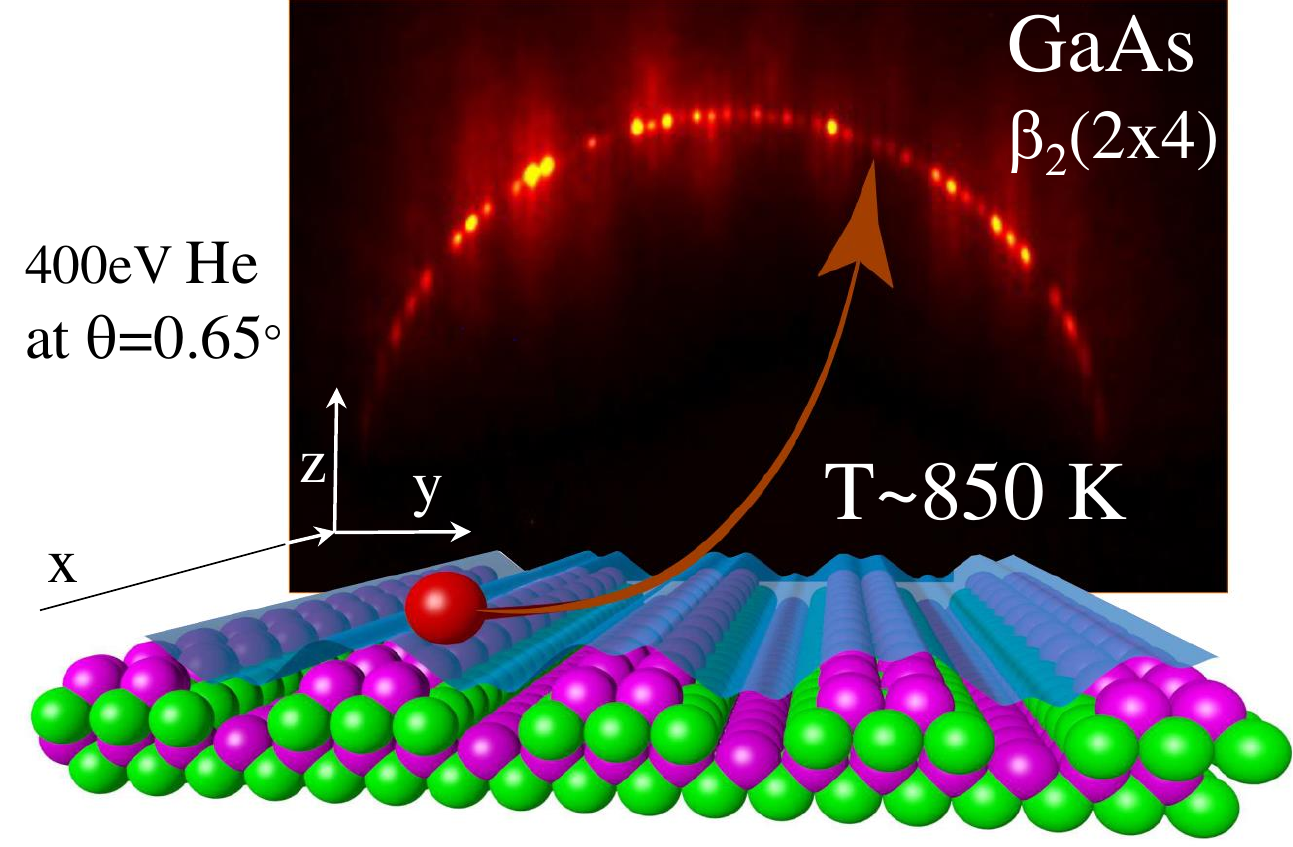}
\caption{Schematic view of GIFAD. The atomic wave of projectile impinging at grazing incidence $\theta$  diffracts along the atomic row of the GaAs target surface (taken from Ref.\cite{Debiossac_PRB_2014}). The image of the scattered particle shows well-defined diffraction spots sitting on a circle associated with energy conservation.
The blue sheet symbolizes the iso-energy surface where the projectile is reflected.}
\label{fgr:schematic}
\end{figure}

Discovered independently in Orsay\cite{Rousseau_2007} and then in Berlin\cite{Schuller_2007}, grazing incidence fast atom diffraction at crystalline surfaces (GIFAD) has developed as a powerful analytic tool in surface science (see Ref.\cite{Winter_PSS_2011, Debiossac_PCCP_2021} for reviews). 
It is a grazing incidence version of thermal energy atom scattering (TEAS), just as RHEED is the high energy and grazing angle version of normal incidence low energy electron diffraction, LEED.
GIFAD provides detailed information on the crystallographic structure of the topmost layer, atomic positions,\cite{Schuller_rumpling} and electron density\cite{Debiossac_PRB_2014,Pan_2021}. 
It does not induce defects or charging of the surface. The decoherence due to thermal motion is reduced, mainly because the average position of the atomic rows has a larger effective mass\cite{Rousseau_2008} than the constituting atoms allowing operation at elevated temperatures, as illustrated in the schematic view in Fig.\ref{fgr:schematic}.
Together with the grazing geometry leaving the volume above the surface free for evaporation cells, GIFAD is well suited for molecular beam epitaxy (MBE)\cite{Debiossac_2016}. It can also count the number of deposited layers \cite{Atkinson_2014}.
GIFAD is a flexible technique that demonstrates diffraction on metal surfaces\cite{Bundaleski_2008,Schueller_2009}, semiconductors\cite{Khemliche_2009,Debiossac_PRB_2014}, and insulators \cite{Busch_2012,Busch_2014,Seifert_2016}.
It also provided clear diffraction from mono-atomic oxide \cite{Winter_2009,Seifert_2016,Busch_2014} or molecular layers grown \textit{in situ} \cite{Seifert_2013,Momeni_2018} and immediately reveals the moir\'e modulation of a deposited graphene layer \cite{Debiossac_graphene,zugarramurdi_2015}.
It is completely insensitive to electric or magnetic fields.

\begin{figure*}[ht]
\centering
\includegraphics[width=0.95\linewidth,trim={0 8mm 0 0},clip,angle =0,draft = false]{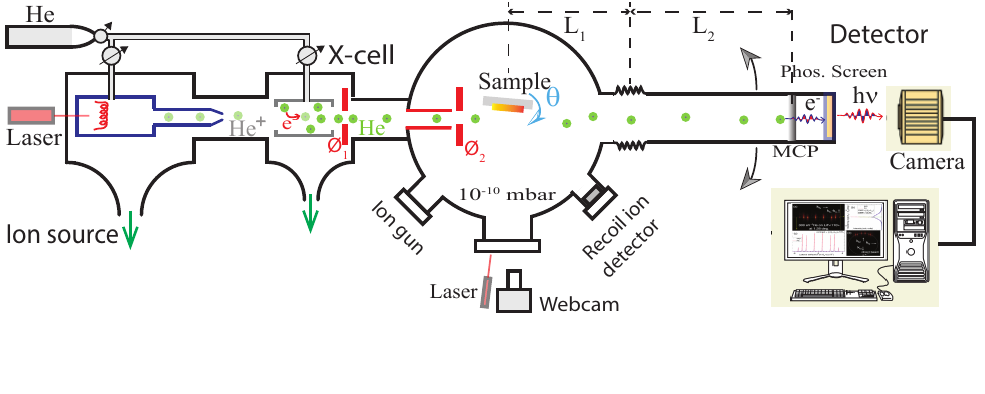}
\caption{ Sketch of the GIFAD setup, with four main parts: ion source, neutralization cell, UHV chamber, and detector system.}
\label{fgr:setup}
\end{figure*}

The general GIFAD setup is presented in Fig.\ref{fgr:setup}, it consists of an ion source combined with a neutralization cell, two divergence limiting slits or diaphragms, a vacuum chamber holding the sample surface, and an imaging detector.
All these elements will be discussed in principle and in view of practical implementation, trying to evaluate what new idea worked efficiently or turned out to be problematic or useless. 
Sec.\ref{sec:det_atom} is entirely devoted to the detector system and the corrections of its aberrations.
The different modes of operation, $\theta$-scan, $\phi$-scan, $E$-scan, $T$-scan, and time-scan are described in sec.\ref{sec:procedure} to \ref{sec:quality} to illustrate typical applications.
In sec.\ref{sec:Goodies}, we discuss additional equipment's taking benefits from the ion or atom beam together with the imaging detector inside the UHV chamber. We also discuss the use of miniature pointing lasers to help face problems specific with small angles geometry.
\section{\label{sec:ion_source} The ion source}
As will be detailed below, GIFAD requires an atom beam injected through tiny diaphragms to reduce its divergence.
In other words, the brightness is most important and, \textit{in fine}, it requires a limited energy dispersion of the ion source. This later is limited by the ionization mechanism but also by the geometry and intensity of the extraction field.
The maximum current is expressed in $\mu$A corresponding to $\sim$ 5 $\times$ 10$^{12}$ ions per second while only a few 10$^{3}$ - 10$^{4}$ atoms per second, but having the best properties, will be needed to record a nice diffraction image in a few seconds.
\subsection{\label{sec:EX05} Filament ion source}
Hot filament ion sources have the reputation of providing bright ion beams in the keV energy range, as required for spatially resolved secondary ion mass spectroscopy (SIMS) or depth profiling \cite{Depth_profiling} applications.
We used the EX05 model from VG Ortec with a differential pumping port allowing a beamline vacuum of better than $5 \times  10^{-7}$ mbar.
The EX05 is equipped with two electrostatic lenses and deflectors, providing flexible adjustment of the beam size, position, and focal distance.
Note that a specific model EX05-N was developed for SIMS or AES with a neutral beam option, but it is not commercially available anymore.
We also tested a non-sequitur ion gun \cite{nonsequitur}, also designed for SIMS applications, which significantly increases beam intensity but does not offer co-axial optical access.
Most ion guns designed for focused ion beams and therefore equipped with ion optics should provide alternate solutions.

\subsection{\label{sec:Polygon}  ECR ions source}
Electron cyclotron ion sources (ECR) use a magnetic structure and microwave to heat a plasma.
They do not have a filament and can offer very long continuous operation and the ability to ionize any material either introduced directly as a gas or as a bulk material indirectly heated inside the source. 
GIFAD was initially discovered on a setup designed with V.A. Morosov from IPM Moscow\cite{Morosov_1996} to investigate grazing incidence scattering of ions on surfaces. 
It was rapidly equipped with a 10 GHz, NanoGan\cite{Pantechnik} Electron Cyclotron Resonance (ECR) ion source  designed to deliver $\mu$A of Ar$^{8+}$ ions.
It operates at low pressure (P$\sim$10$^{-7}$ mbar), and the ionization efficiency is so significant that it acts as an ion pump with the vacuum inside the source decreasing significantly when the extraction field is turned on.
Many different charge states were present in the source, so a high-performance magnet was needed to select the ion charge state of interest resulting in a comparatively long beamline.
We also used a high pressure (P$\sim$10$^{-3}$ mb), low power TES-40 ECR ion source from Polygon Physics\cite{Polygon_Physics} to produce beam of atomic or molecular ion beams of hydrogen.

\subsection{\label{sec:Wien} The Wien Filter}
A mass filter is mandatory when operating the ECR ion source with molecular gas such as H$_2$ to produce H$^+$ or H$_2^+$ ions, but we did not use it when working with high purity noble gas from the hot filament ion sources.
We used a commercial Wien filter from non-sequitur-technologies where the permanent magnet is placed outside the vacuum and can be removed, for instance, for baking purposes.
For a more compact setup, a custom Wien filter taking advantage of the narrow collimation needed for GIFAD can probably be designed.

\section{\label{sec:CXC} The Charge exchange Cell and Beam collimation}
\begin{figure}
\includegraphics[width=0.95\linewidth,angle =0,draft = false]{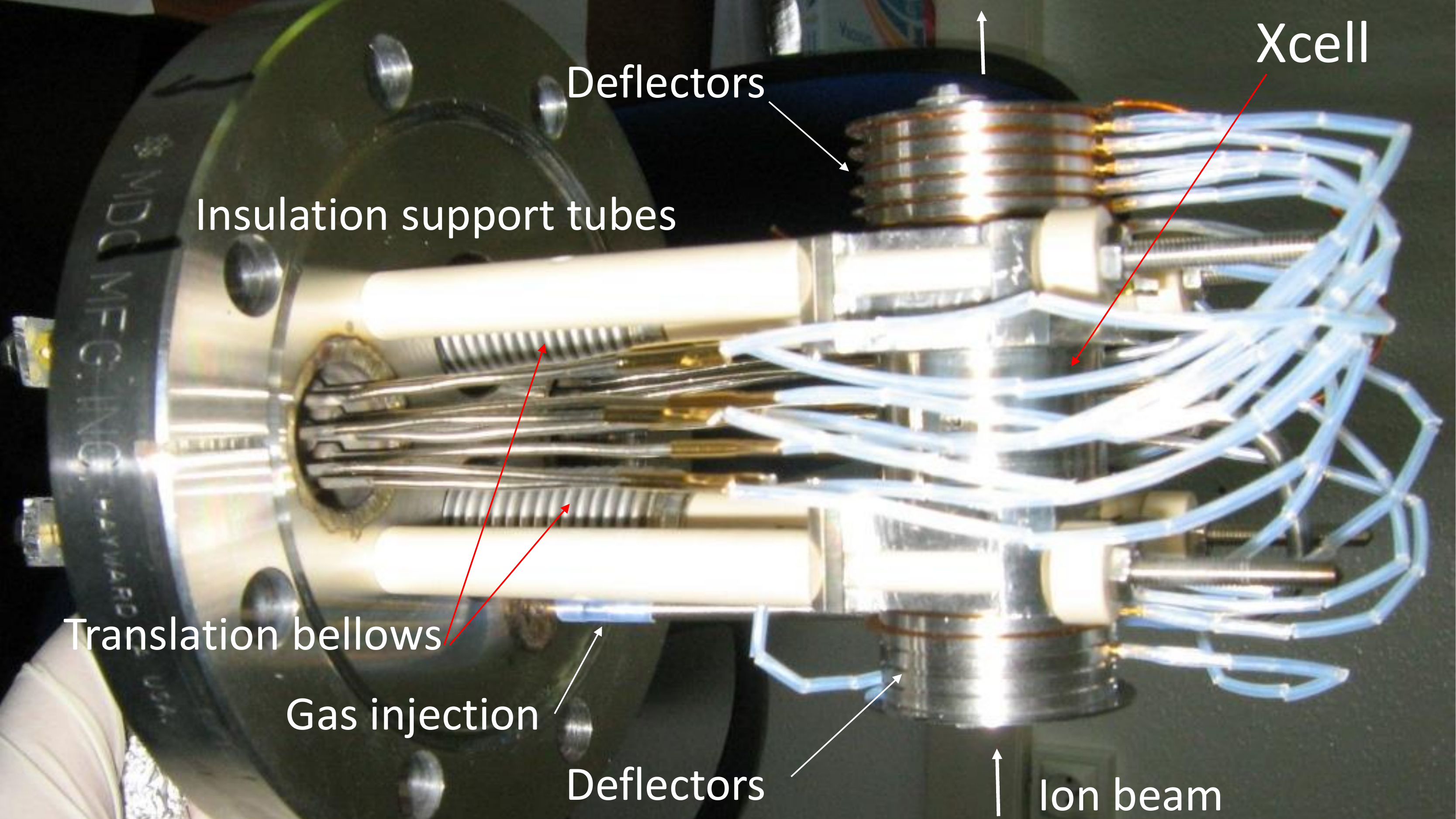}
\caption{\label{fgr:Cell} The neutralization cell in the center is surrounded by two sets of two horizontal and vertical deflectors spot-welded on stacked disks separated by kapton foils. The two bellows are to select and position the entrance and exit diaphragms. Photo taken from \cite{soulisse_2011}}
\end{figure}

The charge exchange cell is a $l=2$ cm long tube with entrance and exit holes designed, so that the internal pressure can be adjusted in the $10^{-3}$ mbar range without a dramatic increase of the background pressure in the beamline.
For resonant neutralization of 1 keV He$^+$ ions on helium, a total cross-section $\sigma \approx 10^{-15}$ cm$^2$ was measured together with a mean scattering angle around 0.1 deg at 1 keV\cite{Gao_1988}.
The optimum pressure is calculated in such that single collisions dominate, i.e. such that 
the probability $P=\sigma n l$ for an He$^+$ ion to capture an electron from the target gas of the cell is around 10\% yielding a target density of $n\approx 5 \times 10^{13}$ particle per cubic cm, \textit{i.e.} a pressure inside the cell around $5 \times 10^{-4}$ mbar \footnote{The same value of the cross-section probably limits the maximum operating pressure of GIFAD around $10^{-4}$mb for diffraction, but higher pressures are possible for triangulation}.
The cell is closed at both ends by fixed holes $\approx$ 1 mm in diameter. 
In front of these holes, a linear translation sealed with a hydro-formed bellows brings a series of 5 fixed pinholes in clear view of the coarse 1 mm hole.
The entrance hole is often left at maximum diameter, while the exit pinhole called $\oslash_1$ is used as the first divergence limiting diaphragm.
The second one $\oslash_2$ is placed at a distance L=0.5 m downstream, at the end of an injection tube, prolonging the beamline inside the UHV chamber until a few cm before the target.
The tube also ends with a coarse $\oslash$= 1 mm hole, ensuring differential pumping. Another system parading a series of 12 pinholes made by electro-erosion is positioned on a miniature DN10CF rotation feed-through.
For this system, where more room is available, the pinhole sizes defining $\oslash_2$ range from 10 $\mu$m to 200 $\mu$m by relative steps of $\sqrt{2}$ allowing an area change by successive factor two, six additional vertical and horizontal slits are also present to allow sheet-shaped beams either parallel or perpendicular to the surface.
For the circular holes, the resulting angular definition is $\delta\theta = (\oslash_1 +\oslash_2)/2L$ that can be adjusted down to 2$\times$ 10$^{-4}$ rad $\approx$ 1 mdeg.
The transverse energy spread of a beam of energy $E_0$=1 keV is $E \delta\theta^2$ which can be as low as 0.5 meV, measured as the width of bound states resonances on the LiF surface \cite{Debiossac_PRL_2014}. 
Needless to say that the intensity passing through both diaphragms is severely reduced.
For most experiments, we use diaphragms between 20 and 50 $\mu$m.
The pinholes are separated by a distance of 1.2 mm, corresponding to an angle of $\delta\alpha$ =2.7 $^\circ$, which is very difficult to read accurately on a miniature rotary drive. We had to implement a more accurate electronic readout using a tiny magnet coaxial to the rotary drive so that its absolute angular position can be read without contact by an AS5140 magnetic sensor. It is read by an Arduino-mini micro-controller updating a 4-digit LED display.
Rather than the absolute angular position $\alpha$ in degree, we chose to display it as a four-digit float Pos=$(\alpha_i - \alpha_0)/\delta\alpha$ where the integer part identifies the pinhole number and the remaining is the actual position of the pinhole in front of the $\oslash$=1 mm diameter.
For instance, a value of Pos=10.12 display indicates the diaphragm number 10, but positioned 0.12 pinhole separation units above the center of the coarse hole.
Note that the factional value is important since the exact position of the pinhole affects the actual beam position.
Fig. \ref{fgr:Cell} shows the charge exchange cell designed to be mounted on a 63CF flange by an insulator, so that it can be floated to change the energy of the ion beam before charge exchange without de-tuning the Wien filter placed before. This allows an ion beam transport at keV energy and deceleration immediately before neutralization.
On both sides of the gas cell, two complete sets of deflectors are positioned co-axially on stacked 2 mm thick plates separated by thin Kapton disks.
Each plate has a 0.8 mm hole drilled on its edge, which serves as a female connector for UHV twist pin connectors having their own spring.
The deflectors before the cell were intended to steer both the position $(y,z)$ and direction $(\theta_y,\theta_z)$ of the ion beam before entering the cell, but in practice, mainly the angular control is used.
The ones placed after the cell do not affect the neutral beam and are used to deflect the residual ion beam away.
Additionally, when connected to a voltage pulse generator, these deflectors can produce a pulsed ion beam useful for surface contaminant analysis, as detailed in section \ref{sec:DirectRecoil}.
Also, it is important to be able to misalign the ion source so that neutral atoms produced at the extraction level and having ill-defined energy, do not contribute.
This can be achieved by forcing mechanically the beamline to be tilted by $\sim$ 1$^\circ$ or by positioning the diaphragms slightly off-axis until this signal disappears and then steering the ion beam inside the cell and experimenting with deflectors. 

As described above, the pressure inside the charge exchange cell lies in the 10$^{-3}$ mbar range, the entrance, and exit diaphragms limit the gas flow so that the pressure just around the cell is below a few 10$^{-7}$ mbar.
The coarse 1 mm hole connecting to the UHV chamber limits the variation of the pressure inside the UHV chamber below 10$^{-9}$ mbar. 
If an additional 60 $l/s$ pump and a coarse 1 mm hole is placed in between, the pressure of a few 10$^{-10}$ mbar inside the UHV chamber is not at all affected by the gas introduced in the ion source or in the gas cell. On the high-pressure side, GIFAD is probably limited around because the mean free path for elastic scattering.
All the turbo-pumps of the beamline and introduction chamber are connected to a single dry primary pump.

\section{\label{sec:manipulator} The UHV manipulator}

GIFAD is a diffraction technique and therefore requires accurate control of two angles, the angle of incidence of the atomic beam $\theta$ and the orientation of the surface $\phi$.
More precisely, $\theta=\pi/2-\angle (\vec{k}_{in}, \vec{S})$ and $\phi=\pi/2-\angle (\vec{k}_{in}\times\vec{S},\vec{u})$, where the vectors $\vec{k}_{in}$, $\vec{S}$ and $\vec{u}_{h,k,0}$ indicate respectively the beam direction, the surface normal and the surface crystallographic axis labelled by its miller indices $h,k$. 

\begin{figure}
\includegraphics[width=0.80\linewidth,draft = false]{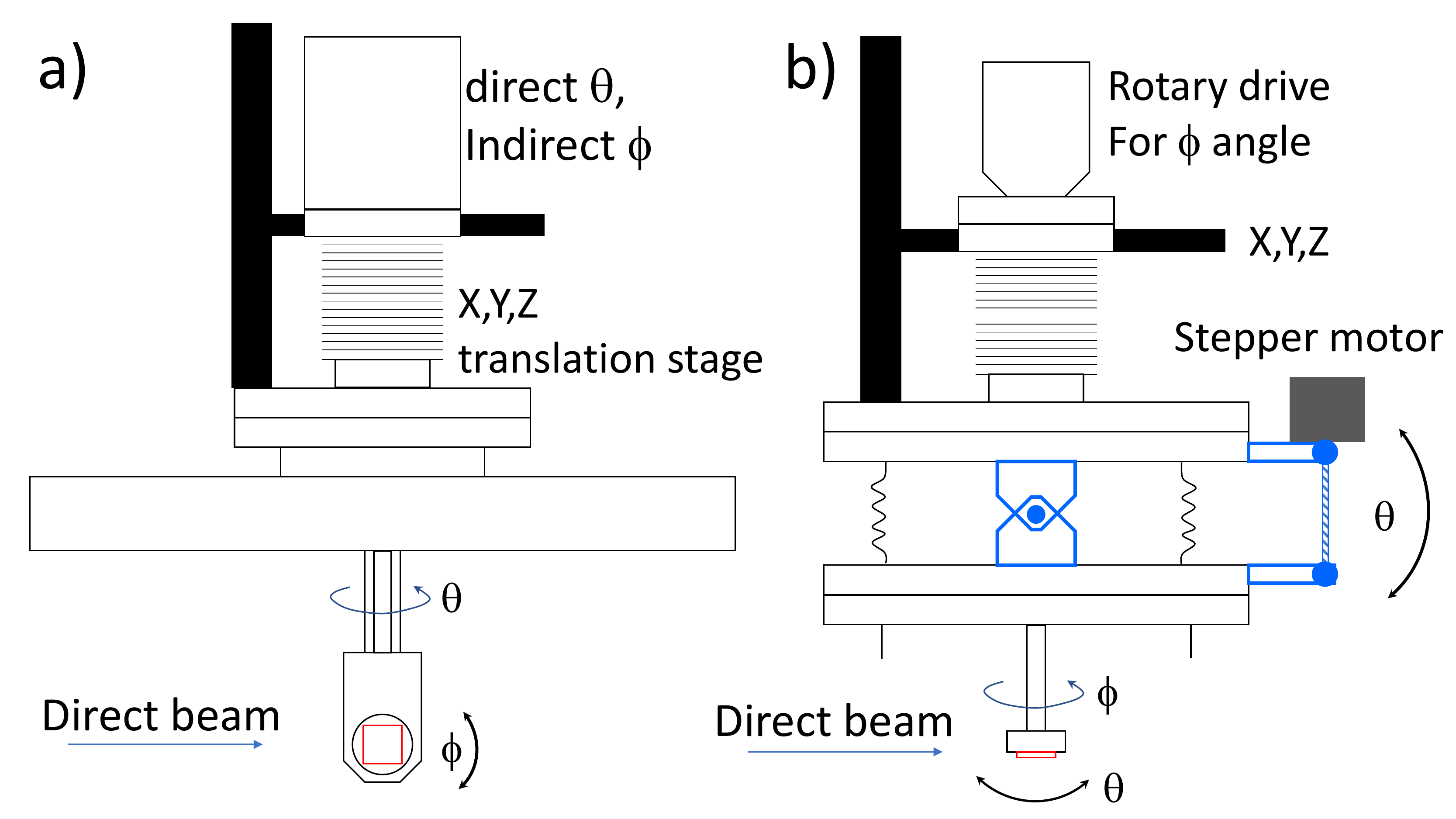}
\caption{\label{fgr:manipulators} Two different manipulators have been used, both fitted on a X,Y,Z platform. a) has two coaxial rotations one direct for $\theta$ and one indirect for $\phi$. b) has only one rotation assigned to $\phi$, the parts in blue have been welded to a DN100CF bellow so that $\theta$ is controlled by changing the parallelism of the flanges.}
\end{figure}

A simple manipulator is characterized by a holding flange with a three-axis $X,Y,Z$ translation stage holding a primary rotation axis perpendicular to this flange and terminated by a sample holder.
More sophisticated models support a co-axial translation or rotation mechanism to perform additional movements, in our case an additional rotation.
We have used two distinct configurations sketched in Fig.\ref{fgr:manipulators}, where the high precision primary rotation axis is affected to the angle of incidence $\theta$ or to the azimuthal angle $\phi$.
In Fig.\ref{fgr:manipulators}a), the additional rotation involves an in-vacuum gear system allowing unlimited $\phi$ movement whereas in Fig.\ref{fgr:manipulators}b) a separate $\pm$ 10 mm movement of the support flange translates into a variation of $\pm$ 4$^\circ$ of the angle of incidence.
The accuracy is excellent but a modification of $\theta$ is associated with a significant shift of the target position mainly along the $x$ direction affecting the angular calibration.

This is not the case with the manipulator MC from UHV-design in Fig.\ref{fgr:manipulators}a), in the vacuum azimuthal movement is not precise enough for crystallographic applications, there is no announced accuracy but a claim for a "reproducibility" of 0.2 deg without any clear definition!

When operating inside a MBE chamber at INSP \cite{Debiossac_PRB_2014, Atkinson_2014, Debiossac_2016}, the local manipulator was equipped with complete azimuthal freedom but limited control of the angle of incidence relative to the beam, so we decided to tune the angle $\theta$ by changing the beam direction.
The ion source and neutralization cell were placed on a motorized bench connected to the MBE chamber by a UHV valve and a flexible bellow.
The combined control of two motors, one placed close to the ion source and the other one after the neutralization cell was used to create a virtual center of rotation at the center of the target wafer in the MBE chamber.
One of the main interests of GIFAD is its compatibility with the MBE environment where it can track online the transformation of the topmost layer with temperature as well as the growth mode of the grown layers and  their specific crystalline structures.
In this respect, a sample holder having the ability to bring the sample at elevated temperatures or to low temperatures was mandatory. 
In studying thermal effects in inelastic diffraction on a LiF surface we could explore temperatures between 140K and 1017K \cite{Pan_2022} where the limitation is given by the noise on the detector, probably due to electronic and ionic emission from the filament.

\section{\label{sec:det_atom} The detector system}

\subsection {Detecting neutral atoms}
Detecting a neutral atom at thermal energy is not straightforward, it carries, in average only $E=kT$=25 meV kinetic energy and is not able to extract an electron. 
For helium, the only solution is to ionize it, and since the ionization potential is 24.58 eV, this is achieved efficiently only by electron impact.
The process is not selective, and ionizes also the residual gas so that helium atom detection is most often associated with an additional mass selective procedure, for instance, a quadrupole mass spectrometer.
This is a major limitation of TEAS where angular resolution requires the use of small apertures, whereas analyzing several diffraction orders requires scanning the detector or sample over a large angular range.

The situation is much more favorable at keV energies where the atoms can be detected with a quantum efficiency above 10\% \cite{Jagutzki_2018} because of the Pauli principle.
When a keV atom tries to penetrate a solid, atomic collision with surface atoms takes place at close internuclear distances so that the overlap of the compact helium ground state electrons will push away the valence electrons of the target atom.
The electrons ejected from the surface can be accelerated and "multiplied" until detection.
Here multiplication relies on the fact that an electron with a few hundred eV energy impacting specific materials will result in the emission of, in average, more than two secondary slow electrons.
These are the processes at work in a photomultiplier tube where after impacting ten plates a burst of a few $10^{7}$count electrons is produced in less than a ns producing a voltage peak of a few tens of mVolts on a 50 $\Omega$ resistor.
In a channeltron, all these $\approx$ 10 successive electronic impacts occur along the tube because it is curved while in micro-channel-plates (MCP) they occur inside each pore because of it very long aspect ratio $L/\phi_{pore}\approx 50$ where $L$ and $\phi_{pore}$ are the pore length and diameter respectively. 
The gain of a single tube is limited to a few thousand electrons, but the use of two staked MCP allows a gain of a few $10^{6}$ electrons per impact.

\subsection{\label{sec:detector} Imaging Detectors}
Imaging detectors are particle detectors with the ability to localize the $(x,y)$ coordinates of the impact of the particle within the detector input area.
This is comparatively simple with MCP, where each pore can be seen as a channeltron tube providing millions of independent detectors. 
Knowing exactly which pore has been hit is possible but still difficult. Several compromises have been proposed (see e.g. \cite{Lapington_2004}).
One approach is to use dedicated electrodes and electronics, and this requires the use of two stacked MCP because the gain of a single MCP is limited by space charge to a few $10^{3}$ electrons.
The two most popular families are either charge division techniques such as continuous resistive anodes \cite{Rousseau_2007} which provide good differential linearity but poor integral linearity, or delay line anodes allowing larger count rate and high resolution \cite{Schuller_2007} and integral linearity, but with a limited differential nonlinearity requiring specific calibration \cite{Hong_2016}.
Even without impact, the detector may trigger randomly, but this can be limited to only a few dark events per second randomly distributed on the MCP active surface so that the signal to noise ratio is usually very large.
The other option we have adopted is to accelerate the electrons of the MCP cloud onto a phosphor screen and image the fluorescence with a camera\cite{Lupone_2018}.
This combination of MCP and phosphor screen, developed initially for military night vision systems, is technologically mature and  offers excellent performances in terms of count rate, spatial resolution, and uniformity as detailed below.
We use a single 80mm MCP placed 1mm ahead of a standard 3 mm thick glass plate coated with an ITO layer and a 4 $\mu m$ P43 Phosphor over-layer.
The light conversion efficiency is around 20 eV of electron energy per emitted photon so that a 2 kV bias between MCP and screen allows each of the $10^3 -10^4$ electrons of the MCP to produce $\approx$ 100 green photons, enough to detect a single impact with a proper lens and camera (see next section).
The screen is stable under UHV conditions and can resist temperatures up to 200 $^{\circ}C$ without losing performance.

We did not investigate carefully the problem of MCP uniformity hoping that our $\sigma =50 ~\mu$m spot size covering at least ten adjacent pores ($\approx (\sigma/\phi_{pore})^2$) would average single pore defects.
We did not experience dark spots or gain variation except at the Molecular Beam Epitaxy chamber, where exposure to Gallium and Arsenic together with the helium impacts was suspected to be responsible for reducing the efficiency in the central zone of the detector.

\subsection{\label{sec:aberration} Position, shape, and intensity aberrations of the lens}
\begin{figure}
\includegraphics[width=0.8\linewidth,draft = false]{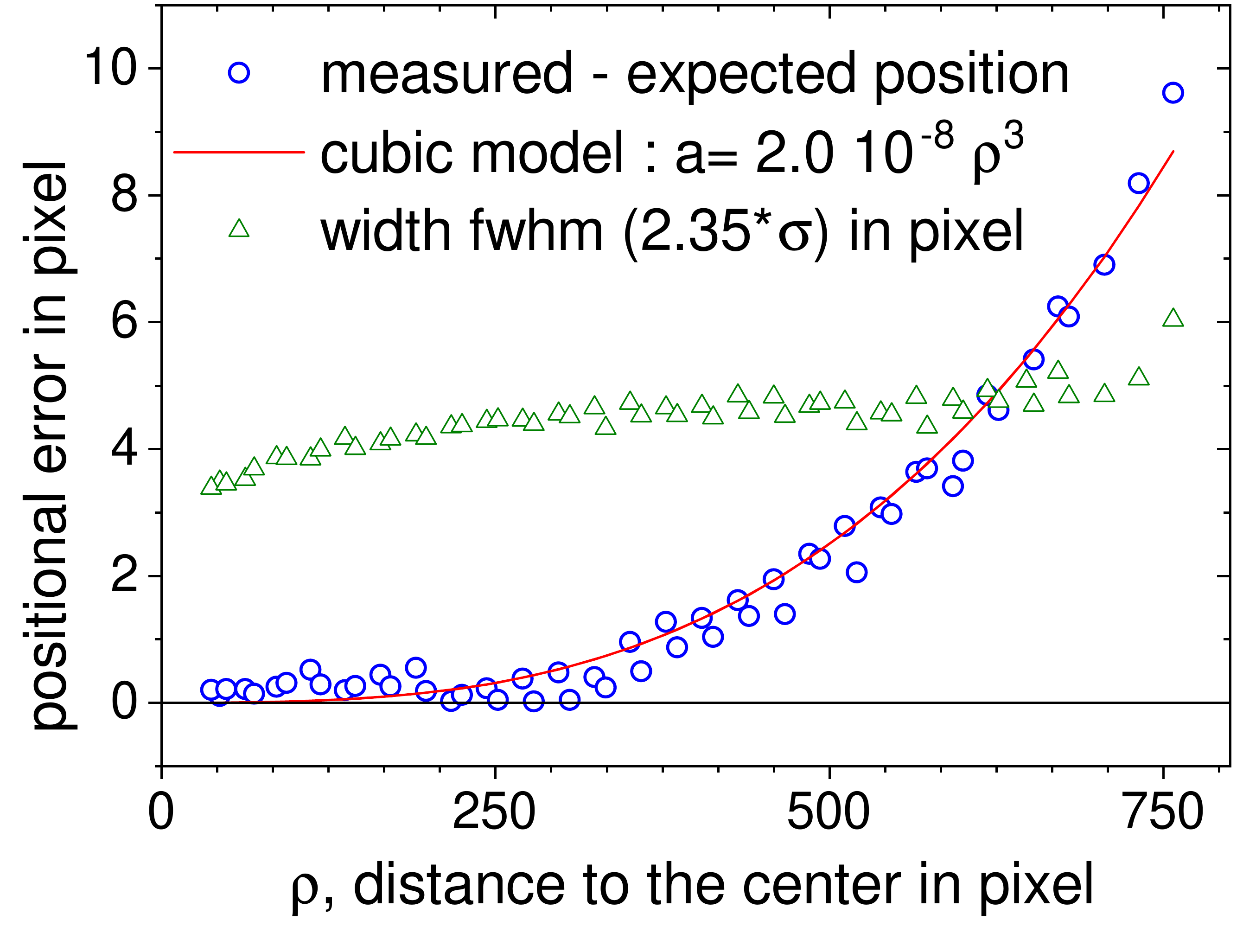}
\caption{\label{fgr:aberration} The lens aberration, measured as the spot position on the camera minus its actual value measured with a caliper is well fitted by a cubic coefficient $a$ (eq.\ref{eq:Lens0}). The spot width is approximately constant. The intensity was found to decrease with a quadratic character.}
\end{figure}

Measuring accurately the scattered positions and intensities is the basis of any diffraction techniques but this is crucial in GIFAD because of the small values of the scattering angles. As an example, the peak separation in Fig.\ref{fgr:schematic}is around 0.05 deg. 
We analyse here the distortions introduced by the lens system imaging the phosphor screen through a UHV window at a distance of 25 cm.
The lens is a high aperture F0.95 "Xenon"  lens from Schneider \cite{Schneider} with a focal lens of 17 mm so that a spot at the edge of the detector hits the lens with a maximum angle below 10 deg, it is operated with its diaphragm wide open.
This choice of the lens and camera position derives from a compromise between light collection efficiency and optical aberrations, we use the pixel of our (1920x1440) digital camera as a unit.
The aberration of the lens has been tested \textit{in-situ} by imaging a fixed beam at different detector position and \textit{ex-situ} by scanning a light spot or imaging a rectangular grid.
We determined the correspondence between actual and measured beam position and track the evolution of the Gaussian width $\sigma$ and the radial asymmetry parameter or skewness displayed in fig\ref{fgr:aberration}.
The defects can be neglected in the central (paraxial) region but can reach almost half a millimeter error in the spot position at the edge. 
This error is easily corrected by a single parameter describing a cubic barrel-type aberration of eq.\ref{eq:Lens0} where $\rho$ and $\rho_c$ are the measured and corrected radial coordinates or distance to the center of the sensor expressed in pixel and $a$ is the measured coefficient. 
Within present settings, the maximum distance to the center can be close to 800 pixels yielding an error of ten channels. 
After this cubic correction, each pixel corresponds to its ideal location on screen with an error less than a pixel ($\approx~50~\mu$m with our magnification) over the whole screen.
The spot width was found approximately constant while the skewness evolved linearly from -0.25 to 0.25 across the whole diameter.

Eq.\ref{eq:Lens1} reports the correction of the measured intensity deriving from the Jacobian of the position correction in eq.\ref{eq:Lens0}.
Finally, the eq.\ref{eq:Lens2} reports the quadratic correction compensating for the vignetting aberration describing the radial intensity decrease, $\sigma_v$ is the measured range in pixel units, indicating a $\sim$25\% drop at the position corresponding to the edge of our detector.
With these simple polynomial radial forms, the position and intensity corrections are performed on the fly in our analysis software.
  \begin{align}
 \rho_c&=\rho (1+a\rho^2)   &a=2.0 \times 10^{-8}  \label{eq:Lens0}\\
  I_c&=I/(1+4a\rho^2)  \label{eq:Lens1}\\
   I_v&=I_c (1+\rho^2/2\sigma_v^2)    &\sigma_v=1080 \label{eq:Lens2}
 \end{align}
\subsection{\label{sec:Camera} The camera aberrations}
The other aberration observed is related to the camera itself, in our case a Hamamatsu C11440 ORCA Flash, CMOS. 
We have noticed that the image of a sharp spot is indeed a quasi-delta limited by the lens resolution, but this delta is sitting on top of a broad circular pedestal decaying slowly from center to an edge with an almost 50 pixel radius. 
Its magnitude immediately at the foot position is only a fraction of a percent, but its integrated intensity is close to 25\%. 
This was not anticipated since we switched from CCD to CMOS technology to avoid the blooming and charge leaking problems.
The observed pedestal does not depend on the exposure time, so we believe that it is due to light scattering in the micro lenses covering each pixel to achieve a very large light collection efficiency.

The problem could be corrected by software but we decided to operate the camera in the single-particle detection mode, which requires that each impact produces enough photons to form a detectable spot in the image. 
To optimize this mode, we raised the MCP bias voltage to 1100V so that each atom impact generates an estimated number of electrons close to $10^4$, each one producing almost 100 photons on the phosphor screen (P43 spec is 20 eV electron energy per emitted photon). Only 0.04\% of these $10^6$ photons will be collected by our lens for each impact but this is significantly above the background level.
The HiPic camera software provided by Hamamatsu offers a so-called "Photon counting mode" perfectly suited for this mode of operation.
In each image, the spots are searched, and a centroid detection is performed. 
Each location were the spot has been detected with an intensity above a given threshold is filled with a one while other pixels are set to zero so that the camera behaves like a single particle detector providing a valuable instant count rate.
The detector also becomes insensitive to moderate stray light because a single photon on the screen is not enough to pass the threshold.
The software embedded inside the camera can sustain such a spot analysis with 10 frames per second without significant frame loss.
The limitation is that the probability to have an overlapping spot should be negligible. 
This implies that the count rate within the surface of a spot is less than 5 to 10 percent per frame.

\begin{figure}
\includegraphics[width=0.7\linewidth,angle =0,draft = false]{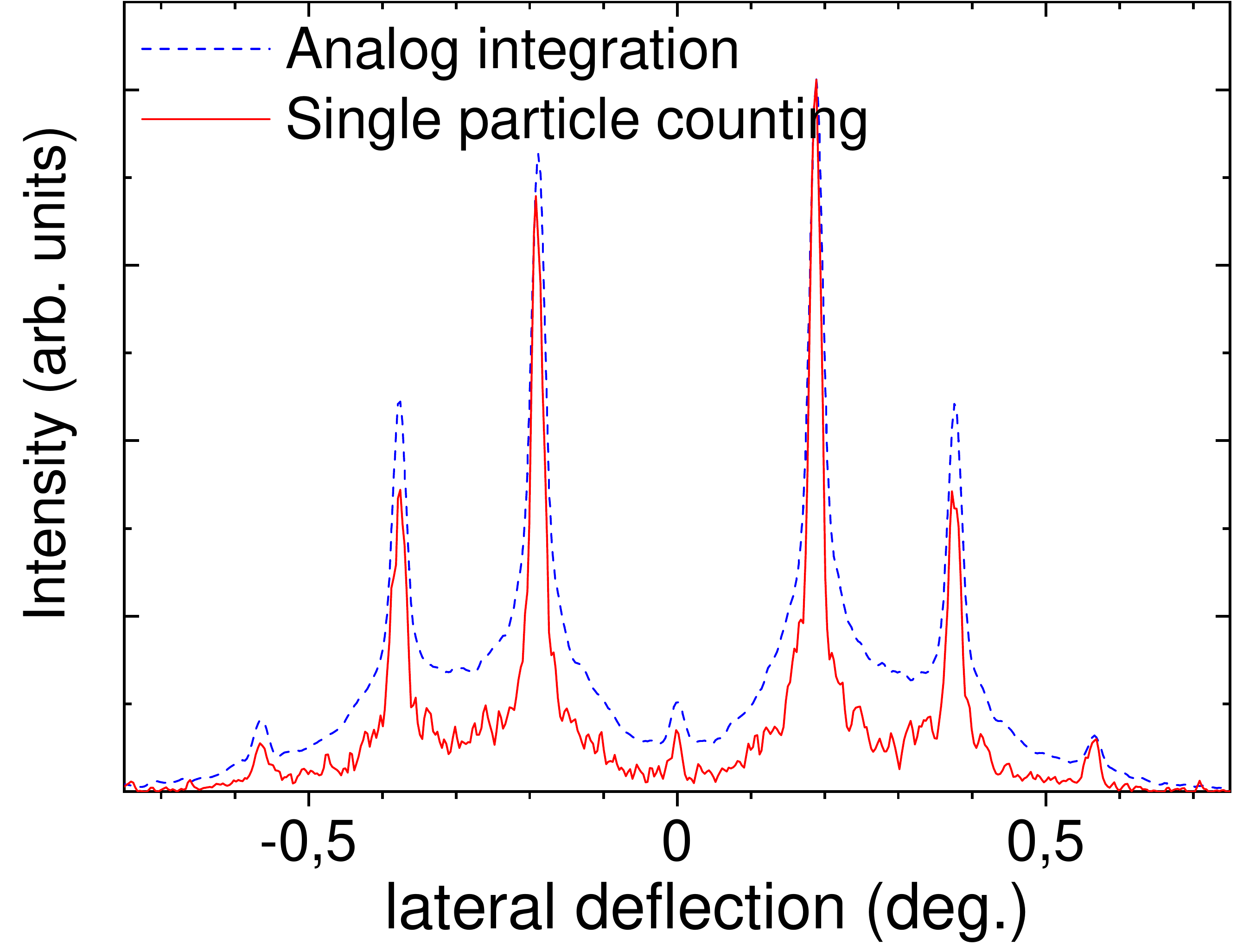}
\caption{\label{fgr:pedestal} Two diffraction profiles recorded on the Laue circle in similar conditions. The dashed blue curve corresponds to the standard analog integration mode while the better contrasted red one is recorded in single particle mode performed directly inside the camera.}
\end{figure}

The improvement in contrast and resolution is illustrated in Fig.\ref{fgr:pedestal} which compares the images recorded in analog and single-particle counting mode. When analyzing the tiny spot of a low-intensity primary beam, the standard deviation $\sigma=2.15$ pixel measured in analog integration mode was reduced to $\sigma=1.65$ in photon counting mode, suggesting a contribution of around 1.5 pixel due to the size of a single-particle. 
This is fully consistent with the value of $\sigma=1.6$ measured from a statistical analysis of a scattering image exposed to 100 ms and where a few hundred individual impacts were visible\cite{Lupone_2018}.
 
\subsection{\label{sec:faithfull} A faithful detector}
The combination presented here of a phosphor screen imaged at medium distance by a high aperture lens corrected by software and a high efficiency CMOS camera operated in the single impact counting mode provide accurate positions and intensities.
It is very close to an ideal detector with high resolution and, most important, an excellent integral and differential linearity.

\section{\label{sec:rotating} Large angle scattering}

\begin{figure}
\includegraphics[width=0.8\linewidth,trim={4cm 2.0cm 6.8cm 5cm},clip,draft = false]{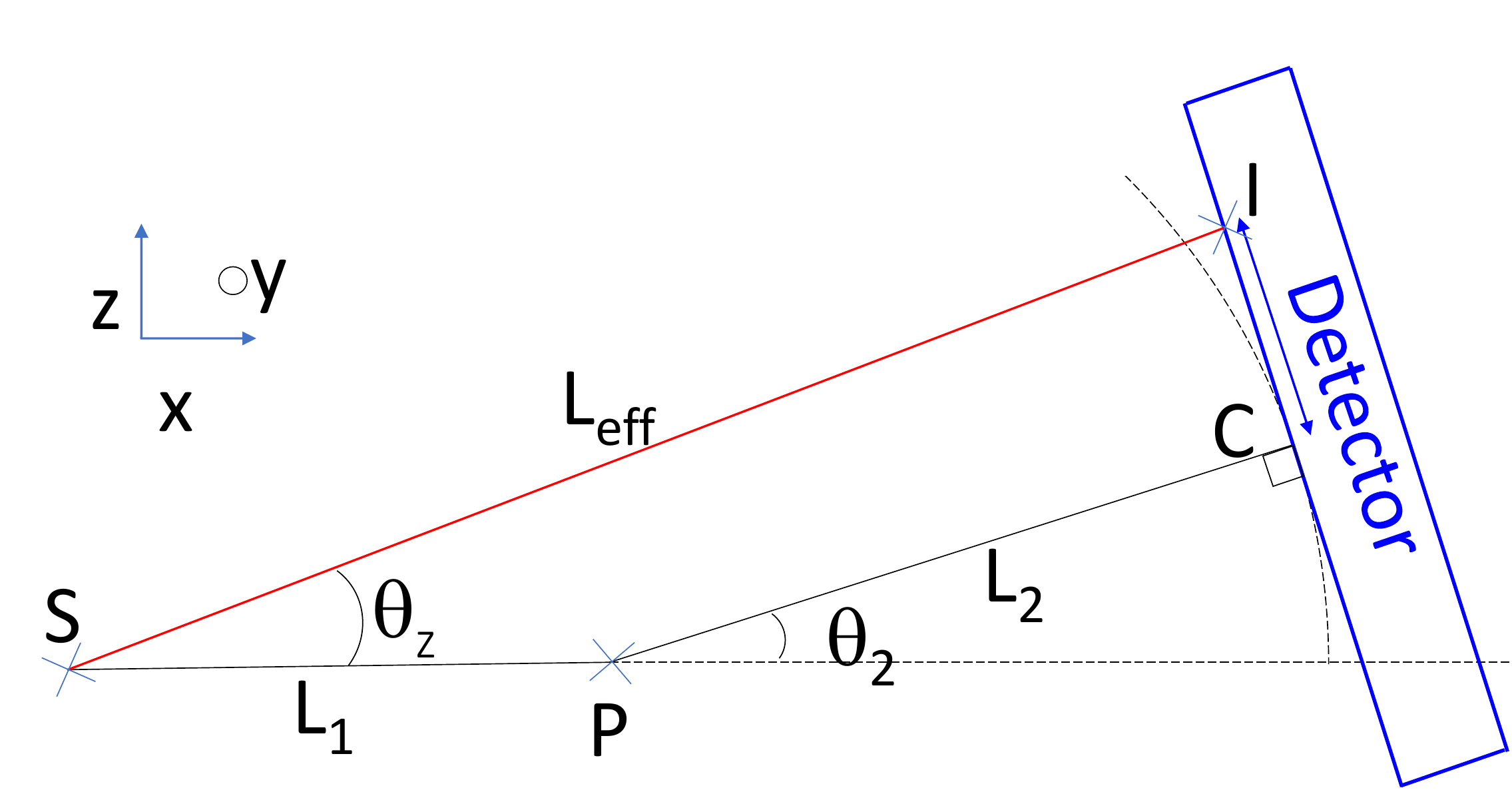}
\caption{\label{fgr:Simulation} The path to the detector is made of two sections (Fig.\ref{fgr:setup}).
A fixed tube of length L$_1$ and a tube of length L$_2$ connected by a below, allowing a relative angle $\theta_2$.
For each impact coordinates $z=IC$, the emission angles $\theta_y,\theta_z$ from the surface are given by eq.\ref{eq:thetaz}}\end{figure}

For beam and detector ports perfectly aligned, our detector\cite{Lupone_2015} with 60 mm visible diameter and a distance to the target center corresponds to a maximum scattering angle of 1.5$^\circ$.
This later could be doubled by giving an angle either of the beam or to the detector, but we have chosen to offer the possibility to explore larger scattering angles simply by putting a commercial DN100CF hydro-formed bellow between the UHV chamber and the tube holding the detector (a DN63CF or DN50CF would have been large enough).
The rotation takes place in the horizontal plane with an axis lying in the middle of the bellow while the detector side sits on a 200 mm long linear rail with motor control and digital caliper.
The camera is attached to the detector flange, and the maximum scattering angle is now close to 4-5 $^\circ$.
Taking the maximum beam energy of 10 keV this corresponds to energy $E_\perp$ of the motion perpendicular to the surface close to 100 eV while the lowest energy $E_\perp$ demonstrated was around 3 meV\cite{Debiossac_PRL_2014}, more than four orders of magnitude lower.
With helium projectile impinging LiF surface, elastic diffraction was observed until $E_\perp \sim$0.5 eV, inelastic diffraction with well-resolved peaks until $E_\perp \sim$1.5 eV and supernumerary rainbows corresponding to unresolved but coherent diffraction peaks until $E_\perp \sim$4.5 eV\cite{Debiossac_PCCP_2021}.
It is likely that quantum effects such as the principal rainbow profile\cite{Miret_Artes_2012} persists above 5-10 eV but that its contribution may be hidden below instrumental resolution.
These comparatively large values bridge the gap between GIFAD and all surface investigations performed with ions \cite{winter_PR2002}. The minimum value of $E_\perp$ was limited above 1 eV due to the image charge attraction before neutralization\cite{Villette_2000}. 
The relation between impact position $(y,z)$ referred to the beam position and scattering angles $(\theta_y,\theta_z)$ is given in eq.\ref{eq:thetaz} and depicted in Fig.\ref{fgr:Simulation}, it is hardly more complex than with a straight tube but requires specific calibrations, as described in Fig.\ref{fgr:Airy}c).

\begin{equation} \label{eq:thetaz}
  \begin{split}
 \theta_z&=atan\frac{z \cos^2 \theta_2 + a}{(z/2) \sin2\theta_2 +a}\\
 a&=(L_2/2)\sin 2\theta_2 \\
  b&=L_1\cos\theta_2 + L_2\sin^2\theta_2 + L_2\\
    L_\text{eff}&=(L_1\cos\theta_2 + L_2)/\cos(\theta_z - \theta_2)\\
    \theta_y&=atan(y/L_\text{eff}) 
  \end{split}
\end{equation}
\section{\label{sec:procedure} Experimental method}
For complete data analysis, it is mandatory to know exactly where the direct beam is located and what is its exact shape. 
This usually requires two separate images, one where the direct beam is reasonably well centered so that the exact shape can be analyzed without distortion. One with the beam is located close to an edge, so that most of the detector is left free to analyze scattered particles.
To avoid damaging the detector with a too intense beam, we have installed a 1\% and a 10\% transmission grid mounted on a translation stage before the detector. 
It is not mandatory when all the diaphragms reducing the beam divergence and intensity are set to a low value, but otherwise, the beam intensity can be significant.
Even with this attenuated beam, the two images of the beam are recorded with a few ms exposure time and are typically recorded only once a day since most of the measurements are associated with variations of the target surface parameters, incidence angle, azimuth angle, or temperature.
As usual for data acquisition with cameras, the signal to noise is improved by recording a reference file without the atomic beam by closing the valve just before the UHV chamber. 
We use the native software to drive the camera and the background file is automatically subtracted from the one recorded with the atomic beam. 
This is particularly important if the atom count rate is low so that long exposures are needed. 
In this case, the nonuniform camera sensitivity and noise would slowly hide the signal. 
These last steps are not needed when the camera is operated in the single-particle detection mode, as described in sec.\ref{sec:detector}.
The data analysis is performed by a homemade software written in Borland C++, allowing customized polar transform, as described in Ref.\cite{Debiossac_Nim_2016} and simple database access to the exact beam position and line-shape associated with each image.
When the interest is on the topology of the potential energy surface, strongly connected to the surface electronic density, only the elastic diffraction is of interest in each image. 
This corresponds to the intensity on the Laue circle of energy conservation clearly visible in Fig.\ref{fgr:schematic}.
It is defined by $k_{in}^2=k_{out}^2=k_y^2 + k_z^2$. 
For convenience, this intensity is transferred onto a straight line by a polar-like transform  \cite{Debiossac_Nim_2016} allowing direct comparison with model surface topology using simplified \cite{Rousseau_2008,Khemliche_2009}  or exact models \cite{Rousseau_2007,Debiossac_2020} to describe the quantum dynamics of the projectile.
This representation on a straight line is also used to plot diffracted intensities as a function of time or angle of incidence or primary energy, as illustrated in the next section.

\section{\label{sec:scans} $\theta$, $\phi$, $E$, $T$, scans and growth monitoring}
\begin{figure}
\includegraphics[width=0.75\linewidth,angle =0,draft = false]{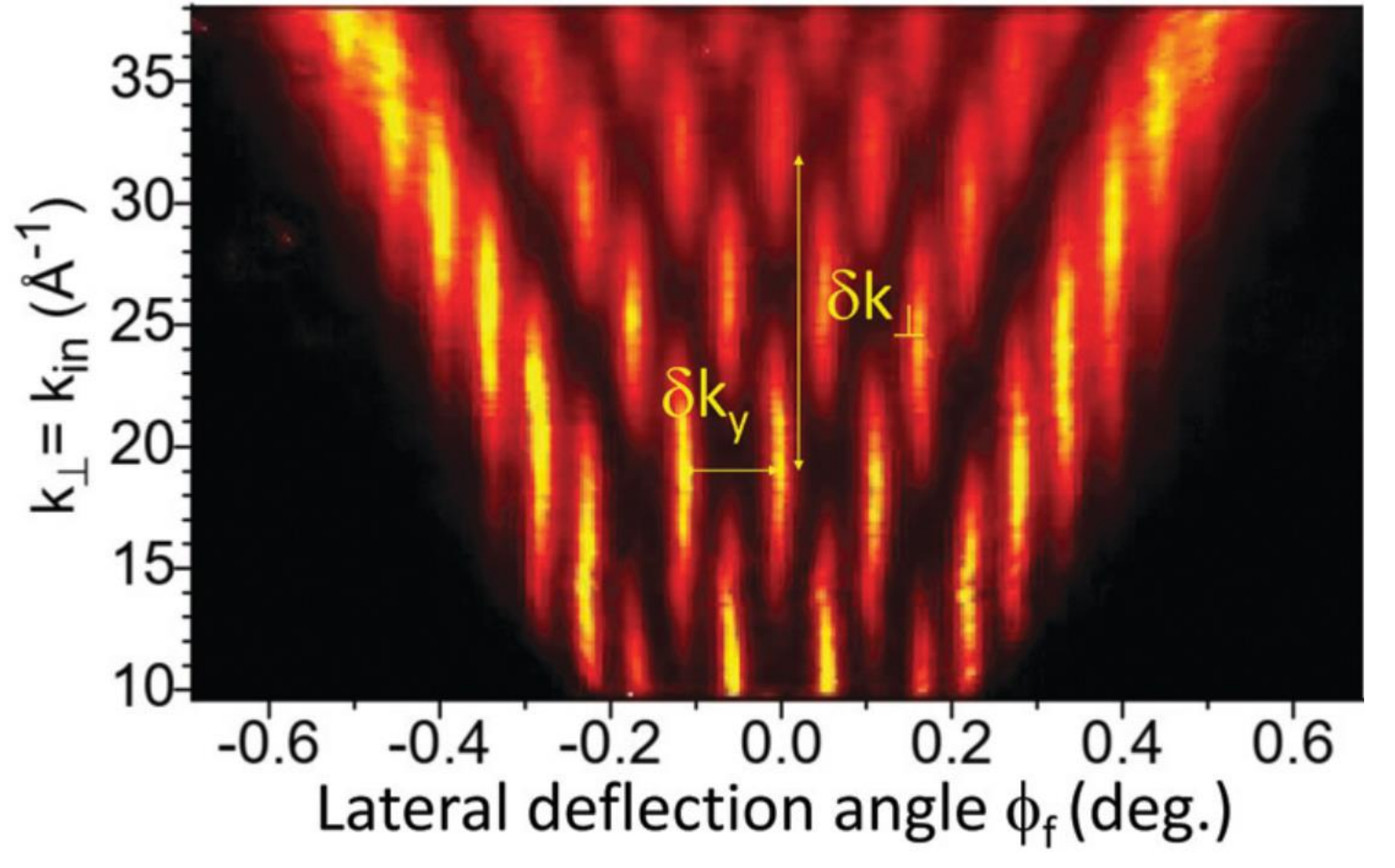}
\caption{\label{fgr:theta_scan} During a variation of the incidence angle $\theta$, the diameter of the Laue circle (Fig.\ref{fgr:schematic}) increases but the spacing between diffraction orders stays fixed.
Here 500 eV Ne projectiles along with LiF[110]\cite{Debiossac_PCCP_2021}.}
\end{figure} 
\begin{figure}
\includegraphics[width=0.78\linewidth,angle =0,draft = false]{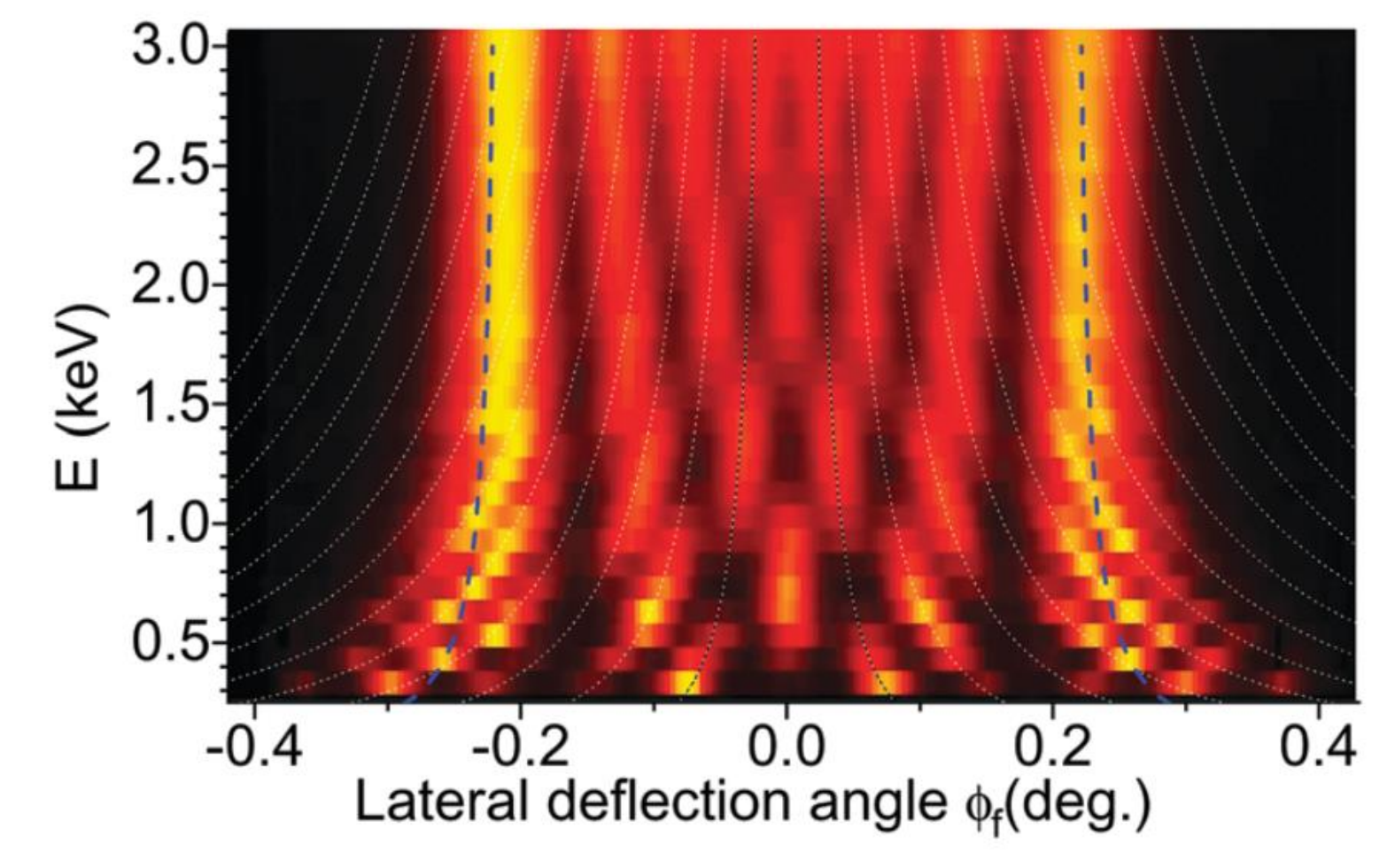}
\caption{\label{fgr:E_scan} During a variation of the primary beam energy $E$ at $\theta$=0.57$^\circ$, the diameter of the Laue circle is fixed while the spacing between diffraction orders varies. The increase of the lateral scattering angle at low primary energy is due to attractive forces (from Ref.\cite{Debiossac_PCCP_2021})}
\end{figure} 
The manipulator is motorized and the automat driving the motors is able to emit a synchronization signal to the camera so that systematic $\phi$-scan, $\theta$-scan can be programmed with small steps without close surveillance. 
For instance, a $z$-scan can be programmed to search the surface region that produces the best diffraction image.
Fig. \ref{fgr:theta_scan}  displays the evolution with the angle of incidence of the intensity recorded on a narrow slice around the Laue circle showing a regular evolution of maxima and minima every second diffraction $\delta k_y=2G_y$.
This indicates a simple unit-cell structure with only one maximum and minimum per lattice unit and also along $z$ with a pseudo period $\delta k_z$ indicating a corrugation amplitude $z_c\approx 2\pi/\delta k_z$ which hardly evolves with the incident value of the angle of incidence $\propto k_z$.
This almost constant corrugation amplitude was understood long ago with thermal atoms. The attractive van der Waals forces create a weak potential energy well, with a depth $D$ and increasing the incoming energy $E_\perp$ to $E_\perp + D$.
Of course, this is not an exact rule, and strong evolution of the corrugation was observed on simple systems due to the evolution from a single to a double maximum situation \cite{Meyer_2016,Bocan_2018}.
For more complex unit-cells, the intensity modulation is also more complex but simple optical models managed to link the observed features to the cell topology with surprising accuracy \cite{Debiossac_PRB_2014,Debiossac_2016}. 
During an energy scan, the angle of incidence is constant, and so is the radius of the Laue circle, while the Bragg angle or peak separation $G_y$ along $y$ is no longer constant.
Fig.\ref{fgr:E_scan} shows a typical increase of the mean scattering width at low energy, which was explained as a refraction effect due to the well-depth $D$ \cite{Debiossac_2020}.
Fig.\ref{fgr:theta_scan} and Fig.\ref{fgr:E_scan} appear different, but the diffracted intensities recorded during an $E$-scan and $\theta$-scan coincide when plotted as a function of the perpendicular energy $E_\perp$=$E\sin^2\theta$  \cite{Debiossac_2020}, as predicted by the Axial Surface Channeling Approximation (ASCA) well established theoretically \cite{Zugarramurdi_2012,Diaz_2016b,Schuller_2005,Danailov_2001,Farias_2004}. 
Along each direction, the 3D potential energy landscape is replaced by its 2D average along this direction and all methods developed to interpret TEAS elastic diffraction can be applied to GIFAD.
Any given potential energy surface, either predicted by theory or simply guessed based on chemical considerations, can be compared with an experimental one using a quantum scattering calculation \cite{Zugarramurdi_2012,Diaz_2016b}.
Alternately, the potential energy surface fitting best to the data can be extracted from a close loop optimization with a fast quantum algorithm \cite{Debiossac_2020}.
Returning to the experiment, one advantage of the $E$-scan compared with $\theta$-scan is that the surface illuminated by the beam does not change, and fine-tuning a 200 eV beam energy by only $\sim 1$ eV gives a sub meV accuracy on $E_\perp$, as needed to explore bound state resonances \cite{Debiossac_PRL_2014}.

\begin{figure}
\includegraphics[width=0.85\linewidth,angle =0,draft = false]{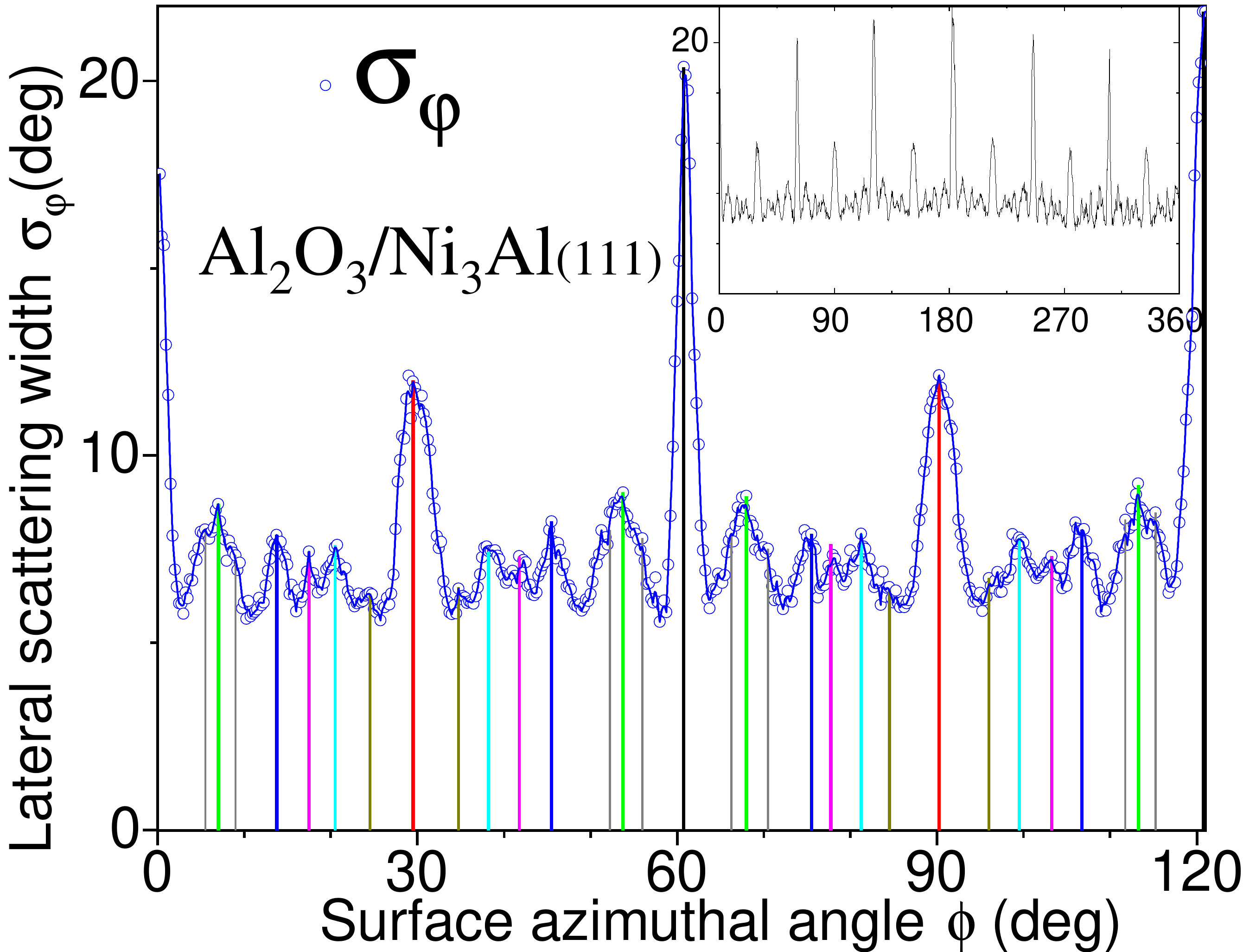}
\caption{\label{fgr:phi_scan_march} Triangulation curve reporting the evolution of the mean width $\sigma_\varphi$ of the scattering profile during a $\phi$-scan or in-plane rotation, with $\varphi=\arctan \frac{k_{fy}}{k_{fz}}$. The insert displays the evolution over a full turn. Several surface channeling directions are readily identified (In collaboration with A.Ouvrard \cite{Alyabyeva_2018})}
\end{figure}

Fig.\ref{fgr:phi_scan_march} displays a $\phi$-scan, also called triangulation curve \cite{Kalashnyk_2016}, where the width $\sigma_\varphi$ of the lateral scattering profile is recorded during an in-plane rotation of the surface. Each peak, here repeated every 60$^\circ$ corresponds to a channeling along a $Al_2O_3/Ni_3Al(111)$ surface low index direction\cite{Alyabyeva_2018}. 
The peak amplitude corresponds to the illuminated section of the Laue circle in Fig.\ref{fgr:schematic} or to the opening angle of  the $V$-structure in  Fig.\ref{fgr:theta_scan} or to the distance between the vertical structures in Fig.\ref{fgr:E_scan}.
Theoretically, this width $\sigma_\varphi=(\Sigma (\varphi-\bar{\varphi})^2)^{1/2}$ can be calculated from quantum mechanics, but in practice, since all quantum contributions are averaged \cite{Zugarramurdi_2013b,Seifert_2011,Debiossac_PRA_2014}, the comparison with classical trajectory calculations \cite{Seifert_2016} readily provides a quantitative estimate of the surface structure. 
In practice, such a $\phi$-scan is the first measurement after introducing a new sample. This can be achieved even with a beam resolution or with a surface coherence that does not allow observation of diffraction. 
After this step, diffraction can be investigated with high resolution along all directions where a peak is observed.

\begin{figure}
\includegraphics[width=1.0\linewidth,trim={0 1cm 0 0},clip,angle =0,draft = false]{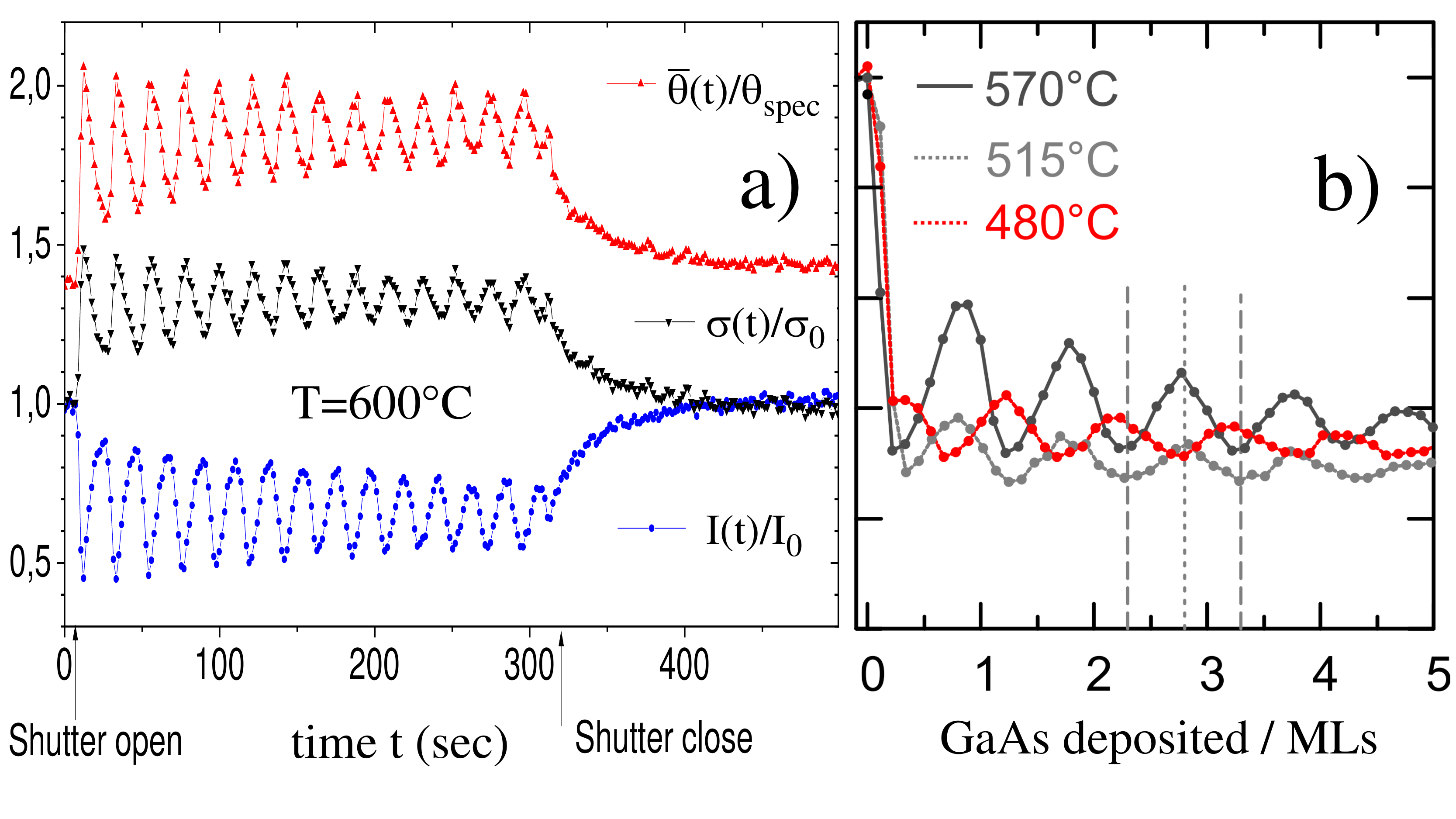}
\caption{\label{fgr:oscillations} The imaging detector allows simple tracking of the number of deposited layers during a layer by layer growth of GaAs/GaAs. Pronounced oscillations are visible for : the intensity I(t) around the Laue circle, the mean polar scattering angle  $\theta_{out}(t)$ and the polar scattering width  $\sigma_\theta (t)$  (from \cite{Debiossac_2016}). The classical nature of these parameters ensures a robust behavior. b) At T=480$^\circ$C the completion of the first layer is delayed by half a growth oscillation due to an initial surface reconstruction consuming Ga (from \cite{Atkinson_2014}). }
\end{figure}

Finally, the simple possible scan is a time evolution of the diffracted intensities at fixed positions of the target surface. 
This can be used to identify the surface reconstructions taking place at different temperatures\cite{debiossac_these} or to track the growth parameters under exposure to molecular beams from evaporation cells.
The Fig.\ref{fgr:oscillations}a) shows pronounced oscillations of all measurable parameters of the diffraction images during a layer by layer, homo-epitaxial growth of $GaAs/GaAs$ at 600$^\circ$C inside a MBE vessel at INSP \cite{Atkinson_2014,Debiossac_2016}.
The mean scattering angle $\langle\theta_{out}\rangle$, the width $\sigma_{\theta}$ or the intensity on the Laue circle.
These oscillations are directly related to the surface reflectivity, which decreases rapidly when ad-atoms are present on top of a new layer under grazing incidence.
For the same system, the Fig.\ref{fgr:oscillations}b) shows that at 480$^\circ$C, the growth oscillation is shifted by half a monolayer revealing the well-documented phase transition from the more arsenic-rich c(4x4) reconstruction, changing to the
(2x4)$\gamma$ at the onset of growth \cite{Atkinson_2014}.
Note that similar results were obtained with RHEED or with keV ions at grazing incidence \cite{Igel_1996,Bernhard_2005}  outlining the dominant role played by ad-atoms on the surface specular reflectivity.
The specific interest of GIFAD lies in the non-destructive behavior of atoms at low effective normal energy $E_\perp$ and the absence of charging effect. 
Also, there are not many techniques able to help online monitoring of thin-film growth of fragile molecular layers\cite{Seifert_2013,Kalashnyk_2016,Momeni_2018}. 
Note that the contrast of the oscillations can be further improved by reporting only the elastic intensity \cite{Debiossac_2016} but this requires a more elaborate image processing.

\section{\label{sec:quality} Probing the surface quality}
Anyone involved in construction knows that a look at grazing incidence will efficiently reveal the superficial defects.
This is a geometry where any protruding defect can perturb the projectile trajectory.
This sensitivity appears at two different levels. The trajectory level and macroscopic level as discussed below.
The elastic diffraction intensity approximately corresponds to the intensity sitting on the Laue circle visible in Fig.\ref{fgr:schematic}. 
It is very sensitive to surface defects at the trajectory level in the sense that probably a single protruding ad-atom along the trajectory is enough to push the atom away from perfect specular reflection.
The length $L$ of the trajectory can be evaluated from classical trajectories but since diffraction is involved, the transverse  ($L_y$) and longitudinal ($L_x=L_y/\theta$) coherence lengths are more relevant to defining the surface $S=L_x \cdot L_y$ needed for coherent reflection.
Neglecting the energy dispersion, these are determined by the momentum spread  associated with the angular spread of the beam $\delta k_\perp= k \delta\theta$.
 Taking into account the angle of incidence $\theta$, the surface coherently illuminated is $S = (k^2 \delta\theta^2 \theta)^{-1}$. 
For 500 eV He and with our resolution of 0.1 mrad this gives a surface $S\sim$5000\AA$^{2}$ at 1$^\circ$ incidence.
In principle, a lack of periodicity in this surface could ruin elastic diffraction.
Since the longitudinal coherence $L_x$ is $1/\theta$ longer than the transverse one $L_y$, it probably means that the mean distance between defects called surface coherence should be larger than $L_x\sim 200$ \AA, otherwise the elastic intensity will be reduced.
On the macroscopic level, the surface illuminated by the beam is $\oslash^2/\theta$ which is closer to a mm$^2$ for $\theta=1^\circ$.
It means that the diffraction signal can be used for online monitoring of the surface coherence length up to several hundred \AA, and that the diagnostic applies over a significant surface around 1 mm$^2$.
Macroscopic defects such as tilt and twist surface mosaicity \cite{Lalmi_2012} as well as wafer curvature induced by surface tension can be tracked \textit{in situ} and online at the mrad level.
More work is needed to identify specific signatures associated with other types of defects such as specific terraces.

\section{Complementary equipments}\label{sec:Goodies}
Each surface science experiment is equipped with specific diagnosing and sample preparation tools, we describe here only complementary detectors and/or equipment that take benefit of the presence of a keV ion or atom beam.
Grazing sputtering \cite{winter_PR2002,hansen_2004}
As described in the previous section, GIFAD relies on the well-collimated neutral beam. 
If no gas is fed in the charge exchange cell, the setup can produce an ion beam with high angular and spatial properties, that can be used for surface analysis.
In addition, from diffraction studies, the presence of a keV ion or atom beam is enough to count the number of deposited layers during growth by tracking the intensity oscillation of the quasi specular beam \cite{Igel_1996,Atkinson_2014,Debiossac_2017}.
This analysis in terms of reflection coefficient or reflectometry is almost identical to atoms or ions. Both have comparable trajectories above the surface, so the probability of encountering obstacles should be comparable.
Triangulation or $\phi$-scan can also be performed with ions \cite{Pfandzelter_2003} instead of atoms.

\subsection{\label{sec:chopping} Chopping the ion beam}
Elastic diffraction occurs only if the successive collisions with the surface atoms are gentle enough and do not trigger any vibrational or electronic excitation.
In general, a quite significant amount of energy can be exchanged with the surface, and measuring it is already a precise diagnosis.
The projectile can also capture or lose electrons from the surface, so that it is useful to measure the energy of neutral atoms, which requires a time of flight measurements and, most often, beam pulsing.
If the moment of impact on the surface is known, then the angular resolved time of flight measure will be a direct signature of the final energy.
 The ion beam can be pulsed either by setting the deflectors in ON and OFF mode generating comparatively long ion pulses of several tens or hundreds of ns.
At variance, by rapidly switching the deflectors from left to right, the time resolution is given by the beam and hole size divided by the switching velocity in front of the last diaphragm \cite{Morosov_1996}.
The timing resolution of around 1 ns in the projectile time of flight can result in a few eV energy resolutions, \cite{Roncin_1999a} but this also requires a time-resolved detector with ns accuracy, difficult with standard cameras.

\begin{figure}
\includegraphics[width=0.70\linewidth,trim={0 0 5cm 0},clip,angle =0,draft = false]{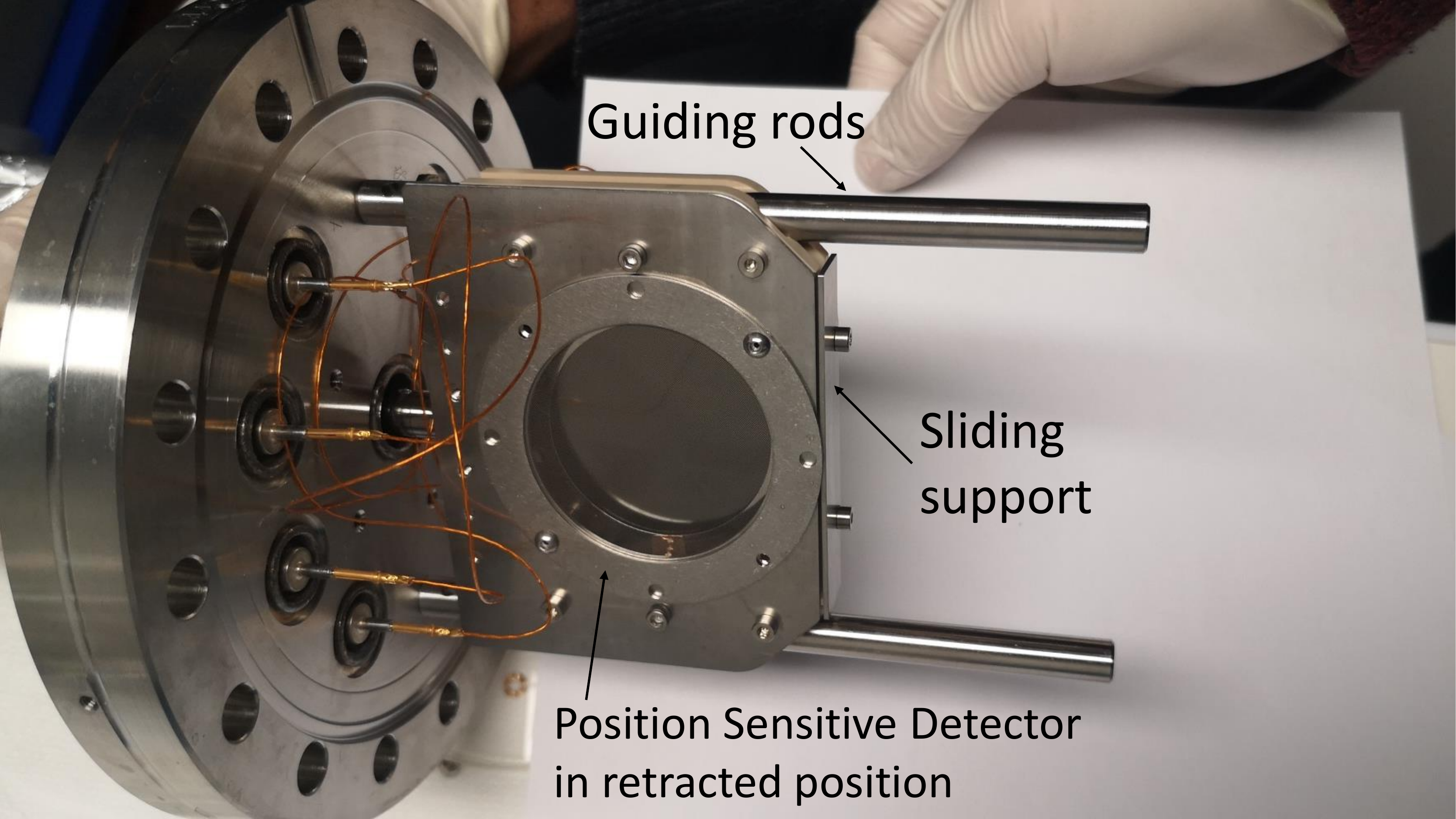}
\caption{\label{fgr:retractable_det} The retractable resistive anode detector has an active diameter of 40 mm and a sub-ns timing accuracy. It is sliding along the two metal rods taped into the 100 CF flange and is tied to a 100 mm linear actuator to be taken in or out of the beam.}
\end{figure}
\subsection{\label{sec:retractable} The retractable resistive anode detector}
The retractable resistive anode detector is a two dimensional position sensitive that can be inserted in the beam.
It consists of two MCP mounted in chevron and a resistive anode from Quantar-tech \cite{Quantar} designed to improve the linear behavior by using a linear edge resistance full-filling the gear conditions.
The detector support is made of peek and slides between two stainless steel rods, it is covered by a stainless steel plate holding a high transparency grid visible in Fig.\ref{fgr:retractable_det}.
The detector is a simple evolution of the similar ones used in previous experiment 
\cite{roncin_1986,Roncin_1987,Flechard_1997} and in particular of the setup where GIFAD was discovered \cite{Rousseau_2007}.
The originality is that it uses a compact low-cost ADC-card controlled with a tiny Arduino microcontroller hooked to the PC via USB.
When used in combination with the beam chopper, the detector can perform energy loss measurements to identify electronic inelastic processes such as surface excitations, either phonons \cite{Auth_1998,Villette_2000} or excitons and trions\cite{khemliche_2001,roncin_2002}.
It also helped understand why ionic insulators emit more electrons than metals\cite{Roncin_1999} and identify a single inelastic collision event among the numerous quasi-elastic ones. 
Also, single-particle detectors are compatible with ratemeters so that the count-rate is not associated with the frame rate or exposure time of the camera making reflectometry and triangulation easier to perform.

\subsection{\label{sec:recoil} The Recoil ion detectors}
\begin{figure}
\includegraphics[width=0.8\linewidth,angle =0,draft = false]{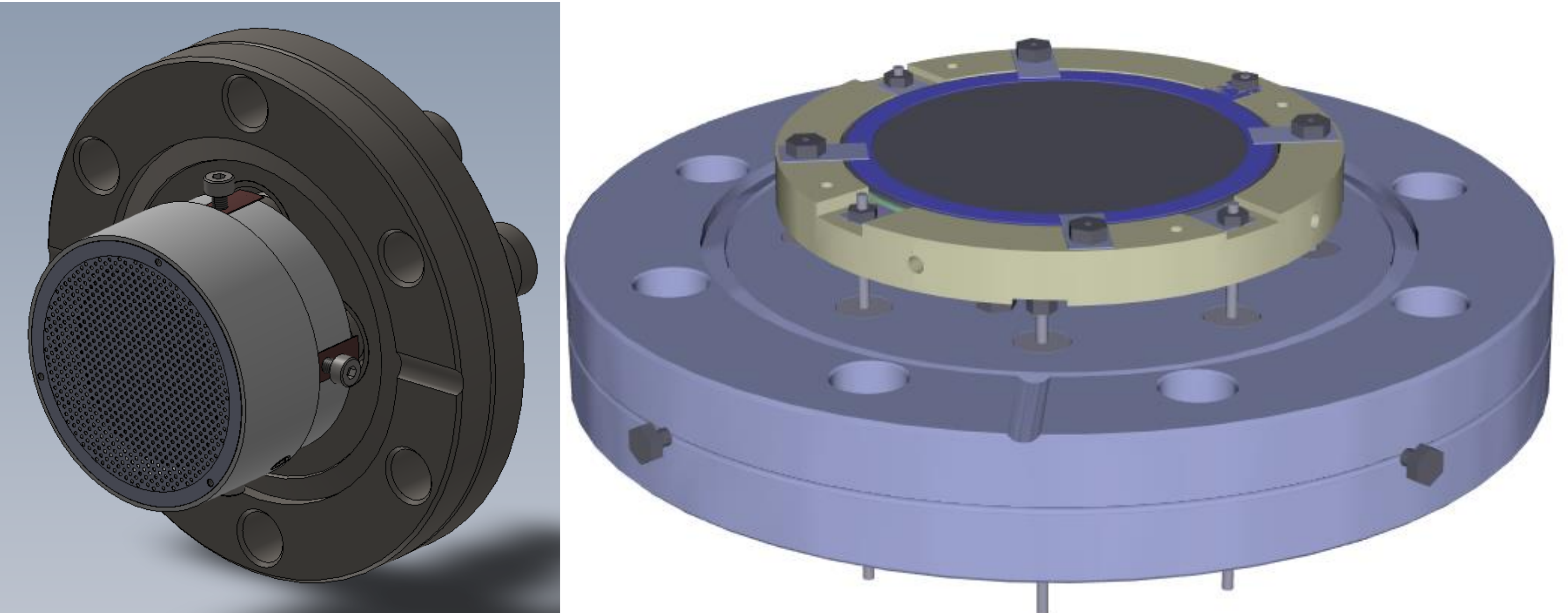}
\caption{\label{fgr:TOF} Schematic view of two double MCP, time of flight detector built in PEEK. The left one fits on a commercial 4-way BNC CF40 flange while the right one, on a 63CF flange is position sensitive\cite{Lupone_2015}.}
\end{figure}

The two recoil ion detectors depicted in Fig \ref{fgr:TOF} are multi-purpose particle detectors built around a pair of MCP.
The large one has a 40 mm active surface and fits on a DN 63 CF flange \cite{Lupone_2015}, it has a homemade resistive anode with significant distortions but these can be corrected to achieve a maximum error of 1 mm localization accuracy while preserving a typical 100 $\mu$m resolution in the center.
The air-side electronic and driving software are therefore similar to the one used for the position sensitive retractable of Fig.\ref{fgr:retractable_det}.

The compact one has an open surface of 30 mm and fits inside a DN 40 CF tube. 
Both detectors have a sub $ns$ timing accuracy.
They can be configured to detect preferentially positive or negative particles by biasing the entrance at a voltage between -5 kV and 3 kV so that the anode voltage does not exceed the 5 kV limit of our electrical feed-throughs.
The detectors are fixed on DN40CF tubes directed to the target and intended to detect particles emitted around 30-45 deg from the surface, either in the forward or backward direction.

\subsection{\label{sec:DirectRecoil} Direct recoil spectroscopy}
Direct Recoil Spectroscopy (DRS) is the generic name for several techniques where atoms or ions with definite energy are sent onto a surface and the ejected or recoiling ions or atoms are analyzed in energy.
If a quasi binary collision took place, then the energy and momentum sharing follows that of the gas phase indicating the mass of the collision partner.
The most general technique is Time of Flight analysis (TOF SARS \cite{Grizzi_1990}) because, most often the particles are ejected as neutrals but low energy ion spectroscopy \cite{Cushman_2016} (LEIS) is presumed to address only the terminal layer.
These analyses can be achieved as a function of the target azimuthal and polar angle at a relatively large incidence angle because only quasi-binary collisions can be identified. 
Under grazing geometry and quasi specular reflection several shallow collisions participate to the deflection so that the energy transferred to the surface atoms is negligible \cite{Winter_2002,Villette_these,Auth_1998}. However, if an ad-atom or terrace edge is encountered, the projectile can undergo a violent binary collision \cite{Pfandzelter_98,hansen_2004}.
The momentum and energy exchange is so large that the binding to the surrounding matrix can be neglected and the TOF becomes a direct signature of the mass $m_t$ of the target atom.
The analysis of the recoils emitted in the forward or backward direction should  help identification of poisoning impurities such as hydrogen contaminants difficult to pump.
More interesting scientifically, the identification of the chemical composition of island edges during growth is a challenging issue where the evolution of the recoil ion time of flight during growth could be helpful.

\subsection{Ion beam triangulation}
When the energy perpendicular to the surface $E_\perp$ is close to or larger than 10 eV, the projectile penetrates the electronic density of the target at distance close to one \AA\ and secondary electrons are emitted.
By tracking the secondary electron yield during an azimuthal scan, some surface structure parameters, such as the direction of the low index axis, can be determined by ion beam triangulation \cite{Pfandzelter_2003}.
This is equivalent to the $\phi$-scan or atom beam triangulation described above, where the increased lateral scattering width identified the channeling direction due to the repeated deflections on the walls of the valley. Here the enhanced electron emission is partly due to the comparatively long zig-zag trajectories inside the valleys. 
This correlation between selected trajectories associated with electron emission can be identified by coincidence experiments using a time-resolved detector similar to the one described in Fig.\ref{fgr:retractable_det}.
At large enough projectile energy, $E_\perp$ reaches a few eV ranges even at a moderate angle of incidence $\theta$. The electron emission can be recorded in coincidence with the scattered atoms or ions \cite{Morosov_1996}.
In our setup, only the retractable detector has a time resolution suitable for such coincidence detection allowing trajectory-dependent electron emission to be identified \cite{Morozov_1999,Morozov_2001}.
The electron emission can be resolved in projectile energy loss with a pulsed beam \cite{Villette_1999}.
So far, fast ion diffraction has never been observed, probably because ions interact too strongly with the surface electrons or with the ions via excitation of optical phonons in the case of insulators \cite{Auth_1998,Villette_1999}.
Diffraction patterns were observed with primary ions but associated with scattered atoms that have neutralized far from the surface, so that they impact the surface as neutral atoms \cite{Rousseau_these}.

\subsection{\label{sec:Sputter} The sputtering ion gun}
Inside the UHV chamber, a commercial ion gun directed to the sample holder is used to sputter clean the surfaces of the samples.
We use an inert gas such as Ne or Ar at an energy between 500 eV and 2 keV. The standard incidence is around 45 deg. but can be adjusted continuously from normal to grazing incidence. 
It is equipped with a focusing lens, and we have added a pair of deflectors in order to scan the target with the ion beam.
By recording  the secondary electron yield with one of the recoil ion detectors described in sec.\ref{sec:Goodies} as a function of the voltage of the deflecting plates, a coarse image of the target can be formed.
For LiF samples, sputter cleaning is associated with the creation of topological defects that we could not remove. Even after thermal treatment and grazing sputtering, the nice diffraction patterns showing sharp elastic spots could be observed only after fresh cleaving of the surface in air and thermal treatment in vacuum.
The situation is the opposite for metals, where the diffraction on the Ag(110) surface could be observed only after repeated cycles of sputtering and annealing \cite{Bundaleski_2008,rubiano_2013}.

\subsection{Evaporation cells}
The main application of GIFAD is probably the monitoring of thin-film growth where its grazing geometry, similar to that of RHEED leaves the volume above the surface free of any instrument for evaporation cells. 
GIFAD was installed inside a MBE vessel to investigate its ability to track surface reconstruction and homo-epitaxial growth monitoring of II-VI  semiconductors \cite{Atkinson_2014}.
For the evaporation of molecules, \cite{Seifert_2013} we have used homemade and commercial retractable evaporation cells allowing the recharging of material without exposing the target surface to air.
So far, for  molecular layers, no detailed topology could be extracted directly from the diffraction pattern, however, the directions where the molecules tend to align are immediately revealed by triangulation curves\cite{Kalashnyk_2016} such as the one displayed in Fig.\ref{fgr:phi_scan_march} and these can be directly compared with classical calculations to select the most appropriate structural model \cite{Seifert_2014}.
The evolution of the reflectivity with time also allows the identification of structural change interpreted as a liquid to solid phase transition occurring when the density of highly mobile molecules at the surface is high enough \cite{Momeni_2018} as could be investigated by STM \cite{Bobrov_2015,Salomon_2021}.


\subsection{\label{sec:tilt_perp} The Webcam and perpendicular laser}
The UHV chamber offers a DN100CF flange directly facing the target surface.
When unused, it is closed by a window flange where we have installed a webcam and a miniature red laser.
The red laser is shifted by $\approx$ 2cm from the center and directed to intercept the surface around its center.
If the target surface is indeed parallel to the flange, the reflected beam lies also 2cm from the center but opposite to the primary spot.
During an azimuthal rotation, the reflected spot describes a circle with a center located at 2$\theta_{laser}$ and a radius 2$\tau$ where $\tau$ is the misalignment of the surface normal with respect to the rotation axis and $\theta_{laser}$ is the arbitrary angle of the laser with the mean surface normal. 
Due to the 30cm long path between the surface and the window, the position of the spot reflected from the surface is easily tracked by the camera. It can be used to follow the target azimuthal angle online and to correct for the possible tilt $\tau$ \cite{Sereno_2016}. 

\begin{figure}
\includegraphics[width=0.85\linewidth,angle =0,draft = false]{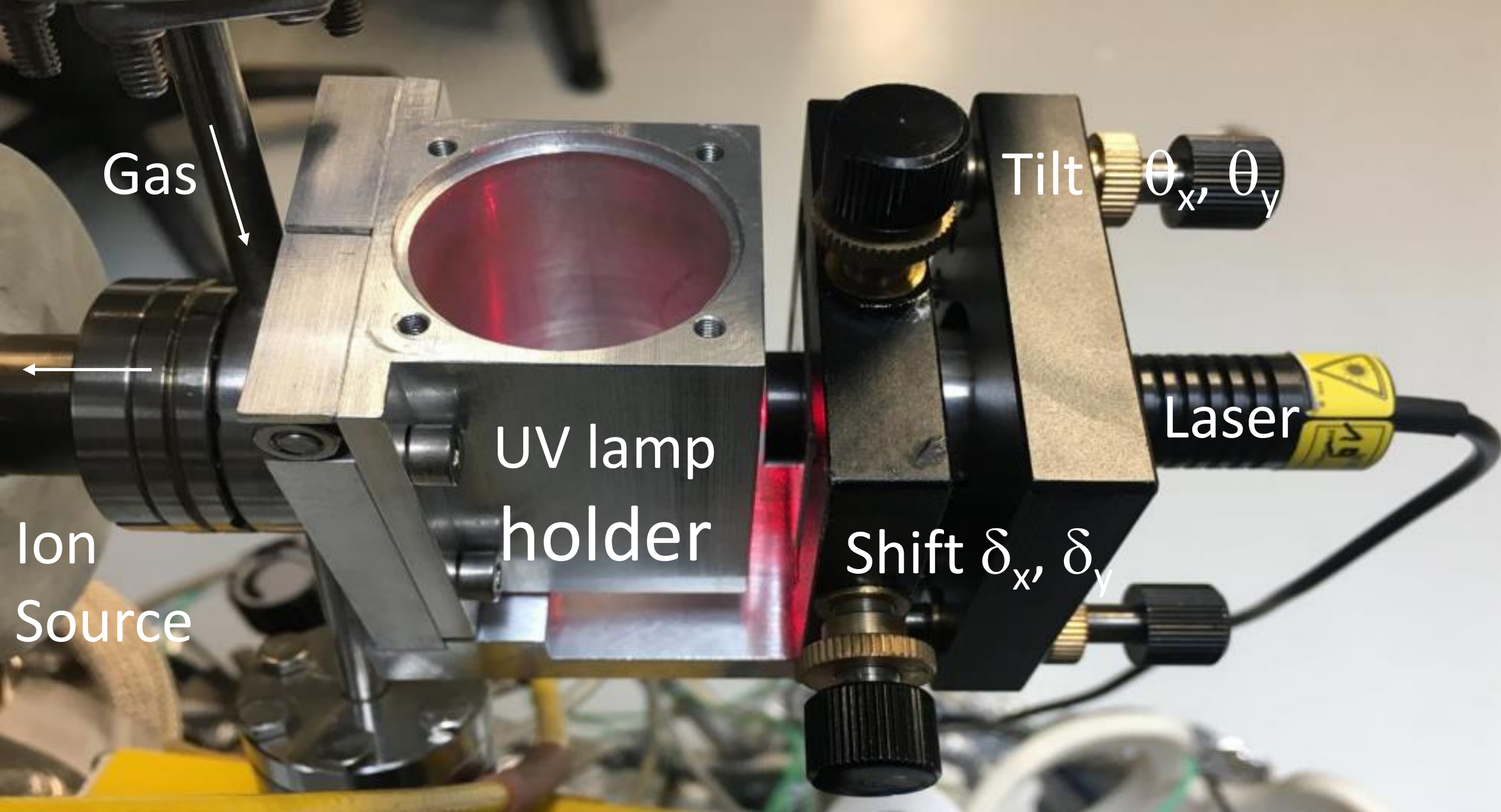}
\caption{\label{fgr:Laser} The colinear laser is held on a miniature $X, \theta_X, Y, \theta_Y$ platform hooked to a view-port at the gas injection side of the ion source.}
\end{figure}
\subsection{\label{sec:Align} The co-axial Laser}

The Alignment procedure of all the diaphragms and differential pumping holes is easier with the miniature $X,Y, \theta_X, \theta_Y$ platform attached directly to the EX05 ion source (section \ref{sec:ion_source}). 
As detailed in Sec.\ref{sec:detector}, when a single MCP is used together with a phosphor screen without aluminum over-layer, there is enough light passing through the (unbiased) MCP to be imaged by the camera allowing quantitative intensity optimization both in vacuum or at atmospheric pressure.

This allows a very simple pre-positioning of the target surface into the beam and direct measurement of the angle of incidence,  which can be different from a crystal plane in case of miscut. In this case, the atoms would reveal the crystal plane while the light would reveal the mean surface plane \cite{Pfandzelter_98}.
Since the laser impact on the surface is visible and can be captured by a webcam, a mark can be "tapped" on the picture of the target surface where a poor or good diffraction is observed.

\begin{figure}
\includegraphics[width=1.0\linewidth,angle =0,draft = false]{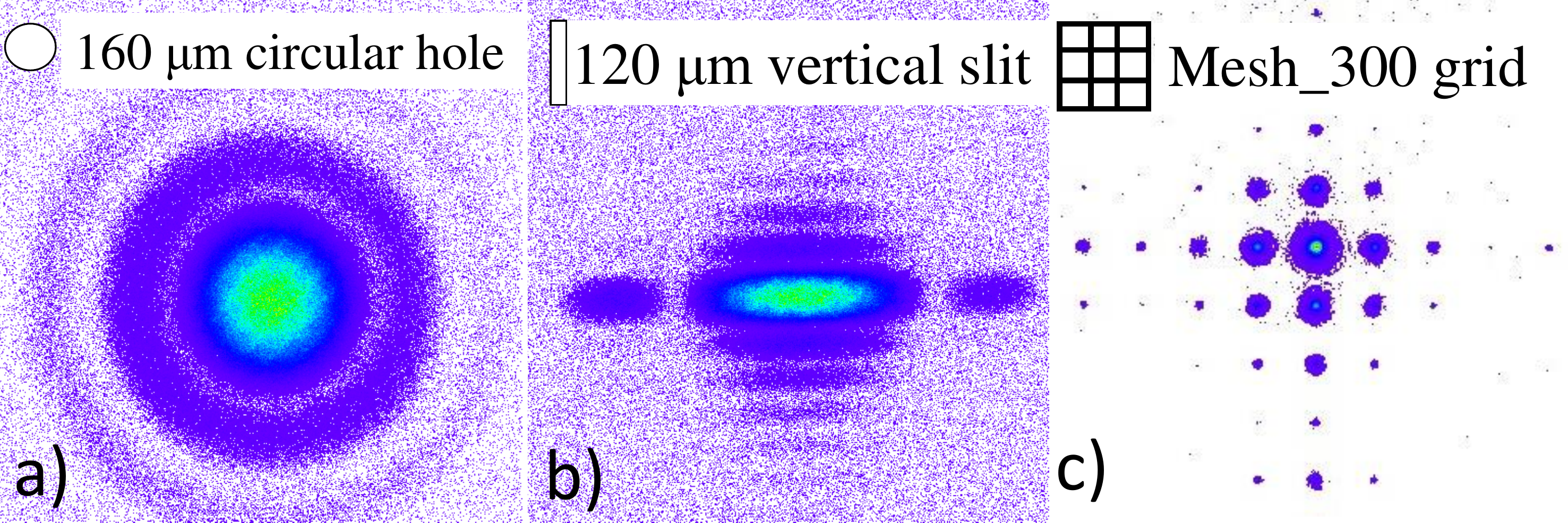}
\caption{\label{fgr:Airy} a) laser diffraction through $\oslash_2$=160 $\mu$m, b) through a 120 $\mu$m wide horizontal slit. c) corresponds to diffraction through the 84$\mu$m pitch of a TEM grid at target position. The Bragg angles are used for \textit{in situ} angular calibration.}
\end{figure}

This is also convenient to calibrate the diaphragms and slits' sizes and positions by analyzing the observed Airy pattern in Fig.\ref{fgr:Airy}a).
It also allows an accurate angular calibration procedure.
A shortcoming attached to our detector is that the calibration is very sensitive to the exact zoom value of the camera lens.
This later is optimized online, with the primary beam to produce a minimum spot size. It is sometimes useful to check the exact overall angular scale depending on actual distances and zoom values, which is not always easy to determine.
We have used a commercial transmission electron microscope grid called mesh-300 with a $d$=84 $\mu$m periodicity, point-welded on the backside of the manipulator's head so that it can be inserted into the laser beam to generate bright spots with $\theta_{Bragg}$=$\lambda/d$ in rad.
The diffracted spots of Fig.\ref{fgr:Airy}c) allow simple calibration of the detection system, whatever the optical elements and actual detector position or orientation.

It also has nice educational potential as many undergraduate students are surprised that visible light diffracts on the collimating diaphragms but, apparently, not on the surface while keV atoms do exactly the opposite. 
Note that using a single free-standing graphene layer, a geometry demonstrated  with highly charger ions\cite{gruber_2016}, diffraction of atoms through graphene was predicted \cite{Brand_2019}.
It also offers a nice opportunity the define the Airy function ubiquitous in wave mechanics, from atmospheric rainbows to atomic collisions \cite{Ford_1959,Roncin_2020}.
Here, in atomic diffraction at surfaces, the Airy function and its companion Bessel function can be derived by the semi-classical intensity modulation observed in diffraction from simple lattice unit and known as supernumerary rainbows\cite{Schueller_2008,Winter_PSS_2011,Miret_Artes_2012,Debiossac_2017,Debiossac_PCCP_2021} visible in each line of Fig.\ref{fgr:E_scan}.

\section{\label{sec:simplified} Conclusions}
The setup presented here is a versatile research setup.
A more compact design of the source section integrating the Wien filter and neutralization cell is possible but has not been attempted.
The energy $E_\perp$ associated with the motion perpendicular to the crystal axis is a key parameter that governs the minimum distance of approach to the surface.
The setup allows operation with atoms from $E_\perp\simeq$ a few meV up to a few 100 eV, opening a direct connection with the keV ion-surface community\cite{Winter_2002} and all its surface sensitive techniques in continuity with the more penetrating, higher energy projectiles of the ion-beam-analysis community\cite{Jalabert_2017}.
We have focused on systematic errors associated with small angles.
These are related to the control of the primary beam, with the target positioning, and to the mechanical, optical and electronic aberrations in the measurement method.
The detector itself can also be customized to use smaller and cheaper micro-channel plates. Since only one MCP is required, the overall benefit/cost trade-off, including the camera, suggests using a large detector.
\textit{A priori} the condition where GIFAD could be helpful in the growth of fragile layers or severe electromagnetic environment. 
It could also help production of high-quality surfaces \textit{i.e.}, with very large coherence length.
In both cases, the ability of GIFAD to provide an online handle to optimize the growth parameters such as surface temperature, evaporation rate, etc... could be decisive.
In many situations, a very simple numerical treatment of the diffraction image can provide "on the fly" quantitative parameters such as the reflectivity, the scattering width, the elastic scattering ratio etc...

\begin{acknowledgments}
We are grateful to the technical staff of ISMO, namely M. Baudier for the C.A.D, J. Guigand for the machining and C. Charrière for the motor-control systems. 
We warmly acknowledge the PhD students and post-docs that have contributed to the grazing incidence setups : V.A. Morosov\cite{Morozov_1999}, A. Kalinin\cite{Villette_1999}, Z. Szilagyi\cite{Morosov_1996},  JP. Atanas\cite{Villette_1999}, J. Villette\cite{Villette_2000},  H. Khemliche\cite{Zugarramurdi_2013b}, A. Momeni\cite{roncin_2002}, N. Bundaleski\cite{Bundaleski_2008}, P. Rousseau\cite{Rousseau_2007}, P. Soulisse\cite{Rousseau_2008}, B. Lalmi\cite{Lalmi_2012}, A. Zugarramurdi\cite{zugarramurdi_2015}, N. Kalashnyk\cite{Kalashnyk_2016}, A. Husseen\cite{Lupone_2015} and M. Debiossac\cite{Debiossac_Nim_2016}.
This work received support from 
the French Agence Nationale de la Recherche (ANR-07-BLAN-0160, ANR-2011-EMMA-003-01) and LabEx PALM (ANR-10-LABX-0039-PALM).
The MBE images in Fig.\ref{fgr:schematic} and Fig.\ref{fgr:oscillations} were recorded at Institut des Nanosciences de Paris in coll. with P. Atkinson, M. Eddrief and V. Etgens, A. Momeni, H. Khemliche and M. Debiossac. One of us was supported by a Chinese Scholarship Council (CSC) Grant reference number 201806180025. 
\end{acknowledgments}

\appendix

\nolinenumbers
 
\bibliography{ref}

\begin{thebibliography}{90}%
\makeatletter
\providecommand \@ifxundefined [1]{%
 \@ifx{#1\undefined}
}%
\providecommand \@ifnum [1]{%
 \ifnum #1\expandafter \@firstoftwo
 \else \expandafter \@secondoftwo
 \fi
}%
\providecommand \@ifx [1]{%
 \ifx #1\expandafter \@firstoftwo
 \else \expandafter \@secondoftwo
 \fi
}%
\providecommand \natexlab [1]{#1}%
\providecommand \enquote  [1]{``#1''}%
\providecommand \bibnamefont  [1]{#1}%
\providecommand \bibfnamefont [1]{#1}%
\providecommand \citenamefont [1]{#1}%
\providecommand \href@noop [0]{\@secondoftwo}%
\providecommand \href [0]{\begingroup \@sanitize@url \@href}%
\providecommand \@href[1]{\@@startlink{#1}\@@href}%
\providecommand \@@href[1]{\endgroup#1\@@endlink}%
\providecommand \@sanitize@url [0]{\catcode `\\12\catcode `\$12\catcode
  `\&12\catcode `\#12\catcode `\^12\catcode `\_12\catcode `\%12\relax}%
\providecommand \@@startlink[1]{}%
\providecommand \@@endlink[0]{}%
\providecommand \url  [0]{\begingroup\@sanitize@url \@url }%
\providecommand \@url [1]{\endgroup\@href {#1}{\urlprefix }}%
\providecommand \urlprefix  [0]{URL }%
\providecommand \Eprint [0]{\href }%
\providecommand \doibase [0]{https://doi.org/}%
\providecommand \selectlanguage [0]{\@gobble}%
\providecommand \bibinfo  [0]{\@secondoftwo}%
\providecommand \bibfield  [0]{\@secondoftwo}%
\providecommand \translation [1]{[#1]}%
\providecommand \BibitemOpen [0]{}%
\providecommand \bibitemStop [0]{}%
\providecommand \bibitemNoStop [0]{.\EOS\space}%
\providecommand \EOS [0]{\spacefactor3000\relax}%
\providecommand \BibitemShut  [1]{\csname bibitem#1\endcsname}%
\let\auto@bib@innerbib\@empty
\bibitem [{\citenamefont {Rousseau}\ \emph {et~al.}(2007)\citenamefont
  {Rousseau}, \citenamefont {Khemliche}, \citenamefont {Borisov},\ and\
  \citenamefont {Roncin}}]{Rousseau_2007}%
  \BibitemOpen
  \bibfield  {author} {\bibinfo {author} {\bibfnamefont {P.}~\bibnamefont
  {Rousseau}}, \bibinfo {author} {\bibfnamefont {H.}~\bibnamefont {Khemliche}},
  \bibinfo {author} {\bibfnamefont {A.~G.}\ \bibnamefont {Borisov}},\ and\
  \bibinfo {author} {\bibfnamefont {P.}~\bibnamefont {Roncin}},\ }\bibfield
  {title} {\enquote {\bibinfo {title} {Quantum scattering of fast atoms and
  molecules on surfaces},}\ }\href
  {https://doi.org/10.1103/PhysRevLett.98.016104} {\bibfield  {journal}
  {\bibinfo  {journal} {Phys. Rev. Lett.}\ }\textbf {\bibinfo {volume} {98}},\
  \bibinfo {pages} {016104} (\bibinfo {year} {2007})}\BibitemShut {NoStop}%
\bibitem [{\citenamefont {Sch\"uller}, \citenamefont {Wethekam},\ and\
  \citenamefont {Winter}(2007)}]{Schuller_2007}%
  \BibitemOpen
  \bibfield  {author} {\bibinfo {author} {\bibfnamefont {A.}~\bibnamefont
  {Sch\"uller}}, \bibinfo {author} {\bibfnamefont {S.}~\bibnamefont
  {Wethekam}},\ and\ \bibinfo {author} {\bibfnamefont {H.}~\bibnamefont
  {Winter}},\ }\bibfield  {title} {\enquote {\bibinfo {title} {Diffraction of
  fast atomic projectiles during grazing scattering from a {LiF}(001)
  surface},}\ }\href {https://doi.org/10.1103/PhysRevLett.98.016103} {\bibfield
   {journal} {\bibinfo  {journal} {Phys. Rev. Lett.}\ }\textbf {\bibinfo
  {volume} {98}},\ \bibinfo {pages} {016103} (\bibinfo {year}
  {2007})}\BibitemShut {NoStop}%
\bibitem [{\citenamefont {Winter}\ and\ \citenamefont
  {Schüller}(2011)}]{Winter_PSS_2011}%
  \BibitemOpen
  \bibfield  {author} {\bibinfo {author} {\bibfnamefont {H.}~\bibnamefont
  {Winter}}\ and\ \bibinfo {author} {\bibfnamefont {A.}~\bibnamefont
  {Schüller}},\ }\bibfield  {title} {\enquote {\bibinfo {title} {Fast atom
  diffraction during grazing scattering from surfaces},}\ }\href
  {https://doi.org/https://doi.org/10.1016/j.progsurf.2011.07.001} {\bibfield
  {journal} {\bibinfo  {journal} {Progress in Surface Science}\ }\textbf
  {\bibinfo {volume} {86}},\ \bibinfo {pages} {169 -- 221} (\bibinfo {year}
  {2011})}\BibitemShut {NoStop}%
\bibitem [{\citenamefont {Debiossac}, \citenamefont {Pan},\ and\ \citenamefont
  {Roncin}(2021)}]{Debiossac_PCCP_2021}%
  \BibitemOpen
  \bibfield  {author} {\bibinfo {author} {\bibfnamefont {M.}~\bibnamefont
  {Debiossac}}, \bibinfo {author} {\bibfnamefont {P.}~\bibnamefont {Pan}},\
  and\ \bibinfo {author} {\bibfnamefont {P.}~\bibnamefont {Roncin}},\
  }\bibfield  {title} {\enquote {\bibinfo {title} {Grazing incidence fast atom
  diffraction, similarities and differences with thermal energy atom scattering
  ({TEAS})},}\ }\href {https://doi.org/10.1039/D0CP05476C} {\bibfield
  {journal} {\bibinfo  {journal} {Phys. Chem. Chem. Phys.}\ }\textbf {\bibinfo
  {volume} {11}},\ \bibinfo {pages} {4564--4569} (\bibinfo {year}
  {2021})}\BibitemShut {NoStop}%
\bibitem [{\citenamefont {Sch\"uller}\ \emph {et~al.}(2010)\citenamefont
  {Sch\"uller}, \citenamefont {Wethekam}, \citenamefont {Blauth}, \citenamefont
  {Winter}, \citenamefont {Aigner}, \citenamefont
  {Simonovi\ifmmode~\acute{c}\else \'{c}\fi{}}, \citenamefont {Solleder},
  \citenamefont {Burgd\"orfer},\ and\ \citenamefont
  {Wirtz}}]{Schuller_rumpling}%
  \BibitemOpen
  \bibfield  {author} {\bibinfo {author} {\bibfnamefont {A.}~\bibnamefont
  {Sch\"uller}}, \bibinfo {author} {\bibfnamefont {S.}~\bibnamefont
  {Wethekam}}, \bibinfo {author} {\bibfnamefont {D.}~\bibnamefont {Blauth}},
  \bibinfo {author} {\bibfnamefont {H.}~\bibnamefont {Winter}}, \bibinfo
  {author} {\bibfnamefont {F.}~\bibnamefont {Aigner}}, \bibinfo {author}
  {\bibfnamefont {N.}~\bibnamefont {Simonovi\ifmmode~\acute{c}\else
  \'{c}\fi{}}}, \bibinfo {author} {\bibfnamefont {B.}~\bibnamefont {Solleder}},
  \bibinfo {author} {\bibfnamefont {J.}~\bibnamefont {Burgd\"orfer}},\ and\
  \bibinfo {author} {\bibfnamefont {L.}~\bibnamefont {Wirtz}},\ }\bibfield
  {title} {\enquote {\bibinfo {title} {Rumpling of {LiF}(001) surface from fast
  atom diffraction},}\ }\href {https://doi.org/10.1103/PhysRevA.82.062902}
  {\bibfield  {journal} {\bibinfo  {journal} {Phys. Rev. A}\ }\textbf {\bibinfo
  {volume} {82}},\ \bibinfo {pages} {062902} (\bibinfo {year}
  {2010})}\BibitemShut {NoStop}%
\bibitem [{\citenamefont {Debiossac}, \citenamefont {Zugarramurdi}\ \emph
  {et~al.}(2014)\citenamefont {Debiossac}, \citenamefont {Zugarramurdi} \emph
  {et~al.}}]{Debiossac_PRB_2014}%
  \BibitemOpen
  \bibfield  {author} {\bibinfo {author} {\bibfnamefont {M.}~\bibnamefont
  {Debiossac}}, \bibinfo {author} {\bibfnamefont {A.}~\bibnamefont
  {Zugarramurdi}}, \emph {et~al.},\ }\bibfield  {title} {\enquote {\bibinfo
  {title} {Combined experimental and theoretical study of fast atom diffraction
  on the ${\ensuremath{\beta}}_{2}(2\ifmmode\times\else\texttimes\fi{}4)$
  reconstructed {GaAs}(001) surface},}\ }\href
  {https://doi.org/10.1103/PhysRevB.90.155308} {\bibfield  {journal} {\bibinfo
  {journal} {Phys. Rev. B}\ }\textbf {\bibinfo {volume} {90}},\ \bibinfo
  {pages} {155308} (\bibinfo {year} {2014})}\BibitemShut {NoStop}%
\bibitem [{\citenamefont {Pan}, \citenamefont {Debiossac},\ and\ \citenamefont
  {Roncin}(2021)}]{Pan_2021}%
  \BibitemOpen
  \bibfield  {author} {\bibinfo {author} {\bibfnamefont {P.}~\bibnamefont
  {Pan}}, \bibinfo {author} {\bibfnamefont {M.}~\bibnamefont {Debiossac}},\
  and\ \bibinfo {author} {\bibfnamefont {P.}~\bibnamefont {Roncin}},\
  }\bibfield  {title} {\enquote {\bibinfo {title} {Polar inelastic profiles in
  fast-atom diffraction at surfaces},}\ }\href
  {https://doi.org/10.1103/PhysRevB.104.165415} {\bibfield  {journal} {\bibinfo
   {journal} {Phys. Rev. B}\ }\textbf {\bibinfo {volume} {104}},\ \bibinfo
  {pages} {165415} (\bibinfo {year} {2021})}\BibitemShut {NoStop}%
\bibitem [{\citenamefont {Rousseau}\ \emph {et~al.}(2008)\citenamefont
  {Rousseau}, \citenamefont {Khemliche}, \citenamefont {Bundaleski},
  \citenamefont {Soulisse}, \citenamefont {Momeni},\ and\ \citenamefont
  {Roncin}}]{Rousseau_2008}%
  \BibitemOpen
  \bibfield  {author} {\bibinfo {author} {\bibfnamefont {P.}~\bibnamefont
  {Rousseau}}, \bibinfo {author} {\bibfnamefont {H.}~\bibnamefont {Khemliche}},
  \bibinfo {author} {\bibfnamefont {N.}~\bibnamefont {Bundaleski}}, \bibinfo
  {author} {\bibfnamefont {P.}~\bibnamefont {Soulisse}}, \bibinfo {author}
  {\bibfnamefont {A.}~\bibnamefont {Momeni}},\ and\ \bibinfo {author}
  {\bibfnamefont {P.}~\bibnamefont {Roncin}},\ }\bibfield  {title} {\enquote
  {\bibinfo {title} {Surface analysis with grazing incidence fast atom
  diffraction ({GIFAD})},}\ }\href
  {https://doi.org/10.1088/1742-6596/133/1/012013} {\bibfield  {journal}
  {\bibinfo  {journal} {Journal of Physics: Conference Series}\ }\textbf
  {\bibinfo {volume} {133}},\ \bibinfo {pages} {012013} (\bibinfo {year}
  {2008})}\BibitemShut {NoStop}%
\bibitem [{\citenamefont {Debiossac}\ and\ \citenamefont
  {Roncin}(2016{\natexlab{a}})}]{Debiossac_2016}%
  \BibitemOpen
  \bibfield  {author} {\bibinfo {author} {\bibfnamefont {M.}~\bibnamefont
  {Debiossac}}\ and\ \bibinfo {author} {\bibfnamefont {P.}~\bibnamefont
  {Roncin}},\ }\bibfield  {title} {\enquote {\bibinfo {title} {Image processing
  for grazing incidence fast atom diffraction},}\ }\href
  {https://doi.org/https://doi.org/10.1016/j.nimb.2016.05.023} {\bibfield
  {journal} {\bibinfo  {journal} {NIM-B}\ }\textbf {\bibinfo {volume} {382}},\
  \bibinfo {pages} {36} (\bibinfo {year} {2016}{\natexlab{a}})}\BibitemShut
  {NoStop}%
\bibitem [{\citenamefont {Atkinson}\ \emph {et~al.}(2014)\citenamefont
  {Atkinson}, \citenamefont {Eddrief}, \citenamefont {Etgens} \emph
  {et~al.}}]{Atkinson_2014}%
  \BibitemOpen
  \bibfield  {author} {\bibinfo {author} {\bibfnamefont {P.}~\bibnamefont
  {Atkinson}}, \bibinfo {author} {\bibfnamefont {M.}~\bibnamefont {Eddrief}},
  \bibinfo {author} {\bibfnamefont {V.~H.}\ \bibnamefont {Etgens}}, \emph
  {et~al.},\ }\bibfield  {title} {\enquote {\bibinfo {title} {Dynamic grazing
  incidence fast atom diffraction during molecular beam epitaxial growth of
  {GaAs}},}\ }\href {https://doi.org/10.1063/1.4890121} {\bibfield  {journal}
  {\bibinfo  {journal} {Applied Physics Letters}\ }\textbf {\bibinfo {volume}
  {105}},\ \bibinfo {pages} {021602} (\bibinfo {year} {2014})}\BibitemShut
  {NoStop}%
\bibitem [{\citenamefont {Bundaleski}\ \emph {et~al.}(2008)\citenamefont
  {Bundaleski}, \citenamefont {Khemliche}, \citenamefont {Soulisse},\ and\
  \citenamefont {Roncin}}]{Bundaleski_2008}%
  \BibitemOpen
  \bibfield  {author} {\bibinfo {author} {\bibfnamefont {N.}~\bibnamefont
  {Bundaleski}}, \bibinfo {author} {\bibfnamefont {H.}~\bibnamefont
  {Khemliche}}, \bibinfo {author} {\bibfnamefont {P.}~\bibnamefont
  {Soulisse}},\ and\ \bibinfo {author} {\bibfnamefont {P.}~\bibnamefont
  {Roncin}},\ }\bibfield  {title} {\enquote {\bibinfo {title} {Grazing
  incidence diffraction of ke{V H}elium atoms on a {Ag}(110) surface},}\ }\href
  {https://doi.org/10.1103/PhysRevLett.101.177601} {\bibfield  {journal}
  {\bibinfo  {journal} {Phys. Rev. Lett.}\ }\textbf {\bibinfo {volume} {101}},\
  \bibinfo {pages} {177601} (\bibinfo {year} {2008})}\BibitemShut {NoStop}%
\bibitem [{\citenamefont {Sch\"uller}\ \emph {et~al.}(2009)\citenamefont
  {Sch\"uller}, \citenamefont {Busch}, \citenamefont {Wethekam},\ and\
  \citenamefont {Winter}}]{Schueller_2009}%
  \BibitemOpen
  \bibfield  {author} {\bibinfo {author} {\bibfnamefont {A.}~\bibnamefont
  {Sch\"uller}}, \bibinfo {author} {\bibfnamefont {M.}~\bibnamefont {Busch}},
  \bibinfo {author} {\bibfnamefont {S.}~\bibnamefont {Wethekam}},\ and\
  \bibinfo {author} {\bibfnamefont {H.}~\bibnamefont {Winter}},\ }\bibfield
  {title} {\enquote {\bibinfo {title} {Fast atom diffraction from
  superstructures on a {Fe}(110) surface},}\ }\href
  {https://doi.org/10.1103/PhysRevLett.102.017602} {\bibfield  {journal}
  {\bibinfo  {journal} {Phys. Rev. Lett.}\ }\textbf {\bibinfo {volume} {102}},\
  \bibinfo {pages} {017602} (\bibinfo {year} {2009})}\BibitemShut {NoStop}%
\bibitem [{\citenamefont {Khemliche}\ \emph {et~al.}(2009)\citenamefont
  {Khemliche}, \citenamefont {Rousseau}, \citenamefont {Roncin}, \citenamefont
  {Etgens},\ and\ \citenamefont {Finocchi}}]{Khemliche_2009}%
  \BibitemOpen
  \bibfield  {author} {\bibinfo {author} {\bibfnamefont {H.}~\bibnamefont
  {Khemliche}}, \bibinfo {author} {\bibfnamefont {P.}~\bibnamefont {Rousseau}},
  \bibinfo {author} {\bibfnamefont {P.}~\bibnamefont {Roncin}}, \bibinfo
  {author} {\bibfnamefont {V.~H.}\ \bibnamefont {Etgens}},\ and\ \bibinfo
  {author} {\bibfnamefont {F.}~\bibnamefont {Finocchi}},\ }\bibfield  {title}
  {\enquote {\bibinfo {title} {Grazing incidence fast atom diffraction: An
  innovative approach to surface structure analysis},}\ }\href
  {https://doi.org/10.1063/1.3246162} {\bibfield  {journal} {\bibinfo
  {journal} {Applied Physics Letters}\ }\textbf {\bibinfo {volume} {95}},\
  \bibinfo {pages} {151901} (\bibinfo {year} {2009})}\BibitemShut {NoStop}%
\bibitem [{\citenamefont {Busch}\ \emph {et~al.}(2012)\citenamefont {Busch},
  \citenamefont {Seifert}, \citenamefont {Meyer},\ and\ \citenamefont
  {Winter}}]{Busch_2012}%
  \BibitemOpen
  \bibfield  {author} {\bibinfo {author} {\bibfnamefont {M.}~\bibnamefont
  {Busch}}, \bibinfo {author} {\bibfnamefont {J.}~\bibnamefont {Seifert}},
  \bibinfo {author} {\bibfnamefont {E.}~\bibnamefont {Meyer}},\ and\ \bibinfo
  {author} {\bibfnamefont {H.}~\bibnamefont {Winter}},\ }\bibfield  {title}
  {\enquote {\bibinfo {title} {Evidence for longitudinal coherence in fast atom
  diffraction},}\ }\href {https://doi.org/10.1103/PhysRevB.86.241402}
  {\bibfield  {journal} {\bibinfo  {journal} {Phys. Rev. B}\ }\textbf {\bibinfo
  {volume} {86}},\ \bibinfo {pages} {241402} (\bibinfo {year}
  {2012})}\BibitemShut {NoStop}%
\bibitem [{\citenamefont {Busch}\ \emph {et~al.}(2014)\citenamefont {Busch},
  \citenamefont {Meyer}, \citenamefont {Irmscher}, \citenamefont {Galazka},
  \citenamefont {G\"{a}rtner},\ and\ \citenamefont {Winter}}]{Busch_2014}%
  \BibitemOpen
  \bibfield  {author} {\bibinfo {author} {\bibfnamefont {M.}~\bibnamefont
  {Busch}}, \bibinfo {author} {\bibfnamefont {E.}~\bibnamefont {Meyer}},
  \bibinfo {author} {\bibfnamefont {K.}~\bibnamefont {Irmscher}}, \bibinfo
  {author} {\bibfnamefont {Z.}~\bibnamefont {Galazka}}, \bibinfo {author}
  {\bibfnamefont {K.}~\bibnamefont {G\"{a}rtner}},\ and\ \bibinfo {author}
  {\bibfnamefont {H.}~\bibnamefont {Winter}},\ }\bibfield  {title} {\enquote
  {\bibinfo {title} {Fast atom diffraction from a $\beta$-{G}a$_2${O}$_3$(100)
  surface},}\ }\href {https://doi.org/10.1063/1.4892350} {\bibfield  {journal}
  {\bibinfo  {journal} {Applied Physics Letters}\ }\textbf {\bibinfo {volume}
  {105}},\ \bibinfo {pages} {051603} (\bibinfo {year} {2014})}\BibitemShut
  {NoStop}%
\bibitem [{\citenamefont {Seifert}\ and\ \citenamefont
  {Winter}(2016)}]{Seifert_2016}%
  \BibitemOpen
  \bibfield  {author} {\bibinfo {author} {\bibfnamefont {J.}~\bibnamefont
  {Seifert}}\ and\ \bibinfo {author} {\bibfnamefont {H.}~\bibnamefont
  {Winter}},\ }\bibfield  {title} {\enquote {\bibinfo {title} {Quantitative
  structure determination using grazing scattering of fast atoms:
  Oxygen-induced missing-row reconstruction of {M}o(112)},}\ }\href
  {https://doi.org/10.1103/PhysRevB.93.205417} {\bibfield  {journal} {\bibinfo
  {journal} {Phys. Rev. B}\ }\textbf {\bibinfo {volume} {93}},\ \bibinfo
  {pages} {205417} (\bibinfo {year} {2016})}\BibitemShut {NoStop}%
\bibitem [{\citenamefont {Winter}\ \emph {et~al.}(2009)\citenamefont {Winter},
  \citenamefont {Seifert}, \citenamefont {Blauth}, \citenamefont {Busch},
  \citenamefont {Schüller},\ and\ \citenamefont {Wethekam}}]{Winter_2009}%
  \BibitemOpen
  \bibfield  {author} {\bibinfo {author} {\bibfnamefont {H.}~\bibnamefont
  {Winter}}, \bibinfo {author} {\bibfnamefont {J.}~\bibnamefont {Seifert}},
  \bibinfo {author} {\bibfnamefont {D.}~\bibnamefont {Blauth}}, \bibinfo
  {author} {\bibfnamefont {M.}~\bibnamefont {Busch}}, \bibinfo {author}
  {\bibfnamefont {A.}~\bibnamefont {Schüller}},\ and\ \bibinfo {author}
  {\bibfnamefont {S.}~\bibnamefont {Wethekam}},\ }\bibfield  {title} {\enquote
  {\bibinfo {title} {Structure of ultrathin oxide layers on metal surfaces from
  grazing scattering of fast atoms},}\ }\href
  {https://doi.org/https://doi.org/10.1016/j.apsusc.2009.05.019} {\bibfield
  {journal} {\bibinfo  {journal} {Applied Surface Science}\ }\textbf {\bibinfo
  {volume} {256}},\ \bibinfo {pages} {365--370} (\bibinfo {year}
  {2009})}\BibitemShut {NoStop}%
\bibitem [{\citenamefont {Seifert}\ \emph {et~al.}(2013)\citenamefont
  {Seifert}, \citenamefont {Busch}, \citenamefont {Meyer},\ and\ \citenamefont
  {Winter}}]{Seifert_2013}%
  \BibitemOpen
  \bibfield  {author} {\bibinfo {author} {\bibfnamefont {J.}~\bibnamefont
  {Seifert}}, \bibinfo {author} {\bibfnamefont {M.}~\bibnamefont {Busch}},
  \bibinfo {author} {\bibfnamefont {E.}~\bibnamefont {Meyer}},\ and\ \bibinfo
  {author} {\bibfnamefont {H.}~\bibnamefont {Winter}},\ }\bibfield  {title}
  {\enquote {\bibinfo {title} {{Surface Structure of Alanine on {C}u(110)
  Studied by Fast Atom Diffraction}},}\ }\href
  {https://doi.org/{10.1103/PhysRevLett.111.137601}} {\bibfield  {journal}
  {\bibinfo  {journal} {{Phys. Rev. Lett.}}\ }\textbf {\bibinfo {volume}
  {{111}}},\ \bibinfo {pages} {137601} (\bibinfo {year} {{2013}})}\BibitemShut
  {NoStop}%
\bibitem [{\citenamefont {Momeni}, \citenamefont {Staicu~Casagrande}\ \emph
  {et~al.}(2018)\citenamefont {Momeni}, \citenamefont {Staicu~Casagrande} \emph
  {et~al.}}]{Momeni_2018}%
  \BibitemOpen
  \bibfield  {author} {\bibinfo {author} {\bibfnamefont {A.}~\bibnamefont
  {Momeni}}, \bibinfo {author} {\bibfnamefont {E.~M.}\ \bibnamefont
  {Staicu~Casagrande}}, \emph {et~al.},\ }\bibfield  {title} {\enquote
  {\bibinfo {title} {Ultrafast crystallization dynamics at an
  organic–inorganic interface revealed in real time by grazing incidence fast
  atom diffraction},}\ }\href {https://doi.org/doi:
  10.1021/acs.jpclett.7b03246} {\bibfield  {journal} {\bibinfo  {journal} {J.
  Phys. Chem. Lett.}\ }\textbf {\bibinfo {volume} {9}},\ \bibinfo {pages} {908}
  (\bibinfo {year} {2018})}\BibitemShut {NoStop}%
\bibitem [{\citenamefont {Debiossac}\ \emph {et~al.}(2016)\citenamefont
  {Debiossac}, \citenamefont {Zugarramurdi}, \citenamefont {Mu} \emph
  {et~al.}}]{Debiossac_graphene}%
  \BibitemOpen
  \bibfield  {author} {\bibinfo {author} {\bibfnamefont {M.}~\bibnamefont
  {Debiossac}}, \bibinfo {author} {\bibfnamefont {A.}~\bibnamefont
  {Zugarramurdi}}, \bibinfo {author} {\bibfnamefont {Z.}~\bibnamefont {Mu}},
  \emph {et~al.},\ }\bibfield  {title} {\enquote {\bibinfo {title} {Helium
  diffraction on {SiC} grown graphene: Qualitative and quantitative
  descriptions with the hard-corrugated-wall model},}\ }\href
  {https://doi.org/10.1103/PhysRevB.94.205403} {\bibfield  {journal} {\bibinfo
  {journal} {Phys. Rev. B}\ }\textbf {\bibinfo {volume} {94}},\ \bibinfo
  {pages} {205403} (\bibinfo {year} {2016})}\BibitemShut {NoStop}%
\bibitem [{\citenamefont {Zugarramurdi}\ \emph {et~al.}(2015)\citenamefont
  {Zugarramurdi}, \citenamefont {Debiossac}, \citenamefont {Lunca-Popa},
  \citenamefont {Mayne} \emph {et~al.}}]{zugarramurdi_2015}%
  \BibitemOpen
  \bibfield  {author} {\bibinfo {author} {\bibfnamefont {A.}~\bibnamefont
  {Zugarramurdi}}, \bibinfo {author} {\bibfnamefont {M.}~\bibnamefont
  {Debiossac}}, \bibinfo {author} {\bibfnamefont {P.}~\bibnamefont
  {Lunca-Popa}}, \bibinfo {author} {\bibfnamefont {A.}~\bibnamefont {Mayne}},
  \emph {et~al.},\ }\bibfield  {title} {\enquote {\bibinfo {title}
  {Determination of the geometric corrugation of graphene on {SiC} (0001) by
  grazing incidence fast atom diffraction},}\ }\href@noop {} {\bibfield
  {journal} {\bibinfo  {journal} {Appl. Phys. Lett}\ }\textbf {\bibinfo
  {volume} {106}},\ \bibinfo {pages} {101902} (\bibinfo {year}
  {2015})}\BibitemShut {NoStop}%
\bibitem [{\citenamefont {Magee}\ and\ \citenamefont
  {Honig}(1982)}]{Depth_profiling}%
  \BibitemOpen
  \bibfield  {author} {\bibinfo {author} {\bibfnamefont {C.~W.}\ \bibnamefont
  {Magee}}\ and\ \bibinfo {author} {\bibfnamefont {R.~E.}\ \bibnamefont
  {Honig}},\ }\bibfield  {title} {\enquote {\bibinfo {title} {Depth profiling
  by sims—depth resolution, dynamic range and sensitivity},}\ }\href
  {https://doi.org/https://doi.org/10.1002/sia.740040202} {\bibfield  {journal}
  {\bibinfo  {journal} {Surf. Interface Anal.}\ }\textbf {\bibinfo {volume}
  {4}},\ \bibinfo {pages} {35--41} (\bibinfo {year} {1982})}\BibitemShut
  {NoStop}%
\bibitem [{non()}]{nonsequitur}%
  \BibitemOpen
  \href {https://nonsequitur-ion-gun.com/Model_1402.html} {\emph {\bibinfo
  {title} {nonsequitur\textregistered, hot filament ion source model 1402 for
  Low Energy Performance}}}\BibitemShut {NoStop}%
\bibitem [{\citenamefont {Morosov}\ \emph {et~al.}(1996)\citenamefont
  {Morosov}, \citenamefont {Kalinin}, \citenamefont {Szilagyi}, \citenamefont
  {Barat},\ and\ \citenamefont {Roncin}}]{Morosov_1996}%
  \BibitemOpen
  \bibfield  {author} {\bibinfo {author} {\bibfnamefont {V.~A.}\ \bibnamefont
  {Morosov}}, \bibinfo {author} {\bibfnamefont {A.}~\bibnamefont {Kalinin}},
  \bibinfo {author} {\bibfnamefont {Z.}~\bibnamefont {Szilagyi}}, \bibinfo
  {author} {\bibfnamefont {M.}~\bibnamefont {Barat}},\ and\ \bibinfo {author}
  {\bibfnamefont {P.}~\bibnamefont {Roncin}},\ }\bibfield  {title} {\enquote
  {\bibinfo {title} {2$\pi$ spectrometer: A new apparatus for the investigation
  of ion surface interaction},}\ }\href {https://doi.org/10.1063/1.1147031}
  {\bibfield  {journal} {\bibinfo  {journal} {Rev. Sci. Instrum.}\ }\textbf
  {\bibinfo {volume} {67}},\ \bibinfo {pages} {2163--2170} (\bibinfo {year}
  {1996})}\BibitemShut {NoStop}%
\bibitem [{Pan()}]{Pantechnik}%
  \BibitemOpen
  \href {https://www.pantechnik.com/wp-content/uploads/2020/07/Nanogan.pdf}
  {\emph {\bibinfo {title} {Nanogan\textregistered, a permanent magnet 10GHz
  electron cyclotron resonance ion source}}}\BibitemShut {NoStop}%
\bibitem [{Pol()}]{Polygon_Physics}%
  \BibitemOpen
  \href {https://polygonphysics.com/products/tes-product-line/} {\emph
  {\bibinfo {title} {Polygon Physics\textregistered TES-40 miniature electron
  cyclotron resonance ion source}}}\BibitemShut {NoStop}%
\bibitem [{\citenamefont {Gao}\ \emph {et~al.}(1988)\citenamefont {Gao},
  \citenamefont {Johnson}, \citenamefont {Schafer}, \citenamefont {Newman},
  \citenamefont {Smith},\ and\ \citenamefont {Stebbings}}]{Gao_1988}%
  \BibitemOpen
  \bibfield  {author} {\bibinfo {author} {\bibfnamefont {R.~S.}\ \bibnamefont
  {Gao}}, \bibinfo {author} {\bibfnamefont {L.~K.}\ \bibnamefont {Johnson}},
  \bibinfo {author} {\bibfnamefont {D.~A.}\ \bibnamefont {Schafer}}, \bibinfo
  {author} {\bibfnamefont {J.~H.}\ \bibnamefont {Newman}}, \bibinfo {author}
  {\bibfnamefont {K.~A.}\ \bibnamefont {Smith}},\ and\ \bibinfo {author}
  {\bibfnamefont {R.~F.}\ \bibnamefont {Stebbings}},\ }\bibfield  {title}
  {\enquote {\bibinfo {title} {Absolute differential cross sections for
  small-angle {He}$^{+}$-{H}e elastic and charge-transfer scattering at ke{V}
  energies},}\ }\href {https://doi.org/10.1103/PhysRevA.38.2789} {\bibfield
  {journal} {\bibinfo  {journal} {Phys. Rev. A}\ }\textbf {\bibinfo {volume}
  {38}},\ \bibinfo {pages} {2789--2793} (\bibinfo {year} {1988})}\BibitemShut
  {NoStop}%
\bibitem [{Note1()}]{Note1}%
  \BibitemOpen
  \bibinfo {note} {The same value of the cross-section probably limits the
  maximum operating pressure of GIFAD around $10^{-4}$mb for diffraction, but
  higher pressures are possible for triangulation}\BibitemShut {NoStop}%
\bibitem [{\citenamefont {Debiossac}\ \emph {et~al.}(2014)\citenamefont
  {Debiossac}, \citenamefont {Zugarramurdi}, \citenamefont {Lunca-Popa},
  \citenamefont {Momeni}, \citenamefont {Khemliche}, \citenamefont {Borisov},\
  and\ \citenamefont {Roncin}}]{Debiossac_PRL_2014}%
  \BibitemOpen
  \bibfield  {author} {\bibinfo {author} {\bibfnamefont {M.}~\bibnamefont
  {Debiossac}}, \bibinfo {author} {\bibfnamefont {A.}~\bibnamefont
  {Zugarramurdi}}, \bibinfo {author} {\bibfnamefont {P.}~\bibnamefont
  {Lunca-Popa}}, \bibinfo {author} {\bibfnamefont {A.}~\bibnamefont {Momeni}},
  \bibinfo {author} {\bibfnamefont {H.}~\bibnamefont {Khemliche}}, \bibinfo
  {author} {\bibfnamefont {A.~G.}\ \bibnamefont {Borisov}},\ and\ \bibinfo
  {author} {\bibfnamefont {P.}~\bibnamefont {Roncin}},\ }\bibfield  {title}
  {\enquote {\bibinfo {title} {Transient quantum trapping of fast atoms at
  surfaces},}\ }\href {https://doi.org/10.1103/PhysRevLett.112.023203}
  {\bibfield  {journal} {\bibinfo  {journal} {Phys. Rev. Lett.}\ }\textbf
  {\bibinfo {volume} {112}},\ \bibinfo {pages} {023203} (\bibinfo {year}
  {2014})}\BibitemShut {NoStop}%
\bibitem [{\citenamefont {Soulisse}(2011)}]{soulisse_2011}%
  \BibitemOpen
  \bibfield  {author} {\bibinfo {author} {\bibfnamefont {P.}~\bibnamefont
  {Soulisse}},\ }\emph {\bibinfo {title} {Développement d'un dispositif
  expérimental pour la diffraction d'atomes rapides et étude de surfaces
  d'isolants ioniques}},\ \href {http://www.theses.fr/2011PA112118/document}
  {Ph.D. thesis},\ \bibinfo  {school} {Université Paris Sud 11} (\bibinfo
  {year} {2011}),\ \bibinfo {note} {thèse dirigée par P.Roncin}\BibitemShut
  {NoStop}%
\bibitem [{\citenamefont {Pan}, \citenamefont {Debiossac},\ and\ \citenamefont
  {Roncin}(2022)}]{Pan_2022}%
  \BibitemOpen
  \bibfield  {author} {\bibinfo {author} {\bibfnamefont {P.}~\bibnamefont
  {Pan}}, \bibinfo {author} {\bibfnamefont {M.}~\bibnamefont {Debiossac}},\
  and\ \bibinfo {author} {\bibfnamefont {P.}~\bibnamefont {Roncin}},\
  }\bibfield  {title} {\enquote {\bibinfo {title} {Temperature dependence in
  fast-atom diffraction at surfaces},}\ }\href
  {https://doi.org/10.1039/D2CP00829G} {\bibfield  {journal} {\bibinfo
  {journal} {Phys. Chem. Chem. Phys.}\ ,\ \bibinfo {pages} {--}} (\bibinfo
  {year} {2022})}\BibitemShut {NoStop}%
\bibitem [{\citenamefont {Fehre}\ \emph {et~al.}(2018)\citenamefont {Fehre},
  \citenamefont {Trojanowskaja}, \citenamefont {Gatzke} \emph
  {et~al.}}]{Jagutzki_2018}%
  \BibitemOpen
  \bibfield  {author} {\bibinfo {author} {\bibfnamefont {K.}~\bibnamefont
  {Fehre}}, \bibinfo {author} {\bibfnamefont {D.}~\bibnamefont
  {Trojanowskaja}}, \bibinfo {author} {\bibfnamefont {J.}~\bibnamefont
  {Gatzke}}, \emph {et~al.},\ }\bibfield  {title} {\enquote {\bibinfo {title}
  {Absolute ion detection efficiencies of microchannel plates and funnel
  microchannel plates for multi-coincidence detection},}\ }\href
  {https://doi.org/10.1063/1.5022564} {\bibfield  {journal} {\bibinfo
  {journal} {Rev. Sci. Instrum.}\ }\textbf {\bibinfo {volume} {89}},\ \bibinfo
  {pages} {045112} (\bibinfo {year} {2018})}\BibitemShut {NoStop}%
\bibitem [{\citenamefont {Lapington}(2004)}]{Lapington_2004}%
  \BibitemOpen
  \bibfield  {author} {\bibinfo {author} {\bibfnamefont {J.}~\bibnamefont
  {Lapington}},\ }\bibfield  {title} {\enquote {\bibinfo {title} {A comparison
  of readout techniques for high-resolution imaging with microchannel plate
  detectors},}\ }\href
  {https://doi.org/https://doi.org/10.1016/j.nima.2004.03.096} {\bibfield
  {journal} {\bibinfo  {journal} {NIM-A}\ }\textbf {\bibinfo {volume} {525}},\
  \bibinfo {pages} {361--365} (\bibinfo {year} {2004})}\BibitemShut {NoStop}%
\bibitem [{\citenamefont {Hong}\ \emph {et~al.}(2016)\citenamefont {Hong},
  \citenamefont {Leredde}, \citenamefont {Bagdasarova}, \citenamefont
  {Fléchard} \emph {et~al.}}]{Hong_2016}%
  \BibitemOpen
  \bibfield  {author} {\bibinfo {author} {\bibfnamefont {R.}~\bibnamefont
  {Hong}}, \bibinfo {author} {\bibfnamefont {A.}~\bibnamefont {Leredde}},
  \bibinfo {author} {\bibfnamefont {Y.}~\bibnamefont {Bagdasarova}}, \bibinfo
  {author} {\bibfnamefont {X.}~\bibnamefont {Fléchard}}, \emph {et~al.},\
  }\bibfield  {title} {\enquote {\bibinfo {title} {High accuracy position
  response calibration method for a micro-channel plate ion detector},}\ }\href
  {https://doi.org/https://doi.org/10.1016/j.nima.2016.08.024} {\bibfield
  {journal} {\bibinfo  {journal} {NIM-A}\ }\textbf {\bibinfo {volume} {835}},\
  \bibinfo {pages} {42} (\bibinfo {year} {2016})}\BibitemShut {NoStop}%
\bibitem [{\citenamefont {Lupone}, \citenamefont {Soulisse},\ and\
  \citenamefont {Roncin}(2018)}]{Lupone_2018}%
  \BibitemOpen
  \bibfield  {author} {\bibinfo {author} {\bibfnamefont {S.}~\bibnamefont
  {Lupone}}, \bibinfo {author} {\bibfnamefont {P.}~\bibnamefont {Soulisse}},\
  and\ \bibinfo {author} {\bibfnamefont {P.}~\bibnamefont {Roncin}},\
  }\bibfield  {title} {\enquote {\bibinfo {title} {A large area high resolution
  imaging detector for fast atom diffraction},}\ }\href
  {https://doi.org/https://doi.org/10.1016/j.nimb.2018.04.030} {\bibfield
  {journal} {\bibinfo  {journal} {NIM-B}\ }\textbf {\bibinfo {volume} {427}},\
  \bibinfo {pages} {95 -- 99} (\bibinfo {year} {2018})}\BibitemShut {NoStop}%
\bibitem [{Sch()}]{Schneider}%
  \BibitemOpen
  \href
  {https://schneiderkreuznach.com/en/industrial-optics/lenses/fast-lenses}
  {\emph {\bibinfo {title} {Xenon 0.95/17-0010 F0.95 aperture C-mount lens from
  Schneider-Kreuznach\textregistered}}}\BibitemShut {NoStop}%
\bibitem [{\citenamefont {Lupone}\ \emph {et~al.}(2015)\citenamefont {Lupone},
  \citenamefont {Damoy}, \citenamefont {Husseen}, \citenamefont {Briand},
  \citenamefont {Debiossac} \emph {et~al.}}]{Lupone_2015}%
  \BibitemOpen
  \bibfield  {author} {\bibinfo {author} {\bibfnamefont {S.}~\bibnamefont
  {Lupone}}, \bibinfo {author} {\bibfnamefont {S.}~\bibnamefont {Damoy}},
  \bibinfo {author} {\bibfnamefont {A.}~\bibnamefont {Husseen}}, \bibinfo
  {author} {\bibfnamefont {N.}~\bibnamefont {Briand}}, \bibinfo {author}
  {\bibfnamefont {M.}~\bibnamefont {Debiossac}}, \emph {et~al.},\ }\bibfield
  {title} {\enquote {\bibinfo {title} {Note: A large open ratio, time, and
  position sensitive detector for time of flight measurements in uhv},}\ }\href
  {https://doi.org/10.1063/1.4939195} {\bibfield  {journal} {\bibinfo
  {journal} {Rev. Sci. Instrum.}\ }\textbf {\bibinfo {volume} {86}},\ \bibinfo
  {pages} {126115} (\bibinfo {year} {2015})}\BibitemShut {NoStop}%
\bibitem [{\citenamefont {Miret-Artés}\ and\ \citenamefont
  {Pollak}(2012)}]{Miret_Artes_2012}%
  \BibitemOpen
  \bibfield  {author} {\bibinfo {author} {\bibfnamefont {S.}~\bibnamefont
  {Miret-Artés}}\ and\ \bibinfo {author} {\bibfnamefont {E.}~\bibnamefont
  {Pollak}},\ }\bibfield  {title} {\enquote {\bibinfo {title} {Classical theory
  of atom–surface scattering: The rainbow effect},}\ }\href
  {https://doi.org/https://doi.org/10.1016/j.surfrep.2012.03.001} {\bibfield
  {journal} {\bibinfo  {journal} {Surface Science Reports}\ }\textbf {\bibinfo
  {volume} {67}},\ \bibinfo {pages} {161--200} (\bibinfo {year}
  {2012})}\BibitemShut {NoStop}%
\bibitem [{\citenamefont {Winter}(2002)}]{winter_PR2002}%
  \BibitemOpen
  \bibfield  {author} {\bibinfo {author} {\bibfnamefont {H.}~\bibnamefont
  {Winter}},\ }\bibfield  {title} {\enquote {\bibinfo {title} {Collisions of
  atoms and ions with surfaces under grazing incidence},}\ }\href
  {https://doi.org/https://doi.org/10.1016/S0370-1573(02)00010-8} {\bibfield
  {journal} {\bibinfo  {journal} {Physics Reports}\ }\textbf {\bibinfo {volume}
  {367}},\ \bibinfo {pages} {387--582} (\bibinfo {year} {2002})}\BibitemShut
  {NoStop}%
\bibitem [{\citenamefont {Villette}\ \emph {et~al.}(2000)\citenamefont
  {Villette}, \citenamefont {Borisov}, \citenamefont {Khemliche}, \citenamefont
  {Momeni},\ and\ \citenamefont {Roncin}}]{Villette_2000}%
  \BibitemOpen
  \bibfield  {author} {\bibinfo {author} {\bibfnamefont {J.}~\bibnamefont
  {Villette}}, \bibinfo {author} {\bibfnamefont {A.~G.}\ \bibnamefont
  {Borisov}}, \bibinfo {author} {\bibfnamefont {H.}~\bibnamefont {Khemliche}},
  \bibinfo {author} {\bibfnamefont {A.}~\bibnamefont {Momeni}},\ and\ \bibinfo
  {author} {\bibfnamefont {P.}~\bibnamefont {Roncin}},\ }\bibfield  {title}
  {\enquote {\bibinfo {title} {Subsurface-channeling-like energy loss structure
  of the skipping motion on an ionic crystal},}\ }\href
  {https://doi.org/10.1103/PhysRevLett.85.3137} {\bibfield  {journal} {\bibinfo
   {journal} {Phys. Rev. Lett.}\ }\textbf {\bibinfo {volume} {85}},\ \bibinfo
  {pages} {3137--3140} (\bibinfo {year} {2000})}\BibitemShut {NoStop}%
\bibitem [{\citenamefont {Debiossac}\ and\ \citenamefont
  {Roncin}(2016{\natexlab{b}})}]{Debiossac_Nim_2016}%
  \BibitemOpen
  \bibfield  {author} {\bibinfo {author} {\bibfnamefont {M.}~\bibnamefont
  {Debiossac}}\ and\ \bibinfo {author} {\bibfnamefont {P.}~\bibnamefont
  {Roncin}},\ }\bibfield  {title} {\enquote {\bibinfo {title} {Image processing
  for grazing incidence fast atom diffraction},}\ }\href
  {https://doi.org/https://doi.org/10.1016/j.nimb.2016.05.023} {\bibfield
  {journal} {\bibinfo  {journal} {NIM-B}\ }\textbf {\bibinfo {volume} {382}},\
  \bibinfo {pages} {36} (\bibinfo {year} {2016}{\natexlab{b}})}\BibitemShut
  {NoStop}%
\bibitem [{\citenamefont {Debiossac}, \citenamefont {Roncin},\ and\
  \citenamefont {Borisov}(2020)}]{Debiossac_2020}%
  \BibitemOpen
  \bibfield  {author} {\bibinfo {author} {\bibfnamefont {M.}~\bibnamefont
  {Debiossac}}, \bibinfo {author} {\bibfnamefont {P.}~\bibnamefont {Roncin}},\
  and\ \bibinfo {author} {\bibfnamefont {A.}~\bibnamefont {Borisov}},\
  }\bibfield  {title} {\enquote {\bibinfo {title} {Refraction of fast {N}e
  atoms in the attractive well of a {LiF}(001) surface},}\ }\href
  {https://doi.org/10.1021/acs.jpclett.0c01157} {\bibfield  {journal} {\bibinfo
   {journal} {J. Phys. Chem. Lett.}\ }\textbf {\bibinfo {volume} {11}},\
  \bibinfo {pages} {4564--4569} (\bibinfo {year} {2020})}\BibitemShut {NoStop}%
\bibitem [{\citenamefont {Meyer}(2016)}]{Meyer_2016}%
  \BibitemOpen
  \bibfield  {author} {\bibinfo {author} {\bibfnamefont {E.}~\bibnamefont
  {Meyer}},\ }\emph {\bibinfo {title} {Strukturuntersuchungen an
  Oxidkristalloberflächen mittels der streifenden Streuung schneller Atome}},\
  \href {https://doi.org/http://dx.doi.org/10.18452/17442} {Ph.D. thesis},\
  \bibinfo  {school} {Humboldt-Universität zu Berlin,
  Mathematisch-Naturwissenschaftliche Fakultät} (\bibinfo {year}
  {2016})\BibitemShut {NoStop}%
\bibitem [{\citenamefont {Bocan}\ and\ \citenamefont
  {Gravielle}(2018)}]{Bocan_2018}%
  \BibitemOpen
  \bibfield  {author} {\bibinfo {author} {\bibfnamefont {G.}~\bibnamefont
  {Bocan}}\ and\ \bibinfo {author} {\bibfnamefont {M.}~\bibnamefont
  {Gravielle}},\ }\bibfield  {title} {\enquote {\bibinfo {title} {{GIFAD} for
  {He/KCl}(001). structure in the pattern for [110] incidence as a measure of
  the projectile-cation interaction},}\ }\href
  {https://doi.org/https://doi.org/10.1016/j.nimb.2018.02.004} {\bibfield
  {journal} {\bibinfo  {journal} {NIM-B}\ }\textbf {\bibinfo {volume} {421}},\
  \bibinfo {pages} {1--6} (\bibinfo {year} {2018})}\BibitemShut {NoStop}%
\bibitem [{\citenamefont {Zugarramurdi}\ and\ \citenamefont
  {Borisov}(2012)}]{Zugarramurdi_2012}%
  \BibitemOpen
  \bibfield  {author} {\bibinfo {author} {\bibfnamefont {A.}~\bibnamefont
  {Zugarramurdi}}\ and\ \bibinfo {author} {\bibfnamefont {A.}~\bibnamefont
  {Borisov}},\ }\bibfield  {title} {\enquote {\bibinfo {title} {Transition from
  fast to slow atom diffraction},}\ }\href
  {https://doi.org/10.1103/PhysRevA.86.062903} {\bibfield  {journal} {\bibinfo
  {journal} {Phys. Rev. A}\ }\textbf {\bibinfo {volume} {86}},\ \bibinfo
  {pages} {062903} (\bibinfo {year} {2012})}\BibitemShut {NoStop}%
\bibitem [{\citenamefont {Muzas}\ \emph {et~al.}(2016)\citenamefont {Muzas},
  \citenamefont {Gatti}, \citenamefont {Mart\'{\i}n},\ and\ \citenamefont
  {D\'{\i}az}}]{Diaz_2016b}%
  \BibitemOpen
  \bibfield  {author} {\bibinfo {author} {\bibfnamefont {A.}~\bibnamefont
  {Muzas}}, \bibinfo {author} {\bibfnamefont {F.}~\bibnamefont {Gatti}},
  \bibinfo {author} {\bibfnamefont {F.}~\bibnamefont {Mart\'{\i}n}},\ and\
  \bibinfo {author} {\bibfnamefont {C.}~\bibnamefont {D\'{\i}az}},\ }\bibfield
  {title} {\enquote {\bibinfo {title} {Diffraction of {H} from {LiF}(001): From
  slow normal incidence to fast grazing incidence},}\ }\href
  {https://doi.org/https://doi.org/10.1016/j.nimb.2016.04.031} {\bibfield
  {journal} {\bibinfo  {journal} {NIM-B}\ }\textbf {\bibinfo {volume} {382}},\
  \bibinfo {pages} {49 -- 53} (\bibinfo {year} {2016})}\BibitemShut {NoStop}%
\bibitem [{\citenamefont {Schüller}\ \emph {et~al.}(2005)\citenamefont
  {Schüller}, \citenamefont {Wethekam}, \citenamefont {Mertens}, \citenamefont
  {Maass}, \citenamefont {Winter},\ and\ \citenamefont
  {Gärtner}}]{Schuller_2005}%
  \BibitemOpen
  \bibfield  {author} {\bibinfo {author} {\bibfnamefont {A.}~\bibnamefont
  {Schüller}}, \bibinfo {author} {\bibfnamefont {S.}~\bibnamefont {Wethekam}},
  \bibinfo {author} {\bibfnamefont {A.}~\bibnamefont {Mertens}}, \bibinfo
  {author} {\bibfnamefont {K.}~\bibnamefont {Maass}}, \bibinfo {author}
  {\bibfnamefont {H.}~\bibnamefont {Winter}},\ and\ \bibinfo {author}
  {\bibfnamefont {K.}~\bibnamefont {Gärtner}},\ }\bibfield  {title} {\enquote
  {\bibinfo {title} {Interatomic potentials from rainbow scattering of ke{V}
  noble gas atoms under axial surface channeling},}\ }\href
  {https://doi.org/https://doi.org/10.1016/j.nimb.2004.12.036} {\bibfield
  {journal} {\bibinfo  {journal} {NIM-B}\ }\textbf {\bibinfo {volume} {230}},\
  \bibinfo {pages} {172--177} (\bibinfo {year} {2005})}\BibitemShut {NoStop}%
\bibitem [{\citenamefont {Danailov}\ \emph {et~al.}(2001)\citenamefont
  {Danailov}, \citenamefont {Pfandzelter}, \citenamefont {Igel}, \citenamefont
  {Winter},\ and\ \citenamefont {Gärtner}}]{Danailov_2001}%
  \BibitemOpen
  \bibfield  {author} {\bibinfo {author} {\bibfnamefont {D.}~\bibnamefont
  {Danailov}}, \bibinfo {author} {\bibfnamefont {R.}~\bibnamefont
  {Pfandzelter}}, \bibinfo {author} {\bibfnamefont {T.}~\bibnamefont {Igel}},
  \bibinfo {author} {\bibfnamefont {H.}~\bibnamefont {Winter}},\ and\ \bibinfo
  {author} {\bibfnamefont {K.}~\bibnamefont {Gärtner}},\ }\bibfield  {title}
  {\enquote {\bibinfo {title} {Test of the interatomic potential in the
  e{V}-region by glancing-angle scattering of he-atoms from {F}e(0 0 1)},}\
  }\href {https://doi.org/https://doi.org/10.1016/S0169-4332(00)00547-X}
  {\bibfield  {journal} {\bibinfo  {journal} {Applied Surface Science}\
  }\textbf {\bibinfo {volume} {171}},\ \bibinfo {pages} {113--119} (\bibinfo
  {year} {2001})}\BibitemShut {NoStop}%
\bibitem [{\citenamefont {Far\'{\i}as}\ \emph {et~al.}(2004)\citenamefont
  {Far\'{\i}as}, \citenamefont {D\'{\i}az}, \citenamefont {Rivi\`ere},
  \citenamefont {Busnengo}, \citenamefont {Nieto}, \citenamefont {Somers},
  \citenamefont {Kroes}, \citenamefont {Salin},\ and\ \citenamefont
  {Mart\'{\i}n}}]{Farias_2004}%
  \BibitemOpen
  \bibfield  {author} {\bibinfo {author} {\bibfnamefont {D.}~\bibnamefont
  {Far\'{\i}as}}, \bibinfo {author} {\bibfnamefont {C.}~\bibnamefont
  {D\'{\i}az}}, \bibinfo {author} {\bibfnamefont {P.}~\bibnamefont
  {Rivi\`ere}}, \bibinfo {author} {\bibfnamefont {H.~F.}\ \bibnamefont
  {Busnengo}}, \bibinfo {author} {\bibfnamefont {P.}~\bibnamefont {Nieto}},
  \bibinfo {author} {\bibfnamefont {M.~F.}\ \bibnamefont {Somers}}, \bibinfo
  {author} {\bibfnamefont {G.~J.}\ \bibnamefont {Kroes}}, \bibinfo {author}
  {\bibfnamefont {A.}~\bibnamefont {Salin}},\ and\ \bibinfo {author}
  {\bibfnamefont {F.}~\bibnamefont {Mart\'{\i}n}},\ }\bibfield  {title}
  {\enquote {\bibinfo {title} {In-plane and out-of-plane diffraction of
  ${\mathrm{h}}_{2}$ from metal surfaces},}\ }\href
  {https://doi.org/10.1103/PhysRevLett.93.246104} {\bibfield  {journal}
  {\bibinfo  {journal} {Phys. Rev. Lett.}\ }\textbf {\bibinfo {volume} {93}},\
  \bibinfo {pages} {246104} (\bibinfo {year} {2004})}\BibitemShut {NoStop}%
\bibitem [{\citenamefont {Kalashnyk}, \citenamefont {Khemliche},\ and\
  \citenamefont {Roncin}(2016)}]{Kalashnyk_2016}%
  \BibitemOpen
  \bibfield  {author} {\bibinfo {author} {\bibfnamefont {N.}~\bibnamefont
  {Kalashnyk}}, \bibinfo {author} {\bibfnamefont {H.}~\bibnamefont
  {Khemliche}},\ and\ \bibinfo {author} {\bibfnamefont {P.}~\bibnamefont
  {Roncin}},\ }\bibfield  {title} {\enquote {\bibinfo {title} {Atom beam
  triangulation of organic layers at 100me{V} normal energy: self-assembled
  perylene on {Ag}(110) at room temperature},}\ }\href
  {https://doi.org/https://doi.org/10.1016/j.apsusc.2015.12.134} {\bibfield
  {journal} {\bibinfo  {journal} {Applied Surface Science}\ }\textbf {\bibinfo
  {volume} {364}},\ \bibinfo {pages} {235} (\bibinfo {year}
  {2016})}\BibitemShut {NoStop}%
\bibitem [{\citenamefont {Alyabyeva}\ \emph {et~al.}(2018)\citenamefont
  {Alyabyeva}, \citenamefont {Ouvrard}, \citenamefont {Zakaria}, \citenamefont
  {Charra},\ and\ \citenamefont {Bourguignon}}]{Alyabyeva_2018}%
  \BibitemOpen
  \bibfield  {author} {\bibinfo {author} {\bibfnamefont {N.}~\bibnamefont
  {Alyabyeva}}, \bibinfo {author} {\bibfnamefont {A.}~\bibnamefont {Ouvrard}},
  \bibinfo {author} {\bibfnamefont {A.-M.}\ \bibnamefont {Zakaria}}, \bibinfo
  {author} {\bibfnamefont {F.}~\bibnamefont {Charra}},\ and\ \bibinfo {author}
  {\bibfnamefont {B.}~\bibnamefont {Bourguignon}},\ }\bibfield  {title}
  {\enquote {\bibinfo {title} {Transition from disordered to long-range ordered
  nanoparticles on {A}l$_2${O}$_3$/{N}i$_3${A}l(111)},}\ }\href
  {https://doi.org/https://doi.org/10.1016/j.apsusc.2018.03.025} {\bibfield
  {journal} {\bibinfo  {journal} {Applied Surface Science}\ }\textbf {\bibinfo
  {volume} {444}},\ \bibinfo {pages} {423--429} (\bibinfo {year}
  {2018})}\BibitemShut {NoStop}%
\bibitem [{\citenamefont {Zugarramurdi}\ \emph {et~al.}(2013)\citenamefont
  {Zugarramurdi}, \citenamefont {Debiossac}, \citenamefont {Lunca-Popa},
  \citenamefont {Alarc\'on}, \citenamefont {Momeni}, \citenamefont {Khemliche},
  \citenamefont {Roncin},\ and\ \citenamefont {Borisov}}]{Zugarramurdi_2013b}%
  \BibitemOpen
  \bibfield  {author} {\bibinfo {author} {\bibfnamefont {A.}~\bibnamefont
  {Zugarramurdi}}, \bibinfo {author} {\bibfnamefont {M.}~\bibnamefont
  {Debiossac}}, \bibinfo {author} {\bibfnamefont {P.}~\bibnamefont
  {Lunca-Popa}}, \bibinfo {author} {\bibfnamefont {L.~S.}\ \bibnamefont
  {Alarc\'on}}, \bibinfo {author} {\bibfnamefont {A.}~\bibnamefont {Momeni}},
  \bibinfo {author} {\bibfnamefont {H.}~\bibnamefont {Khemliche}}, \bibinfo
  {author} {\bibfnamefont {P.}~\bibnamefont {Roncin}},\ and\ \bibinfo {author}
  {\bibfnamefont {A.~G.}\ \bibnamefont {Borisov}},\ }\bibfield  {title}
  {\enquote {\bibinfo {title} {Surface-grating deflection of fast atom
  beams},}\ }\href {https://doi.org/10.1103/PhysRevA.88.012904} {\bibfield
  {journal} {\bibinfo  {journal} {Phys. Rev. A}\ }\textbf {\bibinfo {volume}
  {88}},\ \bibinfo {pages} {012904} (\bibinfo {year} {2013})}\BibitemShut
  {NoStop}%
\bibitem [{\citenamefont {Seifert}\ \emph {et~al.}(2011)\citenamefont
  {Seifert}, \citenamefont {Schüller}, \citenamefont {Winter},\ and\
  \citenamefont {Gärtner}}]{Seifert_2011}%
  \BibitemOpen
  \bibfield  {author} {\bibinfo {author} {\bibfnamefont {J.}~\bibnamefont
  {Seifert}}, \bibinfo {author} {\bibfnamefont {A.}~\bibnamefont {Schüller}},
  \bibinfo {author} {\bibfnamefont {H.}~\bibnamefont {Winter}},\ and\ \bibinfo
  {author} {\bibfnamefont {K.}~\bibnamefont {Gärtner}},\ }\bibfield  {title}
  {\enquote {\bibinfo {title} {Transition from axial to planar surface
  channeling for fast atom diffraction},}\ }\href
  {https://doi.org/https://doi.org/10.1016/j.nimb.2010.11.050} {\bibfield
  {journal} {\bibinfo  {journal} {NIM-B}\ }\textbf {\bibinfo {volume} {269}},\
  \bibinfo {pages} {1212--1215} (\bibinfo {year} {2011})}\BibitemShut {NoStop}%
\bibitem [{\citenamefont {Debiossac}\ and\ \citenamefont
  {Roncin}(2014)}]{Debiossac_PRA_2014}%
  \BibitemOpen
  \bibfield  {author} {\bibinfo {author} {\bibfnamefont {M.}~\bibnamefont
  {Debiossac}}\ and\ \bibinfo {author} {\bibfnamefont {P.}~\bibnamefont
  {Roncin}},\ }\bibfield  {title} {\enquote {\bibinfo {title} {Atomic
  diffraction under oblique incidence: An analytical expression},}\ }\href
  {https://doi.org/10.1103/PhysRevA.90.054701} {\bibfield  {journal} {\bibinfo
  {journal} {Phys. Rev. A}\ }\textbf {\bibinfo {volume} {90}},\ \bibinfo
  {pages} {054701} (\bibinfo {year} {2014})}\BibitemShut {NoStop}%
\bibitem [{\citenamefont {Debiossac}(2014)}]{debiossac_these}%
  \BibitemOpen
  \bibfield  {author} {\bibinfo {author} {\bibfnamefont {M.}~\bibnamefont
  {Debiossac}},\ }\emph {\bibinfo {title} {Diffraction d’atomes rapides sur
  surfaces : des résonances de piégeage à la dynamique de croissance par
  épitaxie}},\ \href {http://www.theses.fr/2014PA112384/document} {Ph.D.
  thesis},\ \bibinfo  {school} {Université Paris Sud 11} (\bibinfo {year}
  {2014}),\ \bibinfo {note} {thèse dirigée par P.Roncin}\BibitemShut
  {NoStop}%
\bibitem [{\citenamefont {Igel}, \citenamefont {Pfandzelter},\ and\
  \citenamefont {Winter}(1996)}]{Igel_1996}%
  \BibitemOpen
  \bibfield  {author} {\bibinfo {author} {\bibfnamefont {T.}~\bibnamefont
  {Igel}}, \bibinfo {author} {\bibfnamefont {R.}~\bibnamefont {Pfandzelter}},\
  and\ \bibinfo {author} {\bibfnamefont {H.}~\bibnamefont {Winter}},\
  }\bibfield  {title} {\enquote {\bibinfo {title} {Intensity oscillations in
  grazing scattering of fast {H}e$^+$ ions during heteroepitaxial growth of
  {Cr} on {F}e(100)},}\ }\href {https://doi.org/10.1209/epl/i1996-00532-1}
  {\bibfield  {journal} {\bibinfo  {journal} {Europhysics Letters}\ }\textbf
  {\bibinfo {volume} {35}},\ \bibinfo {pages} {67--72} (\bibinfo {year}
  {1996})}\BibitemShut {NoStop}%
\bibitem [{\citenamefont {Bernhard}\ and\ \citenamefont
  {Winter}(2005)}]{Bernhard_2005}%
  \BibitemOpen
  \bibfield  {author} {\bibinfo {author} {\bibfnamefont {T.}~\bibnamefont
  {Bernhard}}\ and\ \bibinfo {author} {\bibfnamefont {H.}~\bibnamefont
  {Winter}},\ }\bibfield  {title} {\enquote {\bibinfo {title} {Monitoring
  growth of ultrathin films via ion-induced electron emission},}\ }\href
  {https://doi.org/10.1103/PhysRevB.71.241407} {\bibfield  {journal} {\bibinfo
  {journal} {Phys. Rev. B}\ }\textbf {\bibinfo {volume} {71}},\ \bibinfo
  {pages} {241407} (\bibinfo {year} {2005})}\BibitemShut {NoStop}%
\bibitem [{\citenamefont {Lalmi}\ \emph {et~al.}(2012)\citenamefont {Lalmi},
  \citenamefont {Khemliche}, \citenamefont {Momeni}, \citenamefont {Soulisse},\
  and\ \citenamefont {Roncin}}]{Lalmi_2012}%
  \BibitemOpen
  \bibfield  {author} {\bibinfo {author} {\bibfnamefont {B.}~\bibnamefont
  {Lalmi}}, \bibinfo {author} {\bibfnamefont {H.}~\bibnamefont {Khemliche}},
  \bibinfo {author} {\bibfnamefont {A.}~\bibnamefont {Momeni}}, \bibinfo
  {author} {\bibfnamefont {P.}~\bibnamefont {Soulisse}},\ and\ \bibinfo
  {author} {\bibfnamefont {P.}~\bibnamefont {Roncin}},\ }\bibfield  {title}
  {\enquote {\bibinfo {title} {High resolution imaging of superficial mosaicity
  in single crystals using grazing incidence fast atom diffraction},}\ }\href
  {https://doi.org/10.1088/0953-8984/24/44/442002} {\bibfield  {journal}
  {\bibinfo  {journal} {J. Condens. Matter Phys.}\ }\textbf {\bibinfo {volume}
  {24}},\ \bibinfo {pages} {442002} (\bibinfo {year} {2012})}\BibitemShut
  {NoStop}%
\bibitem [{\citenamefont {Hansen}\ \emph {et~al.}(2004)\citenamefont {Hansen},
  \citenamefont {Polop}, \citenamefont {Michely}, \citenamefont {Friedrich},\
  and\ \citenamefont {Urbassek}}]{hansen_2004}%
  \BibitemOpen
  \bibfield  {author} {\bibinfo {author} {\bibfnamefont {H.}~\bibnamefont
  {Hansen}}, \bibinfo {author} {\bibfnamefont {C.}~\bibnamefont {Polop}},
  \bibinfo {author} {\bibfnamefont {T.}~\bibnamefont {Michely}}, \bibinfo
  {author} {\bibfnamefont {A.}~\bibnamefont {Friedrich}},\ and\ \bibinfo
  {author} {\bibfnamefont {H.~M.}\ \bibnamefont {Urbassek}},\ }\bibfield
  {title} {\enquote {\bibinfo {title} {Step edge sputtering yield at grazing
  incidence ion bombardment},}\ }\href
  {https://doi.org/10.1103/PhysRevLett.92.246106} {\bibfield  {journal}
  {\bibinfo  {journal} {Phys. Rev. Lett.}\ }\textbf {\bibinfo {volume} {92}},\
  \bibinfo {pages} {246106} (\bibinfo {year} {2004})}\BibitemShut {NoStop}%
\bibitem [{\citenamefont {Debiossac}\ \emph {et~al.}(2017)\citenamefont
  {Debiossac}, \citenamefont {Atkinson}, \citenamefont {Zugarramurdi},
  \citenamefont {Eddrief} \emph {et~al.}}]{Debiossac_2017}%
  \BibitemOpen
  \bibfield  {author} {\bibinfo {author} {\bibfnamefont {M.}~\bibnamefont
  {Debiossac}}, \bibinfo {author} {\bibfnamefont {P.}~\bibnamefont {Atkinson}},
  \bibinfo {author} {\bibfnamefont {A.}~\bibnamefont {Zugarramurdi}}, \bibinfo
  {author} {\bibfnamefont {M.}~\bibnamefont {Eddrief}}, \emph {et~al.},\
  }\bibfield  {title} {\enquote {\bibinfo {title} {Fast atom diffraction inside
  a molecular beam epitaxy chamber, a rich combination},}\ }\href
  {https://doi.org/https://doi.org/10.1016/j.apsusc.2016.02.157} {\bibfield
  {journal} {\bibinfo  {journal} {Appl. Surf. Sci.}\ }\textbf {\bibinfo
  {volume} {391}},\ \bibinfo {pages} {53} (\bibinfo {year} {2017})}\BibitemShut
  {NoStop}%
\bibitem [{\citenamefont {Pfandzelter}, \citenamefont {Bernhard},\ and\
  \citenamefont {Winter}(2003)}]{Pfandzelter_2003}%
  \BibitemOpen
  \bibfield  {author} {\bibinfo {author} {\bibfnamefont {R.}~\bibnamefont
  {Pfandzelter}}, \bibinfo {author} {\bibfnamefont {T.}~\bibnamefont
  {Bernhard}},\ and\ \bibinfo {author} {\bibfnamefont {H.}~\bibnamefont
  {Winter}},\ }\bibfield  {title} {\enquote {\bibinfo {title} {Ion beam
  triangulation of ultrathin mn and comn films grown on {C}u(001)},}\ }\href
  {https://doi.org/10.1103/PhysRevLett.90.036102} {\bibfield  {journal}
  {\bibinfo  {journal} {Phys. Rev. Lett.}\ }\textbf {\bibinfo {volume} {90}},\
  \bibinfo {pages} {036102} (\bibinfo {year} {2003})}\BibitemShut {NoStop}%
\bibitem [{\citenamefont {Roncin}\ \emph
  {et~al.}(1999{\natexlab{a}})\citenamefont {Roncin}, \citenamefont {Barat},
  \citenamefont {Atanas}, \citenamefont {Villette},\ and\ \citenamefont
  {Morozov}}]{Roncin_1999a}%
  \BibitemOpen
  \bibfield  {author} {\bibinfo {author} {\bibfnamefont {P.}~\bibnamefont
  {Roncin}}, \bibinfo {author} {\bibfnamefont {M.}~\bibnamefont {Barat}},
  \bibinfo {author} {\bibfnamefont {J.~P.}\ \bibnamefont {Atanas}}, \bibinfo
  {author} {\bibfnamefont {J.}~\bibnamefont {Villette}},\ and\ \bibinfo
  {author} {\bibfnamefont {V.}~\bibnamefont {Morozov}},\ }\bibfield  {title}
  {\enquote {\bibinfo {title} {Energy loss and secondary electron measurements
  in collisions at grazing angle of 10 {keV} {O}$^{q+}$ ions on {LiF}
  surface},}\ }\href {https://doi.org/10.1238/physica.topical.080a00231}
  {\bibfield  {journal} {\bibinfo  {journal} {Physica Scripta}\ }\textbf
  {\bibinfo {volume} {T80}},\ \bibinfo {pages} {231} (\bibinfo {year}
  {1999}{\natexlab{a}})}\BibitemShut {NoStop}%
\bibitem [{Qua()}]{Quantar}%
  \BibitemOpen
  \href {http://www.quantar.com/} {\emph {\bibinfo {title} {Quantar
  technology\textregistered Inc, Santa Cruz CA.}}}\BibitemShut {Stop}%
\bibitem [{\citenamefont {Roncin}, \citenamefont {Laurent},\ and\ \citenamefont
  {Barat}(1986)}]{roncin_1986}%
  \BibitemOpen
  \bibfield  {author} {\bibinfo {author} {\bibfnamefont {P.}~\bibnamefont
  {Roncin}}, \bibinfo {author} {\bibfnamefont {H.}~\bibnamefont {Laurent}},\
  and\ \bibinfo {author} {\bibfnamefont {M.}~\bibnamefont {Barat}},\ }\bibfield
   {title} {\enquote {\bibinfo {title} {An electrostatic spectrometer for a
  simultaneous analysis of energy and scattering angle},}\ }\href
  {https://doi.org/10.1088/0022-3735/19/1/005} {\bibfield  {journal} {\bibinfo
  {journal} {J. Phys. E: Sci. Instr.}\ }\textbf {\bibinfo {volume} {19}},\
  \bibinfo {pages} {37--40} (\bibinfo {year} {1986})}\BibitemShut {NoStop}%
\bibitem [{\citenamefont {Roncin}\ \emph {et~al.}(1987)\citenamefont {Roncin},
  \citenamefont {Gaboriaud}, \citenamefont {Barat},\ and\ \citenamefont
  {Laurent}}]{Roncin_1987}%
  \BibitemOpen
  \bibfield  {author} {\bibinfo {author} {\bibfnamefont {P.}~\bibnamefont
  {Roncin}}, \bibinfo {author} {\bibfnamefont {M.~N.}\ \bibnamefont
  {Gaboriaud}}, \bibinfo {author} {\bibfnamefont {M.}~\bibnamefont {Barat}},\
  and\ \bibinfo {author} {\bibfnamefont {H.}~\bibnamefont {Laurent}},\
  }\bibfield  {title} {\enquote {\bibinfo {title} {Transfer ionization in
  collisions involving multiply charged ions at low {keV} energy},}\ }\href
  {https://doi.org/10.1209/0295-5075/3/1/009} {\bibfield  {journal} {\bibinfo
  {journal} {Europhysics Letters ({EPL})}\ }\textbf {\bibinfo {volume} {3}},\
  \bibinfo {pages} {53--59} (\bibinfo {year} {1987})}\BibitemShut {NoStop}%
\bibitem [{\citenamefont {Flechard}\ \emph {et~al.}(1997)\citenamefont
  {Flechard}, \citenamefont {Duponchel}, \citenamefont {Adoui}, \citenamefont
  {Cassimi} \emph {et~al.}}]{Flechard_1997}%
  \BibitemOpen
  \bibfield  {author} {\bibinfo {author} {\bibfnamefont {X.}~\bibnamefont
  {Flechard}}, \bibinfo {author} {\bibfnamefont {S.}~\bibnamefont {Duponchel}},
  \bibinfo {author} {\bibfnamefont {L.}~\bibnamefont {Adoui}}, \bibinfo
  {author} {\bibfnamefont {A.}~\bibnamefont {Cassimi}}, \emph {et~al.},\
  }\bibfield  {title} {\enquote {\bibinfo {title} {State-selective
  double-electron capture in low-velocity collisions studied by recoil-ion
  momentum spectroscopy},}\ }\href
  {https://doi.org/10.1088/0953-4075/30/16/008} {\bibfield  {journal} {\bibinfo
   {journal} {J. Phys. B: At. Mol. Opt. Phys.}\ }\textbf {\bibinfo {volume}
  {30}},\ \bibinfo {pages} {3697} (\bibinfo {year} {1997})}\BibitemShut
  {NoStop}%
\bibitem [{\citenamefont {Auth}\ \emph {et~al.}(1998)\citenamefont {Auth},
  \citenamefont {Mertens}, \citenamefont {Winter},\ and\ \citenamefont
  {Borisov}}]{Auth_1998}%
  \BibitemOpen
  \bibfield  {author} {\bibinfo {author} {\bibfnamefont {C.}~\bibnamefont
  {Auth}}, \bibinfo {author} {\bibfnamefont {A.}~\bibnamefont {Mertens}},
  \bibinfo {author} {\bibfnamefont {H.}~\bibnamefont {Winter}},\ and\ \bibinfo
  {author} {\bibfnamefont {A.}~\bibnamefont {Borisov}},\ }\bibfield  {title}
  {\enquote {\bibinfo {title} {Threshold in the stopping of slow protons
  scattered from the surface of a wide-band-gap insulator},}\ }\href
  {https://doi.org/10.1103/PhysRevLett.81.4831} {\bibfield  {journal} {\bibinfo
   {journal} {Phys. Rev. Lett.}\ }\textbf {\bibinfo {volume} {81}},\ \bibinfo
  {pages} {4831--4834} (\bibinfo {year} {1998})}\BibitemShut {NoStop}%
\bibitem [{\citenamefont {Khemliche}\ \emph {et~al.}(2001)\citenamefont
  {Khemliche}, \citenamefont {Villette}, \citenamefont {Borisov}, \citenamefont
  {Momeni},\ and\ \citenamefont {Roncin}}]{khemliche_2001}%
  \BibitemOpen
  \bibfield  {author} {\bibinfo {author} {\bibfnamefont {H.}~\bibnamefont
  {Khemliche}}, \bibinfo {author} {\bibfnamefont {J.}~\bibnamefont {Villette}},
  \bibinfo {author} {\bibfnamefont {A.~G.}\ \bibnamefont {Borisov}}, \bibinfo
  {author} {\bibfnamefont {A.}~\bibnamefont {Momeni}},\ and\ \bibinfo {author}
  {\bibfnamefont {P.}~\bibnamefont {Roncin}},\ }\bibfield  {title} {\enquote
  {\bibinfo {title} {Electron bihole complex formation in neutralization of
  ${\mathrm{ne}}^{+}$ on lif(001)},}\ }\href
  {https://doi.org/10.1103/PhysRevLett.86.5699} {\bibfield  {journal} {\bibinfo
   {journal} {Phys. Rev. Lett.}\ }\textbf {\bibinfo {volume} {86}},\ \bibinfo
  {pages} {5699--5702} (\bibinfo {year} {2001})}\BibitemShut {NoStop}%
\bibitem [{\citenamefont {Roncin}\ \emph {et~al.}(2002)\citenamefont {Roncin},
  \citenamefont {Borisov}, \citenamefont {Khemliche}, \citenamefont {Momeni},
  \citenamefont {Mertens},\ and\ \citenamefont {Winter}}]{roncin_2002}%
  \BibitemOpen
  \bibfield  {author} {\bibinfo {author} {\bibfnamefont {P.}~\bibnamefont
  {Roncin}}, \bibinfo {author} {\bibfnamefont {A.~G.}\ \bibnamefont {Borisov}},
  \bibinfo {author} {\bibfnamefont {H.}~\bibnamefont {Khemliche}}, \bibinfo
  {author} {\bibfnamefont {A.}~\bibnamefont {Momeni}}, \bibinfo {author}
  {\bibfnamefont {A.}~\bibnamefont {Mertens}},\ and\ \bibinfo {author}
  {\bibfnamefont {H.}~\bibnamefont {Winter}},\ }\bibfield  {title} {\enquote
  {\bibinfo {title} {Evidence for ${F}^{\ensuremath{-}}$ formation by
  simultaneous double-electron capture during scattering of ${F}^{+}$ from a
  lif(001) surface},}\ }\href {https://doi.org/10.1103/PhysRevLett.89.043201}
  {\bibfield  {journal} {\bibinfo  {journal} {Phys. Rev. Lett.}\ }\textbf
  {\bibinfo {volume} {89}},\ \bibinfo {pages} {043201} (\bibinfo {year}
  {2002})}\BibitemShut {NoStop}%
\bibitem [{\citenamefont {Roncin}\ \emph
  {et~al.}(1999{\natexlab{b}})\citenamefont {Roncin}, \citenamefont {Villette},
  \citenamefont {Atanas},\ and\ \citenamefont {Khemliche}}]{Roncin_1999}%
  \BibitemOpen
  \bibfield  {author} {\bibinfo {author} {\bibfnamefont {P.}~\bibnamefont
  {Roncin}}, \bibinfo {author} {\bibfnamefont {J.}~\bibnamefont {Villette}},
  \bibinfo {author} {\bibfnamefont {J.~P.}\ \bibnamefont {Atanas}},\ and\
  \bibinfo {author} {\bibfnamefont {H.}~\bibnamefont {Khemliche}},\ }\bibfield
  {title} {\enquote {\bibinfo {title} {Energy loss of low energy protons on
  {LiF}(100): Surface excitation and ${H}^{\ensuremath{-}}$ mediated electron
  emission},}\ }\href {https://doi.org/10.1103/PhysRevLett.83.864} {\bibfield
  {journal} {\bibinfo  {journal} {Phys. Rev. Lett.}\ }\textbf {\bibinfo
  {volume} {83}},\ \bibinfo {pages} {864--867} (\bibinfo {year}
  {1999}{\natexlab{b}})}\BibitemShut {NoStop}%
\bibitem [{\citenamefont {Grizzi}\ \emph {et~al.}(1990)\citenamefont {Grizzi},
  \citenamefont {Shi}, \citenamefont {Bu},\ and\ \citenamefont
  {Rabalais}}]{Grizzi_1990}%
  \BibitemOpen
  \bibfield  {author} {\bibinfo {author} {\bibfnamefont {O.}~\bibnamefont
  {Grizzi}}, \bibinfo {author} {\bibfnamefont {M.}~\bibnamefont {Shi}},
  \bibinfo {author} {\bibfnamefont {H.}~\bibnamefont {Bu}},\ and\ \bibinfo
  {author} {\bibfnamefont {J.~W.}\ \bibnamefont {Rabalais}},\ }\bibfield
  {title} {\enquote {\bibinfo {title} {Time‐of‐flight scattering and
  recoiling spectrometer (tof‐sars) for surface analysis},}\ }\href
  {https://doi.org/10.1063/1.1141488} {\bibfield  {journal} {\bibinfo
  {journal} {Rev. Sci. Instrum.}\ }\textbf {\bibinfo {volume} {61}},\ \bibinfo
  {pages} {740--752} (\bibinfo {year} {1990})}\BibitemShut {NoStop}%
\bibitem [{\citenamefont {Cushman}\ \emph {et~al.}(2016)\citenamefont
  {Cushman}, \citenamefont {Brüner}, \citenamefont {Zakel}, \citenamefont
  {Major}, \citenamefont {Lunt}, \citenamefont {Smith}, \citenamefont {Grehl},\
  and\ \citenamefont {Linford}}]{Cushman_2016}%
  \BibitemOpen
  \bibfield  {author} {\bibinfo {author} {\bibfnamefont {C.~V.}\ \bibnamefont
  {Cushman}}, \bibinfo {author} {\bibfnamefont {P.}~\bibnamefont {Brüner}},
  \bibinfo {author} {\bibfnamefont {J.}~\bibnamefont {Zakel}}, \bibinfo
  {author} {\bibfnamefont {G.~H.}\ \bibnamefont {Major}}, \bibinfo {author}
  {\bibfnamefont {B.~M.}\ \bibnamefont {Lunt}}, \bibinfo {author}
  {\bibfnamefont {N.~J.}\ \bibnamefont {Smith}}, \bibinfo {author}
  {\bibfnamefont {T.}~\bibnamefont {Grehl}},\ and\ \bibinfo {author}
  {\bibfnamefont {M.~R.}\ \bibnamefont {Linford}},\ }\bibfield  {title}
  {\enquote {\bibinfo {title} {Low energy ion scattering (leis). a practical
  introduction to its theory{,} instrumentation{,} and applications},}\ }\href
  {https://doi.org/10.1039/C6AY00765A} {\bibfield  {journal} {\bibinfo
  {journal} {Anal. Methods}\ }\textbf {\bibinfo {volume} {8}},\ \bibinfo
  {pages} {3419--3439} (\bibinfo {year} {2016})}\BibitemShut {NoStop}%
\bibitem [{\citenamefont {Winter}\ \emph {et~al.}(2002)\citenamefont {Winter},
  \citenamefont {Mertens}, \citenamefont {Pfandzelter},\ and\ \citenamefont
  {Staemmler}}]{Winter_2002}%
  \BibitemOpen
  \bibfield  {author} {\bibinfo {author} {\bibfnamefont {H.}~\bibnamefont
  {Winter}}, \bibinfo {author} {\bibfnamefont {A.}~\bibnamefont {Mertens}},
  \bibinfo {author} {\bibfnamefont {R.}~\bibnamefont {Pfandzelter}},\ and\
  \bibinfo {author} {\bibfnamefont {V.}~\bibnamefont {Staemmler}},\ }\bibfield
  {title} {\enquote {\bibinfo {title} {Energy transfer of ke{V} {N}e atoms to
  the lattice of a {LiF} (001) surface under channeling},}\ }\href
  {https://doi.org/10.1103/PhysRevA.66.022902} {\bibfield  {journal} {\bibinfo
  {journal} {Phys. Rev. A}\ }\textbf {\bibinfo {volume} {66}},\ \bibinfo
  {pages} {022902} (\bibinfo {year} {2002})}\BibitemShut {NoStop}%
\bibitem [{\citenamefont {Villette}(2000)}]{Villette_these}%
  \BibitemOpen
  \bibfield  {author} {\bibinfo {author} {\bibfnamefont {J.}~\bibnamefont
  {Villette}},\ }\emph {\bibinfo {title} {Etude exp\'{e}rimentale de
  l'int\'{e}raction rasante d'atomes et d'ions sur des surfaces isolantes}},\
  \href {https://tel.archives-ouvertes.fr/tel-00106816/document} {Ph.D.
  thesis},\ \bibinfo  {school} {Universit\'{e} Paris sud-XI} (\bibinfo {year}
  {2000}),\ \bibinfo {note} {thèse dirigée par P. Roncin}\BibitemShut
  {NoStop}%
\bibitem [{\citenamefont {Pfandzelter}\ \emph {et~al.}(1998)\citenamefont
  {Pfandzelter}, \citenamefont {Kowalski}, \citenamefont {Igel},\ and\
  \citenamefont {Winter}}]{Pfandzelter_98}%
  \BibitemOpen
  \bibfield  {author} {\bibinfo {author} {\bibfnamefont {R.}~\bibnamefont
  {Pfandzelter}}, \bibinfo {author} {\bibfnamefont {T.}~\bibnamefont
  {Kowalski}}, \bibinfo {author} {\bibfnamefont {T.}~\bibnamefont {Igel}},\
  and\ \bibinfo {author} {\bibfnamefont {H.}~\bibnamefont {Winter}},\
  }\bibfield  {title} {\enquote {\bibinfo {title} {Rainbow scattering at
  surface steps},}\ }\href
  {https://doi.org/https://doi.org/10.1016/S0039-6028(98)00414-2} {\bibfield
  {journal} {\bibinfo  {journal} {Surface Science}\ }\textbf {\bibinfo {volume}
  {411}},\ \bibinfo {pages} {L894--L899} (\bibinfo {year} {1998})}\BibitemShut
  {NoStop}%
\bibitem [{\citenamefont {Morozov}, \citenamefont {Meyer},\ and\ \citenamefont
  {Roncin}(1999)}]{Morozov_1999}%
  \BibitemOpen
  \bibfield  {author} {\bibinfo {author} {\bibfnamefont {V.}~\bibnamefont
  {Morozov}}, \bibinfo {author} {\bibfnamefont {F.}~\bibnamefont {Meyer}},\
  and\ \bibinfo {author} {\bibfnamefont {P.}~\bibnamefont {Roncin}},\
  }\bibfield  {title} {\enquote {\bibinfo {title} {Multi-coincidence studies as
  a technique for the investigation of ion-surface interaction},}\ }\href
  {https://doi.org/10.1238/Physica.Topical.080a00069} {\bibfield  {journal}
  {\bibinfo  {journal} {Physica Scripta}\ }\textbf {\bibinfo {volume} {T80A}},\
  \bibinfo {pages} {69--72} (\bibinfo {year} {1999})}\BibitemShut {NoStop}%
\bibitem [{\citenamefont {Morozov}\ and\ \citenamefont
  {Meyer}(2001)}]{Morozov_2001}%
  \BibitemOpen
  \bibfield  {author} {\bibinfo {author} {\bibfnamefont {V.~A.}\ \bibnamefont
  {Morozov}}\ and\ \bibinfo {author} {\bibfnamefont {F.~W.}\ \bibnamefont
  {Meyer}},\ }\bibfield  {title} {\enquote {\bibinfo {title} {Path-dependent
  neutralization of multiply charged ar ions incident on {A}u(110)},}\ }\href
  {https://doi.org/10.1103/PhysRevLett.86.736} {\bibfield  {journal} {\bibinfo
  {journal} {Phys. Rev. Lett.}\ }\textbf {\bibinfo {volume} {86}},\ \bibinfo
  {pages} {736--739} (\bibinfo {year} {2001})}\BibitemShut {NoStop}%
\bibitem [{\citenamefont {Villette}\ \emph {et~al.}(1999)\citenamefont
  {Villette}, \citenamefont {Atanas}, \citenamefont {Khemliche}, \citenamefont
  {Barat}, \citenamefont {Morosov},\ and\ \citenamefont
  {Roncin}}]{Villette_1999}%
  \BibitemOpen
  \bibfield  {author} {\bibinfo {author} {\bibfnamefont {J.}~\bibnamefont
  {Villette}}, \bibinfo {author} {\bibfnamefont {J.}~\bibnamefont {Atanas}},
  \bibinfo {author} {\bibfnamefont {H.}~\bibnamefont {Khemliche}}, \bibinfo
  {author} {\bibfnamefont {M.}~\bibnamefont {Barat}}, \bibinfo {author}
  {\bibfnamefont {V.}~\bibnamefont {Morosov}},\ and\ \bibinfo {author}
  {\bibfnamefont {P.}~\bibnamefont {Roncin}},\ }\bibfield  {title} {\enquote
  {\bibinfo {title} {Grazing collision of ke{V} protons on {LiF} correlation
  between energy loss and electron emission},}\ }\href
  {https://doi.org/https://doi.org/10.1016/S0168-583X(99)00434-6} {\bibfield
  {journal} {\bibinfo  {journal} {NIM-B}\ }\textbf {\bibinfo {volume} {157}},\
  \bibinfo {pages} {92 -- 97} (\bibinfo {year} {1999})}\BibitemShut {NoStop}%
\bibitem [{\citenamefont {Rousseau}(2007)}]{Rousseau_these}%
  \BibitemOpen
  \bibfield  {author} {\bibinfo {author} {\bibfnamefont {P.}~\bibnamefont
  {Rousseau}},\ }\emph {\bibinfo {title} {Collisions rasantes d'ions ou
  d'atomes sur les surfaces : de l'échange de charge à la diffraction
  atomique}},\ \href {http://www.theses.fr/2006PA112140/document} {Ph.D.
  thesis},\ \bibinfo  {school} {Univ. Paris Sud 11} (\bibinfo {year} {2007}),\
  \bibinfo {note} {dirigée par P. Roncin}\BibitemShut {NoStop}%
\bibitem [{\citenamefont {Rubiano}\ \emph {et~al.}(2013)\citenamefont
  {Rubiano}, \citenamefont {Bocan}, \citenamefont {Gravielle}, \citenamefont
  {Bundaleski}, \citenamefont {Khemliche},\ and\ \citenamefont
  {Roncin}}]{rubiano_2013}%
  \BibitemOpen
  \bibfield  {author} {\bibinfo {author} {\bibfnamefont {C.~R.}\ \bibnamefont
  {Rubiano}}, \bibinfo {author} {\bibfnamefont {G.~A.}\ \bibnamefont {Bocan}},
  \bibinfo {author} {\bibfnamefont {M.~S.}\ \bibnamefont {Gravielle}}, \bibinfo
  {author} {\bibfnamefont {N.}~\bibnamefont {Bundaleski}}, \bibinfo {author}
  {\bibfnamefont {H.}~\bibnamefont {Khemliche}},\ and\ \bibinfo {author}
  {\bibfnamefont {P.}~\bibnamefont {Roncin}},\ }\bibfield  {title} {\enquote
  {\bibinfo {title} {Ab initio potential for the {He-Ag}(110) interaction
  investigated using grazing-incidence fast-atom diffraction},}\ }\href
  {https://doi.org/10.1103/PhysRevA.87.012903} {\bibfield  {journal} {\bibinfo
  {journal} {Phys. Rev. A}\ }\textbf {\bibinfo {volume} {87}},\ \bibinfo
  {pages} {012903} (\bibinfo {year} {2013})}\BibitemShut {NoStop}%
\bibitem [{\citenamefont {Seifert}\ \emph {et~al.}(2014)\citenamefont
  {Seifert}, \citenamefont {Busch}, \citenamefont {Meyer},\ and\ \citenamefont
  {Winter}}]{Seifert_2014}%
  \BibitemOpen
  \bibfield  {author} {\bibinfo {author} {\bibfnamefont {J.}~\bibnamefont
  {Seifert}}, \bibinfo {author} {\bibfnamefont {M.}~\bibnamefont {Busch}},
  \bibinfo {author} {\bibfnamefont {E.}~\bibnamefont {Meyer}},\ and\ \bibinfo
  {author} {\bibfnamefont {H.}~\bibnamefont {Winter}},\ }\bibfield  {title}
  {\enquote {\bibinfo {title} {Surface structure of alanine on {C}u(110) via
  grazing scattering of fast atoms and molecules},}\ }\href
  {https://doi.org/10.1103/PhysRevB.89.075404} {\bibfield  {journal} {\bibinfo
  {journal} {Phys. Rev. B}\ }\textbf {\bibinfo {volume} {89}},\ \bibinfo
  {pages} {075404} (\bibinfo {year} {2014})}\BibitemShut {NoStop}%
\bibitem [{\citenamefont {Bobrov}, \citenamefont {Kalashnyk},\ and\
  \citenamefont {Guillemot}(2015)}]{Bobrov_2015}%
  \BibitemOpen
  \bibfield  {author} {\bibinfo {author} {\bibfnamefont {K.}~\bibnamefont
  {Bobrov}}, \bibinfo {author} {\bibfnamefont {N.}~\bibnamefont {Kalashnyk}},\
  and\ \bibinfo {author} {\bibfnamefont {L.}~\bibnamefont {Guillemot}},\
  }\bibfield  {title} {\enquote {\bibinfo {title} {True perylene epitaxy on
  {Ag}(110) driven by site recognition effect},}\ }\href
  {https://doi.org/10.1063/1.4913325} {\bibfield  {journal} {\bibinfo
  {journal} {The Journal of Chemical Physics}\ }\textbf {\bibinfo {volume}
  {142}},\ \bibinfo {pages} {101929} (\bibinfo {year} {2015})}\BibitemShut
  {NoStop}%
\bibitem [{\citenamefont {Salomon}\ \emph {et~al.}(2021)\citenamefont
  {Salomon}, \citenamefont {Minissale}, \citenamefont {Lairado}, \citenamefont
  {Coussan}, \citenamefont {Rousselot-Pailley}, \citenamefont {Dulieu},\ and\
  \citenamefont {Angot}}]{Salomon_2021}%
  \BibitemOpen
  \bibfield  {author} {\bibinfo {author} {\bibfnamefont {E.}~\bibnamefont
  {Salomon}}, \bibinfo {author} {\bibfnamefont {M.}~\bibnamefont {Minissale}},
  \bibinfo {author} {\bibfnamefont {F.}~\bibnamefont {Lairado}}, \bibinfo
  {author} {\bibfnamefont {S.}~\bibnamefont {Coussan}}, \bibinfo {author}
  {\bibfnamefont {P.}~\bibnamefont {Rousselot-Pailley}}, \bibinfo {author}
  {\bibfnamefont {F.}~\bibnamefont {Dulieu}},\ and\ \bibinfo {author}
  {\bibfnamefont {T.}~\bibnamefont {Angot}},\ }\bibfield  {title} {\enquote
  {\bibinfo {title} {Pyrene adsorption on a {Ag}(111) surface},}\ }\href
  {https://doi.org/10.1021/acs.jpcc.1c01350} {\bibfield  {journal} {\bibinfo
  {journal} {The Journal of Physical Chemistry C}\ }\textbf {\bibinfo {volume}
  {125}},\ \bibinfo {pages} {11166--11174} (\bibinfo {year}
  {2021})}\BibitemShut {NoStop}%
\bibitem [{\citenamefont {Sereno}\ \emph {et~al.}(2016)\citenamefont {Sereno},
  \citenamefont {Lupone}, \citenamefont {Debiossac}, \citenamefont
  {Kalashnyk},\ and\ \citenamefont {Roncin}}]{Sereno_2016}%
  \BibitemOpen
  \bibfield  {author} {\bibinfo {author} {\bibfnamefont {M.}~\bibnamefont
  {Sereno}}, \bibinfo {author} {\bibfnamefont {S.}~\bibnamefont {Lupone}},
  \bibinfo {author} {\bibfnamefont {M.}~\bibnamefont {Debiossac}}, \bibinfo
  {author} {\bibfnamefont {N.}~\bibnamefont {Kalashnyk}},\ and\ \bibinfo
  {author} {\bibfnamefont {P.}~\bibnamefont {Roncin}},\ }\bibfield  {title}
  {\enquote {\bibinfo {title} {Active correction of the tilt angle of the
  surface plane with respect to the rotation axis during azimuthal scan},}\
  }\href {https://doi.org/https://doi.org/10.1016/j.nimb.2016.05.001}
  {\bibfield  {journal} {\bibinfo  {journal} {NIM-B}\ }\textbf {\bibinfo
  {volume} {382}},\ \bibinfo {pages} {123 -- 126} (\bibinfo {year}
  {2016})}\BibitemShut {NoStop}%
\bibitem [{\citenamefont {Gruber}\ \emph {et~al.}(2016)\citenamefont {Gruber},
  \citenamefont {Wilhelm}, \citenamefont {P{\'e}tuya}, \citenamefont {Smejkal},
  \citenamefont {Kozubek}, \citenamefont {Hierzenberger}, \citenamefont {Bayer}
  \emph {et~al.}}]{gruber_2016}%
  \BibitemOpen
  \bibfield  {author} {\bibinfo {author} {\bibfnamefont {E.}~\bibnamefont
  {Gruber}}, \bibinfo {author} {\bibfnamefont {R.}~\bibnamefont {Wilhelm}},
  \bibinfo {author} {\bibfnamefont {R.}~\bibnamefont {P{\'e}tuya}}, \bibinfo
  {author} {\bibfnamefont {V.}~\bibnamefont {Smejkal}}, \bibinfo {author}
  {\bibfnamefont {R.}~\bibnamefont {Kozubek}}, \bibinfo {author} {\bibfnamefont
  {A.}~\bibnamefont {Hierzenberger}}, \bibinfo {author} {\bibfnamefont
  {B.}~\bibnamefont {Bayer}}, \emph {et~al.},\ }\bibfield  {title} {\enquote
  {\bibinfo {title} {Ultrafast electronic response of graphene to a strong and
  localized electric field},}\ }\href {https://doi.org/10.1038/ncomms13948}
  {\bibfield  {journal} {\bibinfo  {journal} {Nat. Commun.}\ }\textbf {\bibinfo
  {volume} {7}},\ \bibinfo {pages} {1--7} (\bibinfo {year} {2016})}\BibitemShut
  {NoStop}%
\bibitem [{\citenamefont {Brand}\ \emph {et~al.}(2019)\citenamefont {Brand},
  \citenamefont {Debiossac}, \citenamefont {Susi}, \citenamefont {Aguillon},
  \citenamefont {Kotakoski}, \citenamefont {Roncin},\ and\ \citenamefont
  {Arndt}}]{Brand_2019}%
  \BibitemOpen
  \bibfield  {author} {\bibinfo {author} {\bibfnamefont {C.}~\bibnamefont
  {Brand}}, \bibinfo {author} {\bibfnamefont {M.}~\bibnamefont {Debiossac}},
  \bibinfo {author} {\bibfnamefont {T.}~\bibnamefont {Susi}}, \bibinfo {author}
  {\bibfnamefont {F.}~\bibnamefont {Aguillon}}, \bibinfo {author}
  {\bibfnamefont {J.}~\bibnamefont {Kotakoski}}, \bibinfo {author}
  {\bibfnamefont {P.}~\bibnamefont {Roncin}},\ and\ \bibinfo {author}
  {\bibfnamefont {M.}~\bibnamefont {Arndt}},\ }\bibfield  {title} {\enquote
  {\bibinfo {title} {Coherent diffraction of hydrogen through the 246 pm
  lattice of graphene},}\ }\href {https://doi.org/10.1088/1367-2630/ab05ed}
  {\bibfield  {journal} {\bibinfo  {journal} {New Journal of Physics}\ }\textbf
  {\bibinfo {volume} {21}},\ \bibinfo {pages} {033004} (\bibinfo {year}
  {2019})}\BibitemShut {NoStop}%
\bibitem [{\citenamefont {Ford}\ and\ \citenamefont
  {Wheeler}(1959)}]{Ford_1959}%
  \BibitemOpen
  \bibfield  {author} {\bibinfo {author} {\bibfnamefont {K.}~\bibnamefont
  {Ford}}\ and\ \bibinfo {author} {\bibfnamefont {J.}~\bibnamefont {Wheeler}},\
  }\bibfield  {title} {\enquote {\bibinfo {title} {Semiclassical description of
  scattering},}\ }\href
  {https://doi.org/https://doi.org/10.1016/0003-4916(59)90026-0} {\bibfield
  {journal} {\bibinfo  {journal} {Annals of Physics}\ }\textbf {\bibinfo
  {volume} {7}},\ \bibinfo {pages} {259--286} (\bibinfo {year}
  {1959})}\BibitemShut {NoStop}%
\bibitem [{\citenamefont {Roncin}(2020)}]{Roncin_2020}%
  \BibitemOpen
  \bibfield  {author} {\bibinfo {author} {\bibfnamefont {P.}~\bibnamefont
  {Roncin}},\ }\bibfield  {title} {\enquote {\bibinfo {title} {Revisiting
  atomic collisions physics with highly charged ions, a tribute to michel
  barat},}\ }\href {https://doi.org/10.1088/1361-6455/abaaf9} {\bibfield
  {journal} {\bibinfo  {journal} {J. Phys. B: At., Mol. and Opt. Physics}\
  }\textbf {\bibinfo {volume} {53}},\ \bibinfo {pages} {202001} (\bibinfo
  {year} {2020})}\BibitemShut {NoStop}%
\bibitem [{\citenamefont {Sch\"uller}\ and\ \citenamefont
  {Winter}(2008)}]{Schueller_2008}%
  \BibitemOpen
  \bibfield  {author} {\bibinfo {author} {\bibfnamefont {A.}~\bibnamefont
  {Sch\"uller}}\ and\ \bibinfo {author} {\bibfnamefont {H.}~\bibnamefont
  {Winter}},\ }\bibfield  {title} {\enquote {\bibinfo {title} {Supernumerary
  rainbows in the angular distribution of scattered projectiles for grazing
  collisions of fast atoms with a {LiF}(001) surface},}\ }\href
  {https://doi.org/10.1103/PhysRevLett.100.097602} {\bibfield  {journal}
  {\bibinfo  {journal} {Phys. Rev. Lett.}\ }\textbf {\bibinfo {volume} {100}},\
  \bibinfo {pages} {097602} (\bibinfo {year} {2008})}\BibitemShut {NoStop}%
\bibitem [{\citenamefont {Jalabert}, \citenamefont {Vickridge},\ and\
  \citenamefont {Chabli}(2017)}]{Jalabert_2017}%
  \BibitemOpen
  \bibfield  {author} {\bibinfo {author} {\bibfnamefont {D.}~\bibnamefont
  {Jalabert}}, \bibinfo {author} {\bibfnamefont {I.}~\bibnamefont
  {Vickridge}},\ and\ \bibinfo {author} {\bibfnamefont {A.}~\bibnamefont
  {Chabli}},\ }\href
  {https://www.wiley.com/en-us/Swift+Ion+Beam+Analysis+in+Nanosciences-p-9781119008675}
  {\emph {\bibinfo {title} {Swift Ion Beam Analysis in Nanosciences}}}\
  (\bibinfo  {publisher} {John Wiley \& Sons},\ \bibinfo {year}
  {2017})\BibitemShut {NoStop}%
\end{thebibliography}%

\end{document}